\documentclass[12pt]{article}

\usepackage{amssymb,amsfonts,amsthm}
\usepackage{amsmath}
\usepackage{bm}             
\usepackage[mathscr]{eucal}
\usepackage{cancel}
\usepackage{wasysym}

\usepackage{graphicx}
\usepackage{subfigure}
\def\be{\begin{equation}}
\def\ee{\end{equation}}
\def\bea{\begin{eqnarray}}
\def\eea{\end{eqnarray}}

\def\hsp5{\hspace{5mm}}

\theoremstyle{remark}

\newcommand{\sfrac}[2]{{\textstyle{#1\over#2}}}

\title{\sc Quintessential $\alpha$-attractor inflation:\\
A dynamical systems analysis}

\begin{document}

\author{
\sc Artur Alho,$^{1}$\thanks{Electronic address:{\tt
aalho@math.ist.utl.pt}}\,, Claes Uggla,$^{2}$\thanks{Electronic address:{\tt
claes.uggla@kau.se}}\\
$^{1}${\small\em Center for Mathematical Analysis, Geometry and Dynamical Systems,}\\
{\small\em Instituto Superior T\'ecnico, Universidade de Lisboa,}\\
{\small\em Av. Rovisco Pais, 1049-001 Lisboa, Portugal.}\\
$^{2}${\small\em Department of Physics, Karlstad University,}\\
{\small\em S-65188 Karlstad, Sweden.}}


\date{}
\maketitle

\begin{abstract}

The equations for quintessential $\alpha$-attractor inflation
with a single scalar field, radiation and matter in a spatially flat
FLRW spacetime are recast into a regular dynamical system on a
compact state space. This enables a complete description of the
solution space of these models. The inflationary attractor solution
is shown to correspond to the unstable center manifold of a de
Sitter fixed point, and we describe connections between slow-roll
and dynamical systems approximations for this solution, including
Pad\'e approximants. We also introduce a new method for
systematically obtaining initial data for quintessence evolution by
using dynamical systems properties; in particular, this method exploits
that there exists a radiation dominated line of fixed points with
an unstable quintessence attractor submanifold, which plays a role
that is reminiscent of that of the inflationary attractor solution
for inflation.

\end{abstract}

\newpage

\section{Introduction\label{sec:intro}}

It is commonly believed that the observable Universe is almost spatially
homogeneous, isotropic and flat on sufficiently large spatial scales.
The most popular explanation for this is that the Universe has had
an early brief period of accelerated expansion, inflation. 
Surprisingly, in 1998 observations of type Ia supernovae suggested that
the expansion of the Universe is once more
accelerating~\cite{rieetal98,peretal99}.
The simplest explanation for these latter observations, which
have gained further support from observations of the
cosmic microwave background 
and the large-scale structure 
of the Universe~\cite{planck18,plaX18,alametal17}, seems to be a constant
energy density $\Lambda$, although this `dark energy' is tiny when
compared to the vacuum energy during the inflationary epoch.
Moreover, apart from $\Lambda$, the matter content today is believed
to be dominated by a `cold dark matter' (CDM) component, which leads
to $\Lambda$CDM cosmology. This paradigm, especially when combined with
inflation, yields a remarkably consistent description
of a growing number of increasingly precise
observations~\cite{planck18,plaX18,alametal17}. There are, however,
tensions between some data sets, which is a cause for
concern~\cite{AbbDES20,rieetal19}, but it remains to be seen
if this eventually will pose a serious threat to
the standard scenario.

The resemblance between the two accelerating epochs
tantalizingly suggests that they may be connected:
Can the present dark energy be a remnant from the inflationary epoch?
This prompted Peebles and Vilenkin~\cite{peevil99} in 1998 to
propose that the `constant' energy densities should be replaced with
a unifying dynamical description, quintessential inflation, {\it i.e.},
a scalar field with a potential driving the two periods of
accelerated expansion --- an inflaton field ending as
`dark energy' quintessence.

Scalar field potentials are often presented with more or less
\emph{ad hoc} motivations, as is illustrated by the many examples
of potentials in inflationary cosmology~\cite{maretal14}, but
it is desirable to have some theoretical basis for them.
An example of this are the $\alpha$-attractor models
arising from string theory motivated
supergravity~\cite{galetal15,kallin15b}. In this framework
the underlying hyperbolic geometry of the moduli space and the
flatness of the K{\"a}hler potential in the inflation direction is
described by a single scalar field $\phi$, which leads
to the following phenomenological Lagrangian density:
\begin{equation}\label{phiaction}
\mathcal{L} = \sqrt{-g}\left(\frac12 R -
\frac12\frac{\partial_\mu\phi\partial^\mu\phi}{\left(1 - \frac{\phi^2}{6\alpha}\right)^2} - V(\phi)\right)
+ \mathcal{L}_\mathrm{matter},
\end{equation}
where $g$ is the determinant of the space-time metric $g_{\mu\nu}$, $R$ is the
associated Ricci scalar, and $\phi\in (-\sqrt{6\alpha},\sqrt{6\alpha})$
is a scalar field with a potential $V(\phi)$, while $\mathcal{L}_\mathrm{matter}$
denotes the matter Lagrangian density.\footnote{We use reduced Planck units: $c=1$,
$\hbar = 1$, $8\pi G = M_\mathrm{Pl}^{-2} = 1$, where
$M_\mathrm{Pl} = m_\mathrm{Pl}/\sqrt{8\pi}$ is the reduced Planck mass.}
The positive parameter $\alpha$ can take any value, but extended supergravity,
M-theory, and string theory suggest that $3\alpha = 1, 2, 3, 4, 5, 6, 7$ are
preferred~\cite{ferkal16,kaletal17,kaletal17b}. The transformation
\begin{equation}\label{varphidef}
\phi = \sqrt{6\alpha}\tanh\frac{\varphi}{\sqrt{6\alpha}}
\end{equation}
yields a canonical normalized scalar field $\varphi$ and the
scalar field Lagrangian density
\begin{equation}\label{varphiaction}
\mathcal{L}_\varphi = -\sqrt{-g}\left[\frac12(\partial\varphi)^2 +
V\left(\sqrt{6\alpha}\tanh\frac{\varphi}{\sqrt{6\alpha}}\right)\right],
\end{equation}
where $\phi$ and $\varphi$ coincide in the limit $\phi\rightarrow 0$,
while a tiny vicinity of the moduli space boundary $|\phi| = \sqrt{6\alpha}$
becomes stretched to large $|\varphi|$ extending to infinity.

The $\alpha$-attractor models obtain their name from that the
slow-roll approximation yields certain universal properties in $n_s-r$ diagrams, which result in the universal $\alpha$-attractor prediction (see {\it e.g.}, \cite{akretal20})
\begin{equation}\label{nrRel}
r = 3\alpha(1-n_s)^2,
\end{equation}
where $n_s$ is the spectral index of primordial scalar curvature perturbations and
$r$ is the primordial tensor-to-scalar ratio.
Although originally designed for inflation, the $\alpha$-attractor models
have lately been adapted to describe quintessential
inflation~\cite{dimowe17,dimetal18,akretal18,akretal20},
where stretching near the moduli space boundary yields an inflationary
potential plateau and a lower quintessential plateaux or exponential tail.
This leads to quite intriguing results, but there still remain many challenges.
For example, the underlying physics is far from being
understood: there is no unambiguous physical understanding of
reheating after inflation,\footnote{Also, reheating conditions are
different for traditional and quintessential inflation.
In standard inflation reheating occurs after inflation when the
field oscillates in a potential minimum, yielding an average
scalar field equation of state $w_\varphi \approx 0$. For quintessential
$\alpha$-attractor inflation the scalar field potential is
monotonic and has no minimum, instead the scalar field goes
through a `kinaton' phase at the end of inflation where $w_\varphi\approx 1$.}
and no one has detected a quintessential inflaton particle.
There are also several fine-tuning issues: fine tuning of (i) model
parameters, (ii) initial data for a given model,
(iii) possible couplings of the scalar field to other
fields (although stretching near the inflationary moduli space
boundary provides some predictive robustness).\footnote{For
discussions about couplings to other fields, which become
exponentially small for large fields $|\varphi|$,
see~\cite{kaletal14,kallin3b,galetal15,caretal15a,caretal15b,lin15,lin17,lin17b},
where, in addition, connections with the Starobinski model~\cite{Sta80}
in the Einstein frame~\cite{BC88} and the Higgs inflation
models~\cite{BS08} are pointed out and where other $\alpha$-attractor
aspects are also considered.}

Moreover, heuristic approximations and arguments are
used extensively in scalar field cosmology, sometimes
supported by numerical explorations that not always are as
systematic as one might want, thereby clouding
issues such as observational viability and fine-tuning of model
parameters and initial data. To address such topics in a systematic
manner, and to provide clarity and rigour, we have embarked on a program
studying models in a Friedmann-Lema\^{i}tre-Robertson-Walker
(FLRW) spacetime background, and perturbations thereof, from
a dynamical systems perspective. This entails formulating
the equations for a hierarchy of models, see~\cite{alhetal19b},
as useful dynamical systems, {\it i.e.}, as dynamical systems that
make it possible to apply powerful local and global dynamical
systems methods, and systematic quantitative numerical
investigations~\cite{alhetal19b,alhetal15,alhugg15b,alhugg17,alhetal19a,alhetal23}.

Dynamical systems and dynamical systems methods were introduced in
cosmology in 1971 by Collins~\cite{collins71} who treated 2-dimensional
dynamical systems while Bogoyavlensky and Novikov (1973) \cite{BN73}
used dynamical systems techniques for higher dimensional dynamical
systems in relativistic cosmology. This early work has subsequently
been followed up and extended by many researchers,
see {\it e.g.}~\cite{waiell97,col03,bahetal18} and references therein.
In the present paper we recast the field equations for the present
models into a \emph{regular} dynamical system on a \emph{compact state space}.
For brevity, we restrict considerations to the Einstein and matter
equations for a spatially flat spacetime; perturbations
will be considered elsewhere. We will present figures that illustrate
key features of the solution space, which is accomplished
by numerically computing solutions based on initial data obtained
systematically from dynamical systems properties.
We will also show that the `kinaton' epoch corresponds to solutions
that are transiently shadowing an invariant kinaton boundary subset
in the dynamical systems state space. Furthermore, we show that
the inflationary attractor solution corresponds to the unstable center
manifold of a de Sitter fixed point in this framework, and we also
describe connections between slow-roll and dynamical systems
approximations, including Pad\'e approximants, for this center manifold.

The paper is structured as follows. In section~\ref{dynsys} we
describe the quintessential $\alpha$-attractor inflation models
(with further asymptotic details about the quintessential $\alpha$-attractor
potentials given in Appendix~\ref{app:asymp}) and
derive a new dynamical system, which provides a new context for these models.
Section~\ref{sec:gendynfeatures} identifies general dynamical system features,
including an integral and monotonic functions that show that all solution trajectories
originate and end at fixed points on certain boundaries of the dynamical system,
which leads to a description of the entire solution space of these models.
Section~\ref{sec:inflation} describes the solution space on the scalar field
boundary, on which the radiation and matter content is zero,
but focus on the inflationary attractor solution and
connections between slow-roll based approximations and center manifold
approximations for this solution. Section~\ref{sec:freezing} connects the end
of inflation with the freezing of the scalar field and derives a new
improved approximation for the freezing value of
the scalar field, obtained by dynamical systems considerations.
Section~\ref{sec:quintevol} introduces a new systematic method for
obtaining suitable initial data for quintessence evolution, where
formulas for comparisons with $\Lambda$CDM and radiation
are found in Appendix~\ref{app:LCDMrad}. The paper concludes in
section~\ref{sec:disc} with a discussion about some of the results
and future developments.

\section{Dynamical systems formulation}\label{dynsys}

In this section we derive a dynamical system suitable for
quintessential $\alpha$-attractor inflation, but first we
characterize what the latter entails.

\subsection{Asymmetric cosmological $\alpha$-attractors\label{subsec:asym}}

In agreement with~\cite{peevil99,dimowe17,dimetal18},
we consider a monotonically \emph{decreasing} scalar field potential
$V(\phi)>0$ in~\eqref{phiaction} where $V(\phi)$ and its derivatives
are non-singular when $\phi \in [-\sqrt{6\alpha},\sqrt{6\alpha}]$, enabling
Taylor series expansions at $\phi= \pm \sqrt{6\alpha}$ of $V(\phi)$:
%
%
%
\begin{equation}\label{phiseries}
V(\phi) = V_\pm + (\phi \mp \sqrt{6\alpha})V_\pm^\prime
+{\cal O}\left((\phi \mp \sqrt{6\alpha})^2\right),
\end{equation}
where
%
%
%
\begin{equation}\label{Vpmprimedef}
V_\pm := \left. V(\phi)\right|_{\phi=\pm\sqrt{6\alpha}}, \qquad
V_\pm^\prime := \left. \partial_\phi V\right|_{\phi=\pm\sqrt{6\alpha}}.
\end{equation}
Near $\phi=\pm\sqrt{6\alpha}$, and thereby when $\varphi\rightarrow \pm \infty$,
the variable $\phi$ in~\eqref{varphidef} can be expanded in
$\exp(\mp 2\varphi/\sqrt{6\alpha})$ as follows:
%
%
%
\begin{equation}\label{phivarphi}
\phi = \pm\sqrt{6\alpha}\left[1 - 2e^{\mp 2\varphi/\sqrt{6\alpha}} +
{\cal O}\left(e^{\mp 4\varphi/\sqrt{6\alpha}}\right)\right],
\end{equation}
which leads to that potential 
takes the following asymptotic form in $\varphi$:
%
%
%
%
\begin{equation}\label{exptail}
V = V_\pm \mp 2\sqrt{6\alpha}V_\pm^\prime e^{\mp 2\varphi/\sqrt{6\alpha}}
+ {\cal O}\left(e^{\mp 4\varphi/\sqrt{6\alpha}}\right).
\end{equation}
%
%
%
%
%

In our dynamical systems treatment we will use a bounded scalar field variable as one of our variables.
The variable $\phi$ would do for our purposes, but there is an advantage to have a boundary that does
not depend on any parameters. For this reason we scale $\phi$ and introduce the following scalar
field variable:
\begin{equation}
\bar{\varphi} := \frac{\phi}{\sqrt{6\alpha}} = \tanh\frac{\varphi}{\sqrt{6\alpha}}.\label{barvarphidef}
\end{equation}
As we will see, it is not the potential that enters our dynamical system but
$\lambda(\bar{\varphi})$, which is defined by
\begin{equation}
\lambda(\bar{\varphi}) := - \frac{V_{,\varphi}}{V} =
-\frac{1}{\sqrt{6\alpha}}(1 - \bar{\varphi}^2)\left(\frac{V_{,\bar{\varphi}}}{V}\right).\label{lambdadef1}
\end{equation}
%
%
%
In agreement with the above discussion we require that (with a slight
abuse of notation) $V(\bar{\varphi})=V(\varphi(\bar{\varphi}))>0$,
$\lambda(\bar{\varphi})>0$ (since the potential is monotonically decreasing)
when $\bar{\varphi} \in (-1,1)$, and that
$\lambda(\bar{\varphi})$ and its derivatives are non-singular when
$\bar{\varphi} \in [-1,1]$. Since asymptotics turn out to be associated
with the $\bar{\varphi}=\pm 1$ boundaries, it is of special interest to perform Taylor
expansions of $\lambda(\bar{\varphi})$ at
$\bar{\varphi}=\pm 1$, which can be expressed in the quantities
\begin{equation}
\lambda_\pm^{(n)} = \partial_{\bar{\varphi}}^n\lambda|_{\bar{\varphi} = \pm 1},
\end{equation}
where we for simplicity write 
$\lambda_\pm^{(0)} = \lambda_\pm$.

Quintessential $\alpha$-attractor inflation models have an inflationary plateaux
\begin{equation}\label{plateaux}
V_- > 0, \qquad \lambda_- = 0
\end{equation}
when $\varphi \rightarrow - \infty$
(recall that $V_\pm = \left. V(\phi)\right|_{\phi=\pm\sqrt{6\alpha}}$, which implies
that $V_\pm = \left. V(\bar{\varphi})\right|_{\bar{\varphi}=\pm 1}$)
and an asymptotic constant or exponential tail when $\varphi \rightarrow + \infty$,
which leads to
\begin{subequations}\label{futureasymp}
\begin{alignat}{3}
0 < &\,\, V_+\ll V_-, &\qquad \lambda_+ &=0 &\qquad &(\text{asymptotic constant}, V_+),\label{asympconst}\\
V_+ &=0, &\qquad \lambda_+ &= 2/\sqrt{6\alpha} &\qquad &(\text{exponential tail}, V \propto e^{-\lambda_+\varphi}).\label{asymptail}
\end{alignat}
\end{subequations}

Since the inflationary plateaux
value $V_-$ is a common feature of all quintessential $\alpha$-attractor
inflationary models, it is natural to define the dimensionless
inflationary plateaux-normalized potential
\begin{equation}\label{barV}
\bar{V}(\bar{\varphi}) = \frac{V(\bar{\varphi})}{V_-}.
\end{equation}
As specific examples of quintessential $\alpha$-attractor inflationary
potentials, consider the following monotonically decreasing
potentials~\cite{dimowe17,dimetal18,akretal18,akretal20}:
\begin{subequations}\label{pot2exp}
\begin{align}
\text{LC}\!:\quad \bar{V} &= \frac{1 + \nu\left(1 - \frac{\phi}{\sqrt{6\alpha}}\right)}{1 + 2\nu} =
\frac{1 + \nu\left(1 - \tanh\frac{\varphi}{\sqrt{6\alpha}}\right)}{1 + 2\nu},\\
\text{LT}\!:\quad \bar{V} &= \frac12\left(1 - \frac{\phi}{\sqrt{6\alpha}}\right) =
\frac12\left(1 - \tanh\frac{\varphi}{\sqrt{6\alpha}}\right),\\
\text{EC}\!:\quad \bar{V} &= e^{-\nu(1 + \frac{\phi}{\sqrt{6\alpha}})} = e^{-\nu(1 + \tanh\frac{\varphi}{\sqrt{6\alpha}})},\\
\text{ET}\!:\quad \bar{V} &= \frac{e^{-\nu(1 + \frac{\phi}{\sqrt{6\alpha}})} - e^{-2\nu}}{1 - e^{-2\nu}} =
\frac{e^{-\nu(1 + \tanh\frac{\varphi}{\sqrt{6\alpha}})} - e^{-2\nu}}{1 - e^{-2\nu}}.
\end{align}
\end{subequations}
The LC, EC, ET models are characterized by the 3 parameters
$V_->0$, $\alpha >0$ and $\nu >0$, while LT has 2 parameters, $V_->0$
and $\alpha >0$. The nomenclature L, E in LC, LT, EC, ET stands for
a potential that is linear respectively exponential in $\phi$ and $\bar{\varphi}$;
C represents a potential that is asymptotically a positive quintessential constant,
while T denotes a potential with an exponential quintessential tail in
$\varphi$, when $\varphi \rightarrow \infty$.\footnote{Further details about
the asymptotic features of the potentials of these models, used later in the
paper, are found in Appendix~\ref{app:asymp}.} In~\cite{akretal18,akretal20}
the EC and ET models where referred to as Exp-model I and Exp-model II,
respectively.\footnote{In contrast
to~\cite{akretal18,akretal20}, which use a monotonically increasing
scalar field potential, we choose the potential to be monotonically
decreasing, {\it i.e.}, a comparison entails
$(\phi,\varphi) \rightarrow -(\phi, \varphi)$. In addition, the relation
between the various constants in~\cite{akretal18,akretal20} and those used here
are given by $\Lambda = V_+ = V_-/(1+2\nu)$, $\gamma = V_-\nu/\sqrt{6\alpha}$
for the LC case, while $\gamma=V_-/(2\sqrt{6\alpha})$ for the LT models;
$\gamma=\nu$, $M^2 = V_-$ for the EC case, while $\gamma=\nu$,
$M^2=V_-/(1 - e^{-2\nu})$ for the ET models.}

\begin{figure}[ht!]
	\begin{center}
		\subfigure[$\bar{V}(\varphi)$ for an $\text{EC}$ model.]{\label{}
			\includegraphics[width=0.3\textwidth]{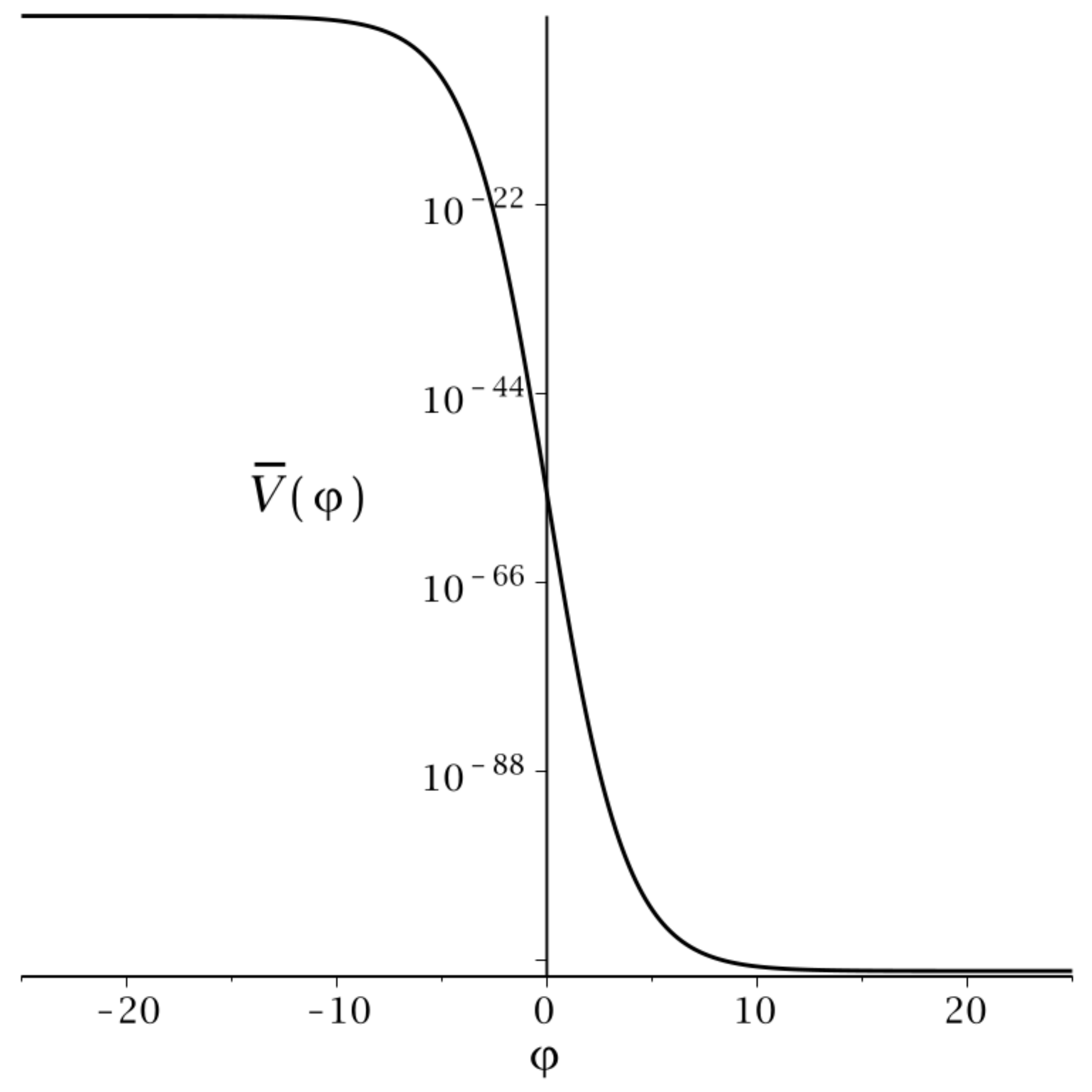}}
		\hspace{0.2cm}
		\subfigure[$\bar{V}(\varphi)$ for an $\text{ET}$ model.]{\label{}
			\includegraphics[width=0.3\textwidth]{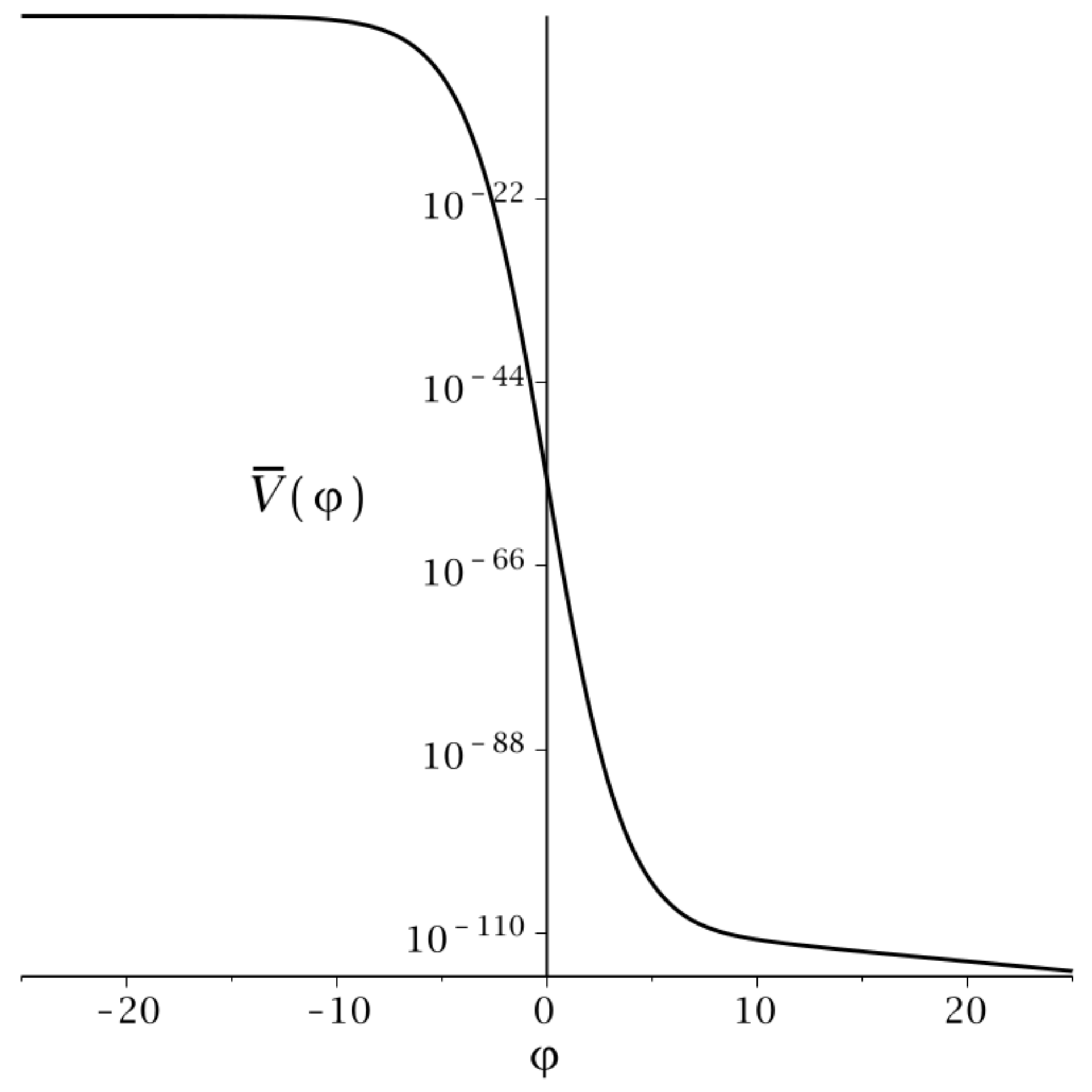}}
				\hspace{0.2cm}
				\subfigure[$\bar{\varphi}(\varphi)$]{\label{}
			\includegraphics[width=0.3\textwidth]{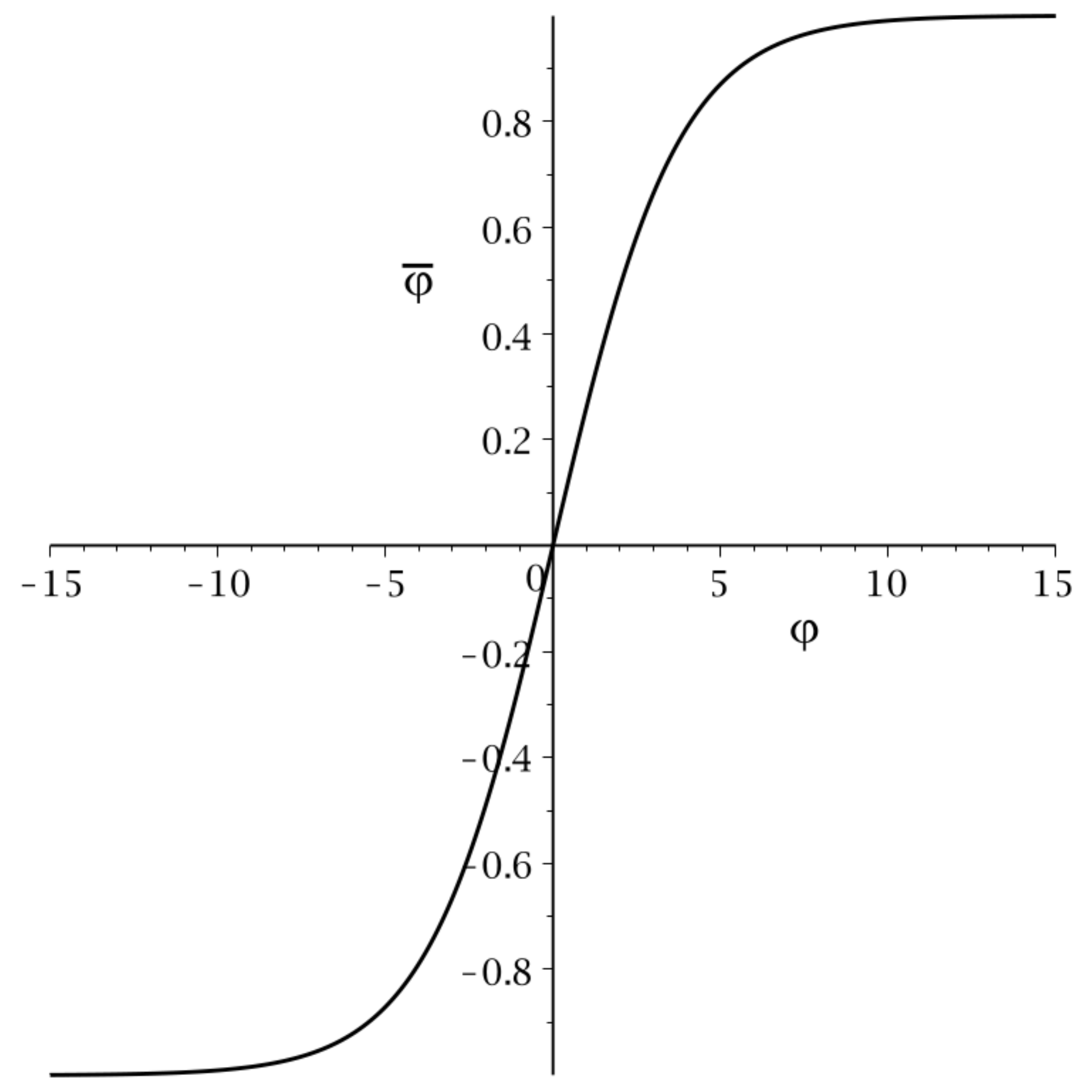}}
		\vspace{-0.5cm}
	\end{center}
	\caption{Graphs $\bar{V}(\varphi) = V(\varphi)/V_-$ for the $\text{EC}$ and $\text{ET}$ models
     and a graph $\bar{\varphi}(\varphi)$, with $\alpha = \frac73$ and $\nu = 128$. Note that these
     figures illustrate that the potential plateaus occur when $|\bar{\varphi}|$ is very close to one.}
	\label{Fig_VBar}
\end{figure}

\begin{figure}[ht!]
	\begin{center}
		\subfigure[$\lambda(\bar{\varphi})$: $\text{EC}$ models.]{\label{EClambda}
			\includegraphics[width=0.3\textwidth]{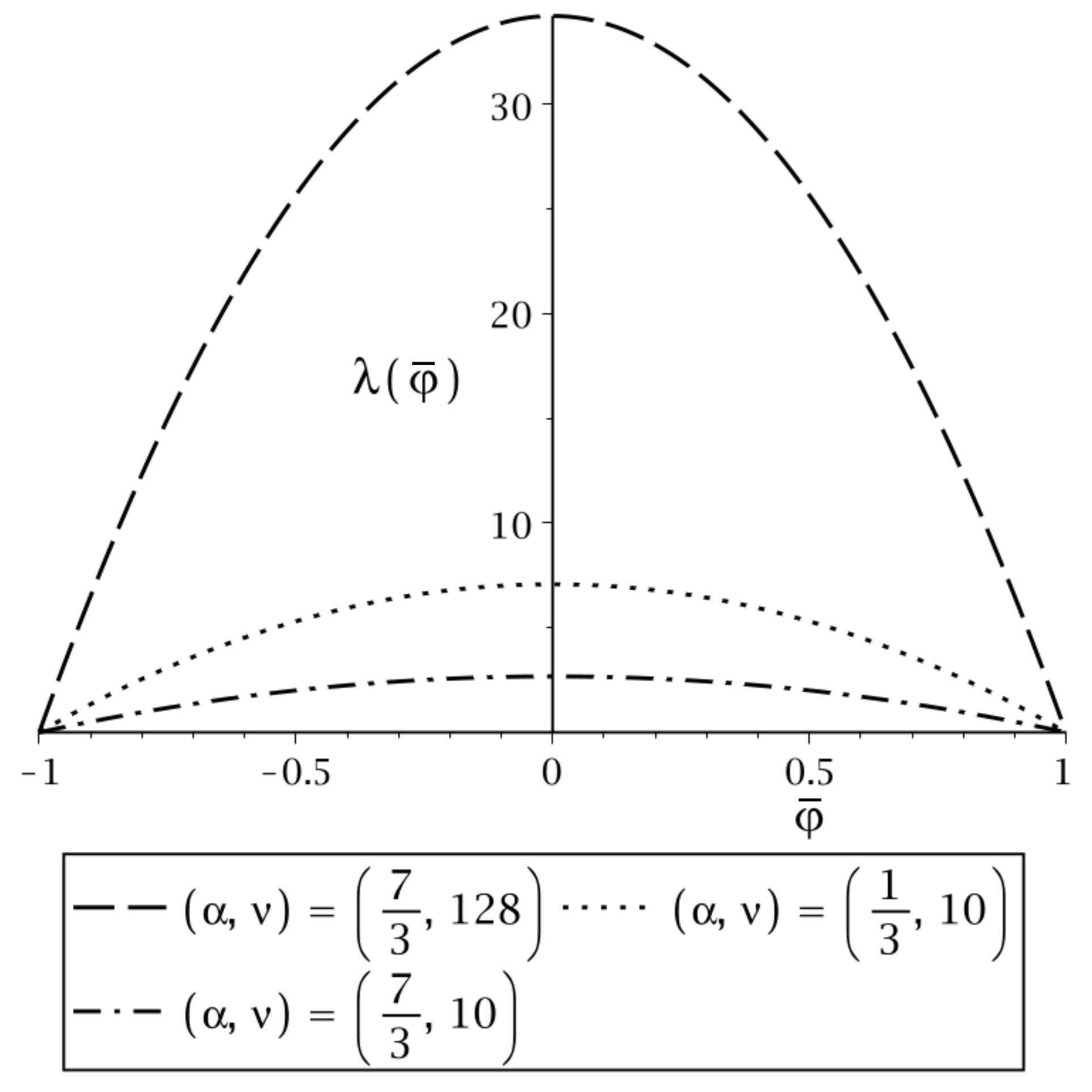}}
		\hspace{1.5cm}
		\subfigure[$\lambda(\bar{\varphi})$: $\text{ET}$ models.]{\label{ETlambda}
			\includegraphics[width=0.3\textwidth]{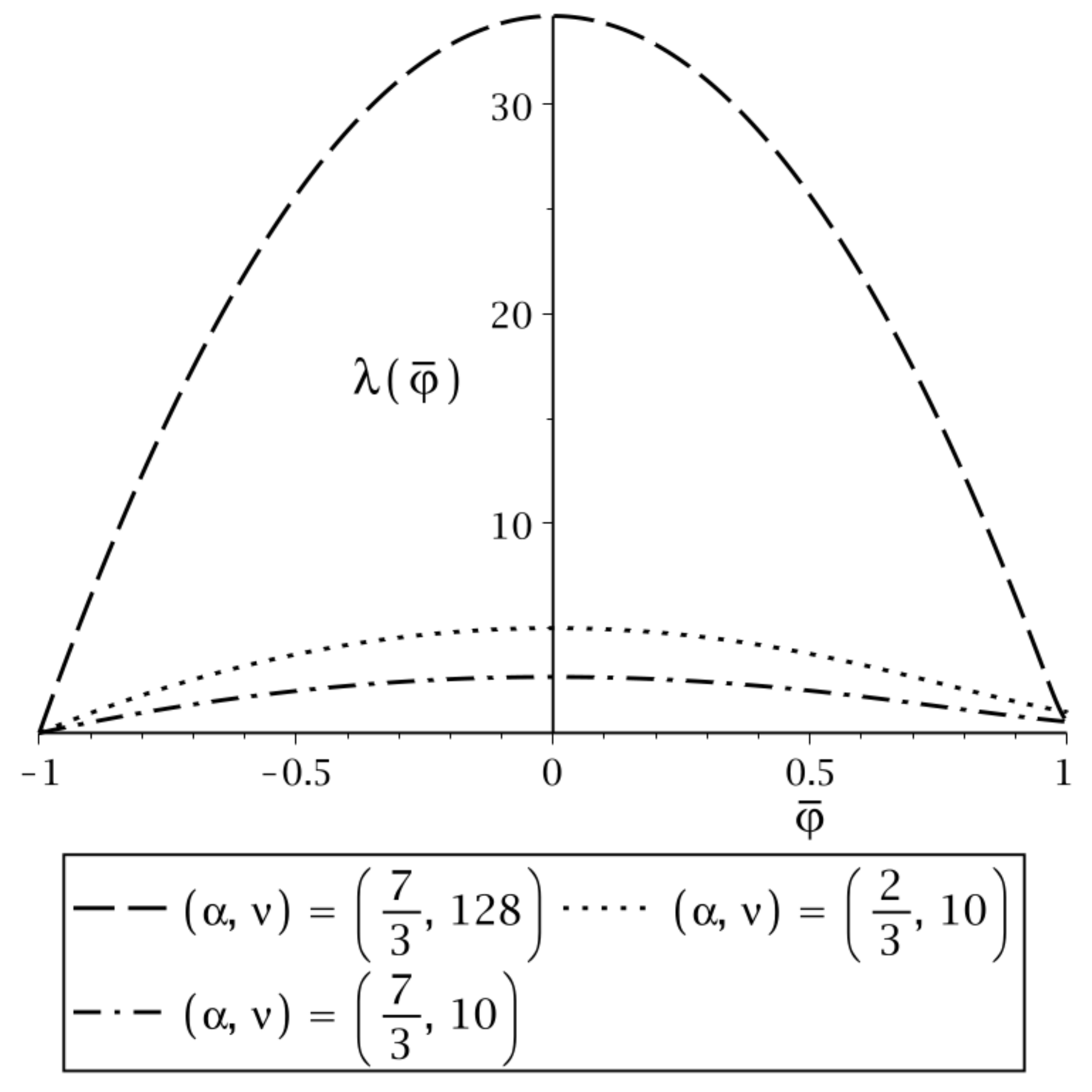}}
	\vspace{-0.5cm}
	\end{center}
	\caption{
            Figures (a) and (b) contain graphs $\lambda(\bar{\varphi})$ for the
            $\text{EC}$ and $\text{ET}$ models for a variety of values of $\nu$ and $\alpha$. The
            difference between the $\text{EC}$ and $\text{ET}$ models for $\bar{\varphi}\leq 0$ is negligible for the present
            values of $\nu$ since $e^{-\nu}\ll 1$, where $\lambda(\bar{\varphi})$ for $\bar{\varphi}\leq 0$
            therefore is governed by $\nu/\sqrt{6\alpha}$; distinguishable differences only occur when
            $\bar{\varphi}>0$, where $\lambda_+=0$ for the $\text{EC}$ models while
            $\lambda_+ = 2/\sqrt{6\alpha}$ in the $\text{ET}$ case.}
	\label{Fig_Vlambda_barVarphi}
\end{figure}

For brevity we will focus on the EC and ET models in the main text,
where $\bar{V}(\varphi)$ is illustrated for these models, as is $\bar{\varphi}(\varphi)$,
in Figure~\ref{Fig_VBar}.
Expressed in $\bar{\varphi}$, the $V_-$-normalized potential $\bar{V}(\bar{\varphi})$, $V_+$, and
the associated $\lambda(\bar{\varphi}) = -V_{,\varphi}/V = -\bar{V}_{,\varphi}/\bar{V}$,
for the EC and ET examples are given by:
%
%
%
\begin{subequations}\label{barvexamplesECET}
\begin{alignat}{3}
\text{EC}\!:\quad \bar{V} &= e^{-\nu(1 + \bar{\varphi})}, &\qquad
V_+ &= e^{-2\nu}V_-, &\qquad
\lambda &= \frac{\nu}{\sqrt{6\alpha}}\left(1 - \bar{\varphi}^2\right), \label{lambdaECmain}\\
\text{ET}\!:\quad \bar{V} &= \frac{e^{-\nu(1 + \bar{\varphi})} - e^{-2\nu}}{1 - e^{-2\nu}}, &\qquad
V_+ &= 0, &\qquad \lambda &= \frac{\nu}{\sqrt{6\alpha}}
\left(\frac{1 - \bar{\varphi}^2}{1 - e^{-\nu(1 - \bar{\varphi})}}\right),\label{lambdaETmain}
\end{alignat}
\end{subequations}
Note that, in contrast to $V(\phi)$ and $V(\varphi)$, $\bar{V}(\bar{\varphi})$
is independent of $\alpha$. The function $\lambda(\bar{\varphi})$ for a variety
of EC and ET models is illustrated in Figure~\ref{Fig_Vlambda_barVarphi}.
%

%

Both $V({\bar\varphi})$ and $\lambda({\bar\varphi})$
are differentiable (even analytic) when $\bar{\varphi}\in[-1,1]$, where
$\lambda_-=0$ for the EC and ET models while $\lambda_+=0$ for the EC
models and $\lambda_+=2/\sqrt{6\alpha}$ for the ET models.
%
%
%
%

As argued in~\cite{dimetal18}, $\alpha \sim {\cal O}(1)$ where, as discussed in {\it e.g.}~\cite{akretal18},
maximal supergravity, string theory, and M-theory suggest that $3\alpha = 1, 2, 3, 4, 5, 6, 7$
are particularly interesting, where the choice $\alpha = 1$ corresponds to the Starobinski model and
Higgs inflation in the Einstein frame. Note also that the recent upper limit of the tensor-to-scalar ratio $r<0.028$ in G. Galloni \emph{et al.}~\cite{galetal23}, in combination with relation~\eqref{nrRel} 
and the value of the spectral index $n_s=0.966$
results in the upper limit $\alpha\lesssim8$.
The large difference between the energy density of the inflationary plateaux and
the present dark energy density leads to a much larger value of $\nu$. Based on arguments
in~\cite{akretal20}, one expects $\nu \sim {\cal O}(100)$. In the case of the EC models we can obtain
a more precise result: As we will see, the Hubble variable $H$ is monotonically decreasing,
while $\lim_{N\rightarrow +\infty}(3H^2) = V_+ = e^{-2\nu}V_-$, from which it follows that
$3H_0^2 > V_+ = e^{-2\nu}V_-$, where $H_0$ is the present value of $H$.
This in turn yields $\nu > \ln\left(\sqrt{\frac{V_-}{3H_0^2}}\right)$.
Using the result in footnote~\ref{H0} for $V_-/3H_0^2$, which
uses $\alpha = \frac73$ (the bound is approximately proportional to $\alpha$),
results in $\nu > 127$. We are henceforth therefore concerned with
\begin{equation}\label{alphanuapprox}
\alpha \sim {\cal O}(1),\qquad \nu \sim {\cal O}(100),
\end{equation}
Note, however, that, $\lambda_+$, as will be shown, must satisfy the inequality
$\lambda_+ = 2/\sqrt{6\alpha} < \sqrt{2}$ to yield eternal future acceleration,
which corresponds to
\begin{equation}
\alpha > \frac13.
\end{equation}
For illustrative purposes we will typically
choose $\alpha = \frac73$ and $\nu = 128$ as specific examples.

\subsection{Field equations and heuristics}\label{subsec:fieldheur}

Consider a spatially flat FLRW spacetime,
\begin{equation}
ds^2 = -dt^2 + a^2(t)\delta_{ij}dx^idx^j,
\end{equation}
where $a(t)$ is the cosmological scale factor, and a source consisting of a
canonically normalized minimally coupled scalar field, $\varphi$, with a
potential $V(\varphi)>0$, and thereby the following
energy density and pressure,
\begin{equation}\label{rhophipphi}
\rho_\varphi = \frac12\dot{\varphi}^2 + V(\varphi),\qquad
p_\varphi = \frac12\dot{\varphi}^2 - V(\varphi),
\end{equation}
and two non-interacting fluids: radiation with $\rho_\gamma>0$,
$p_\gamma= \rho_\gamma/3$, and matter with
$\rho_\mathrm{m}>0$, $p_\mathrm{m}=0$.

The definition of the Hubble variable $H$,
the (Landau–) Raychaudhuri equation, the Gauss/Hamiltonian constraint (often referred to
as the Friedmann equation in FLRW cosmology), the non-linear Klein-Gordon
equation, and the energy conservation laws for radiation and matter,
are given by
\begin{subequations}\label{Mainsysdim}
\begin{align}
\dot{a} &= aH,\label{adotH}\\
\dot{H} + H^2 &= -\frac16(\rho + 3p), \label{Ray}\\
3H^2 &= \rho, \label{Gauss}\\
\ddot{\varphi} &=-3H\dot{\varphi} - V_{,\varphi}, \label{KG}\\
\dot{\rho}_\gamma &= -4H\rho_\gamma,\\
\dot{\rho}_\mathrm{m} &= -3H\rho_\mathrm{m},
\end{align}
\end{subequations}
where an overdot represents the derivative with respect to the cosmic
clock time $t$, and where the total energy
density $\rho$ and pressure $p$ are given by
\begin{equation}
\rho = \rho_\varphi + \rho_\gamma + \rho_\mathrm{m},\qquad
p = p_\varphi + \frac13\rho_\gamma.
\end{equation}
Since $\rho>0$ implies that $H^2>0$ in~\eqref{Gauss}, it follows that
the Hubble variable $H$ satisfies $H>0$ for expanding models.

Thus, both $\rho_\gamma$ and $\rho_\mathrm{m}$ monotonically decrease
and go to zero toward the future while they blow up asymptotically toward the past.
Furthermore, $\rho_\gamma/\rho_\mathrm{m}$
tends to zero toward the future and infinity toward the past, {\it i.e.}, radiation dominates
over matter toward the past while matter dominates over radiation toward the future.
Heuristically, $-3H\dot{\varphi}$ acts as friction in the non-linear Klein-Gordon
equation~\eqref{KG}, which makes $\rho_\varphi$ decrease, as follows from
\begin{equation}\label{dotrhophi}
\dot{\rho}_\varphi = -3H(\rho_\varphi + p_\varphi) = -3H\dot{\varphi}^2.
\end{equation}

Monotonically decreasing inflationary
quintessence potentials, $V_{,\varphi}<0$, with an $\alpha$-attractor
potential plateaux when $\varphi\rightarrow-\infty$,
yield solutions with different asymptotic behaviour in $\varphi$:
\begin{itemize}
\item[(i)] Solutions for which the scalar field originates from
$\varphi\rightarrow + \infty$ and then decreases, {\it i.e.}, $\dot{\varphi}<0$,
while decelerating, since then $\ddot{\varphi}>0$; throughout this motion
the scalar field energy $\rho_\varphi$ is decreasing and hence
the scalar field eventually hits the potential and bounces, since $\dot{\varphi}=0$
implies that $\ddot{\varphi}|_{\dot{\varphi} = 0} = - V_{,\varphi}>0$, which
is followed by that the scalar field increases
until $\varphi\rightarrow\infty$.
\item[(ii)] Solutions for which the scalar field originates from
$\varphi\rightarrow - \infty$  and then forever increases until
$\varphi\rightarrow\infty$. Moreover, in the pure scalar field case with
$\rho_\gamma = \rho_\mathrm{m} = 0$ there exists a single solution,
the inflationary `attractor' solution, for which $\varphi$ is forever
increasing until $\varphi\rightarrow\infty$, with an asymptotic origin
toward the past at the asymptotic inflationary plateaux, where
$\dot{\varphi}=0$ and $\rho_\varphi=V_-$ when $\varphi\rightarrow - \infty$.
\item[(iii)] Solutions
for which the scalar field originates from finite values $\varphi_*$,
which happens when the asymptotic energy density content
is dominated by the radiation energy density toward the past, and then
increases until $\varphi\rightarrow\infty$.
\end{itemize}

Cases (ii) and (iii) may not be heuristically clear, but
the dynamical systems analysis below shows that this description is correct;
moreover, this analysis establishes that cases (i) and (ii) correspond to open sets
of solutions while case (iii) corresponds to a set of co-dimension one of solutions.
We will also show that the limits $\varphi_*$ in case (iii) is
intimately connected with the phenomenon of scalar field `freezing,' which
plays a significant role for cosmologically viable quintessence solutions.

\subsection{Derivation of the dynamical system}

To obtain a useful dynamical system, we first define the following
quantities
\begin{subequations}\label{vardef1}
\begin{alignat}{2}
\Omega_\gamma &:= \frac{\rho_\gamma}{3H^2}, &\qquad\qquad
\Omega_\mathrm{m} &:= \frac{\rho_\mathrm{m}}{3H^2},\label{Omdef}\\
\Omega_\varphi &:= \frac{\rho_\varphi}{3H^2}, &\qquad\qquad
\Omega_V &:= \frac{V}{3H^2},\label{Omphidef}\\
 w_{\gamma\mathrm{m}}&:= \frac{p_\gamma + p_\mathrm{m}}{\rho_\gamma + \rho_\mathrm{m}}
= \frac13\left(\frac{\rho_\gamma}{\rho_\gamma + \rho_\mathrm{m}}\right), &\qquad\qquad
w_\varphi &:= \frac{p_\varphi}{\rho_\varphi}.\label{wdef}
\end{alignat}
\end{subequations}
%
%
%
%
%
We then follow~\cite{alhetal23} and introduce the variables $u$ and $v$
according to
\begin{subequations}
\begin{align}
u &:= \frac{\varphi^\prime}{\sqrt{3\Omega_\varphi}},\label{udef}\\
v &:= \sqrt{\frac{\Omega_\varphi}{3}},\label{vdef}
\end{align}
\end{subequations}
where a ${}^\prime$ henceforth denotes the derivative
with respect to $e$-fold time\footnote{The $e$-fold time derivative
${}^\prime$ should not be confused with the ${}^\prime$ in the
definition of $V_\pm^\prime$ in~\eqref{Vpmprimedef}.}
\begin{equation}\label{Ndef}
N = \ln(a/a_0),
\end{equation}
where $a_0 = a(t_0)$. In agreement with this notation, quantities with
values at the present time, $t=t_0 \Rightarrow N=0$,
are denoted by a subscript ${}_0$. The definition~\eqref{Ndef}
implies that $N\rightarrow - \infty$ and $N\rightarrow + \infty$
when $a\rightarrow 0$ and $a \rightarrow\infty$, respectively.
We then recall from~\eqref{barvarphidef} that we use the following bounded
scalar field variable,
\begin{equation}
\bar{\varphi} := \frac{\phi}{\sqrt{6\alpha}} = \tanh\frac{\varphi}{\sqrt{6\alpha}}.\label{barvarphidef2}
\end{equation}
The final step to obtain our new dynamical
system is to change the time variable from the clock time
$t$ to the $e$-fold time $N$ by using that
\begin{equation}
\frac{d}{dt}=H\frac{d}{dN},\qquad \frac{d^2}{dt^2}=
H^2\left(\frac{d^2}{dN^2} - (1+q)\frac{d}{dN}\right),
\end{equation}
where $q$ is the \emph{deceleration parameter}, defined by
\begin{equation}\label{qdef1}
q := -\frac{a\ddot{a}}{\dot{a}^2} = -1 - \frac{H^\prime}{H}.
\end{equation}
Using the above definition for the state vector $(\bar{\varphi},u,v,w_{\gamma\mathrm{m}})$
and the field equations in~\eqref{Mainsysdim} then leads to:
%
%
%
\begin{subequations}\label{Dynsys}
\begin{align}
\bar{\varphi}^\prime &= \frac{3}{\sqrt{6\alpha}}(1 - \bar{\varphi}^2)uv,\label{barvarphieq}\\
u^\prime &= \frac32\left(2 - u^2\right)\!\left(v\lambda(\bar{\varphi}) - u\right), \label{ueq} \\
v^\prime &= \frac32\left(w_{\gamma\mathrm{m}} + 1 - u^2\right)\!\left(1 - 3v^2\right)v, \label{veq}\\
w_{\gamma\mathrm{m}}^\prime &= -(1 - 3w_{\gamma\mathrm{m}})w_{\gamma\mathrm{m}}, \label{wrmeq}
\end{align}
\end{subequations}
where $\lambda(\bar{\varphi})$ was defined in equation~\eqref{lambdadef1}.
%
%
%
%
The deceleration parameter $q$ is given by the expression
\begin{equation}\label{qdef}
q = -1 + \frac32\left[3u^2v^2 + (w_{\gamma\mathrm{m}} + 1)(1 - 3v^2)\right],
\end{equation}
as follows from the definition~\eqref{qdef1} and the Raychaudhuri
equation~\eqref{Ray}. It is also of interest to define an effective
equation of state parameter for the entire matter content
\begin{equation}
w_\mathrm{eff} := \frac{p}{\rho} = w_{\gamma\mathrm{m}} - 3(w_{\gamma\mathrm{m}} + 1 - u^2)v^2,
\end{equation}
and hence $q = \frac12(1 + 3w_\mathrm{eff})$. Note also that it follows that
\begin{equation}\label{wphiu}
w_\varphi = u^2 - 1,\qquad \Omega_V = \frac32\left(2-u^2\right)v^2,
\end{equation}
and hence that it is easy to visualize $w_\varphi$ and $\Omega_\varphi$
since $w_\varphi$ is constant when $u=\mathrm{constant}$ while
$\Omega_\varphi$ is constant when $v=\mathrm{constant}$, due to the
definition~\eqref{vdef}.

Once the scalar field potential has been specified,
$\lambda(\bar{\varphi})$ follows, which implies
that~\eqref{Dynsys} is a closed system of autonomous
ordinary differential equations, {\it i.e.}, a dynamical system.


\subsection{State space structure}

The state space $(\bar{\varphi},u,v,w_{\gamma\mathrm{m}})$ for
quintessential $\alpha$-attractor inflation with
a monotonically decreasing potential $V>0$, $\lambda >0$,
radiation, $\rho_\gamma>0$, and matter, $\rho_\mathrm{m}>0$,
is determined by the inequalities
\begin{equation}\label{ineq1}
-1 < \bar{\varphi} < 1,\qquad -\sqrt{2} < u < \sqrt{2}, \qquad
0 < v < \frac{1}{\sqrt{3}}, \qquad
0 < w_{\gamma\mathrm{m}} < \frac13.
\end{equation}
It follows that the boundaries of the state space are given by
\begin{equation}\label{3Dboundarysets}
\bar{\varphi} = \pm 1, \qquad u = \pm\sqrt{2}, \qquad
v = 0, \qquad v = \frac{1}{\sqrt{3}},\qquad w_{\gamma\mathrm{m}} = 0,\qquad w_{\gamma\mathrm{m}} = \frac13.
\end{equation}
Since $\lambda(\bar{\varphi})$
and its derivatives are non-singular for $\bar{\varphi}\in [-1,1]$, and
$\lambda_-=0$, $\lambda_+$, $\lambda_\pm^{(n)}$ thereby exist,
this implies that \emph{all boundaries are invariant sets}. Due to the
differentiability assumptions for the potential, the dynamical system~\eqref{Dynsys}
is \emph{regular} everywhere, including on the boundaries~\eqref{3Dboundarysets}.
This makes it possible to include the 3-dimensional invariant boundary sets
and obtain a \emph{compact} state space. This is highly desirable, since
all asymptotics for the solutions for the state vector
$(\bar{\varphi},u,v,w_{\gamma\mathrm{m}})$
turn out to reside on these boundaries. We therefore from now on consider
the compact state space for which
\begin{equation}\label{ineq2}
-1 \leq \bar{\varphi} \leq 1,\qquad -\sqrt{2} \leq u \leq \sqrt{2}, \qquad
0 \leq v \leq \frac{1}{\sqrt{3}}, \qquad
0 \leq w_{\gamma\mathrm{m}} \leq \frac13,
\end{equation}
while the state space defined by the inequalities~\eqref{ineq1} is
referred to as the \emph{interior} state space.

It is straightforward to show that the deceleration parameter $q$
in~\eqref{qdef} satisfies
\begin{equation}\label{qineq}
-1 \leq q \leq 2.
\end{equation}
Here $q=2$ on the invariant boundary subset
$v=1/\sqrt{3}$, $u=\pm\sqrt{2}$, which corresponds to $
\Omega_\gamma = \Omega_\mathrm{m} = \Omega_V = 0$. 
The lower bound, $q=-1$, which corresponds to a
de Sitter state, occurs when
$v=1/\sqrt{3}$, $u=0$. This, due to~\eqref{ueq}, requires that
$\lambda=0$, which is only possible on the $\bar{\varphi}=-1$ boundary,
where $\lambda_-=0$, and on the $\bar{\varphi}=1$ boundary when
$\lambda_+=0$, {\it i.e.}, for the quintessential models with an
asymptotically constant potential $V \rightarrow V_+>0$
when $\varphi\rightarrow \infty$ ($\bar{\varphi}\rightarrow 1$).

\section{General dynamical systems features}\label{sec:gendynfeatures}

\begin{figure}[ht!]
	\begin{center}
			\subfigure[Potential independent boundary structures on the $w_{\gamma\mathrm{m}}=1/3$ state space.]{\label{fig:Boxstatespace}
			\includegraphics[width=0.35\textwidth]{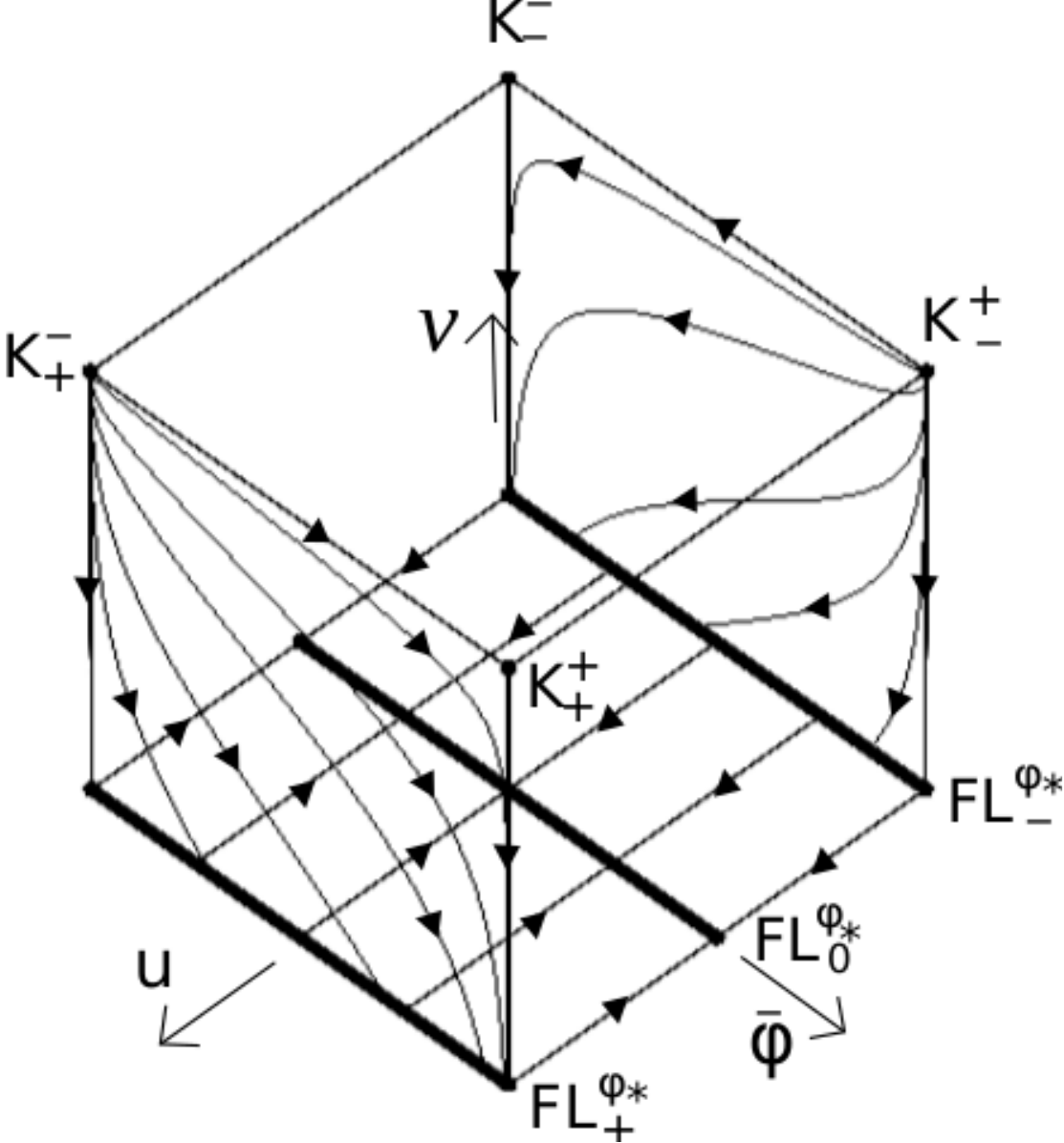}}
		\hspace{1.5cm}
			\subfigure[Projection onto $(\bar{\varphi},w_{\gamma\mathrm{m}})$-space of the $v=0$
            subset, where $\bar{\varphi} = \bar{\varphi}_* = \mathrm{constant}$.]{\label{}
	\includegraphics[width=0.3\textwidth]{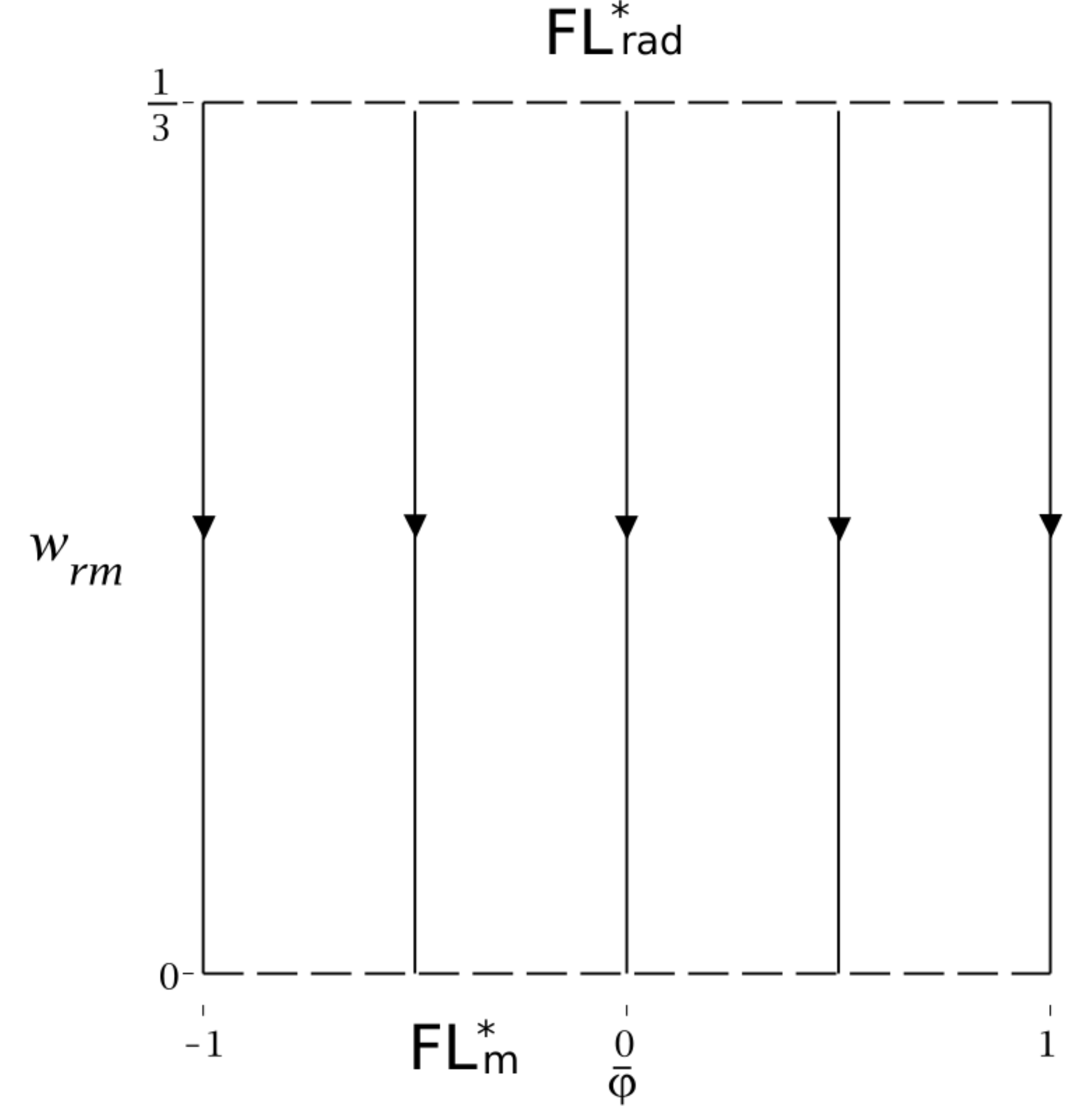}}\\
\subfigure[$\lambda=0$.]{\label{L0}
\includegraphics[width=0.30\textwidth]{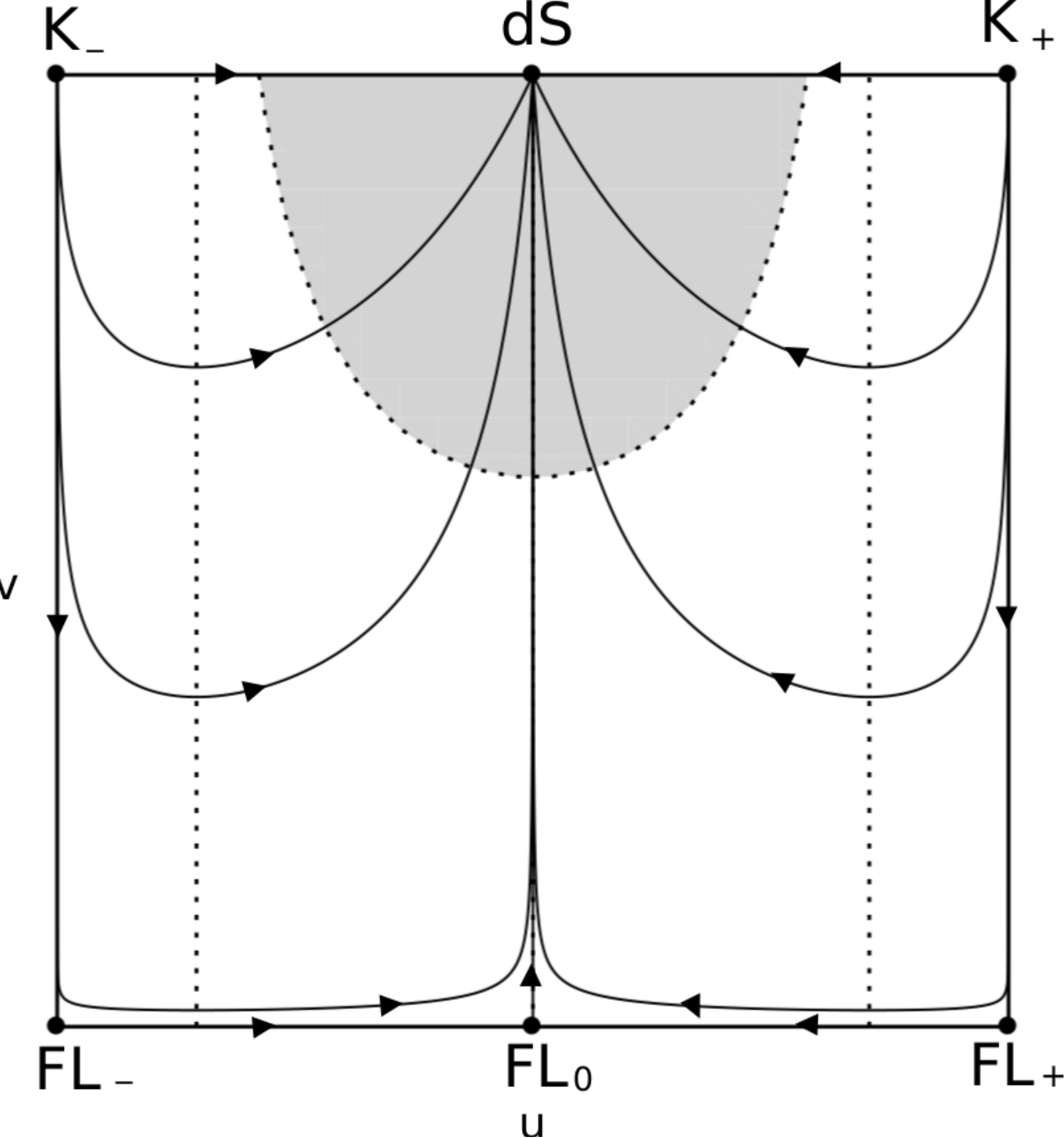}}
\hspace{1.5cm}
\subfigure[ $0<\lambda\leq\sqrt{2}$.]{\label{exp.pot1}
\includegraphics[width=0.30\textwidth]{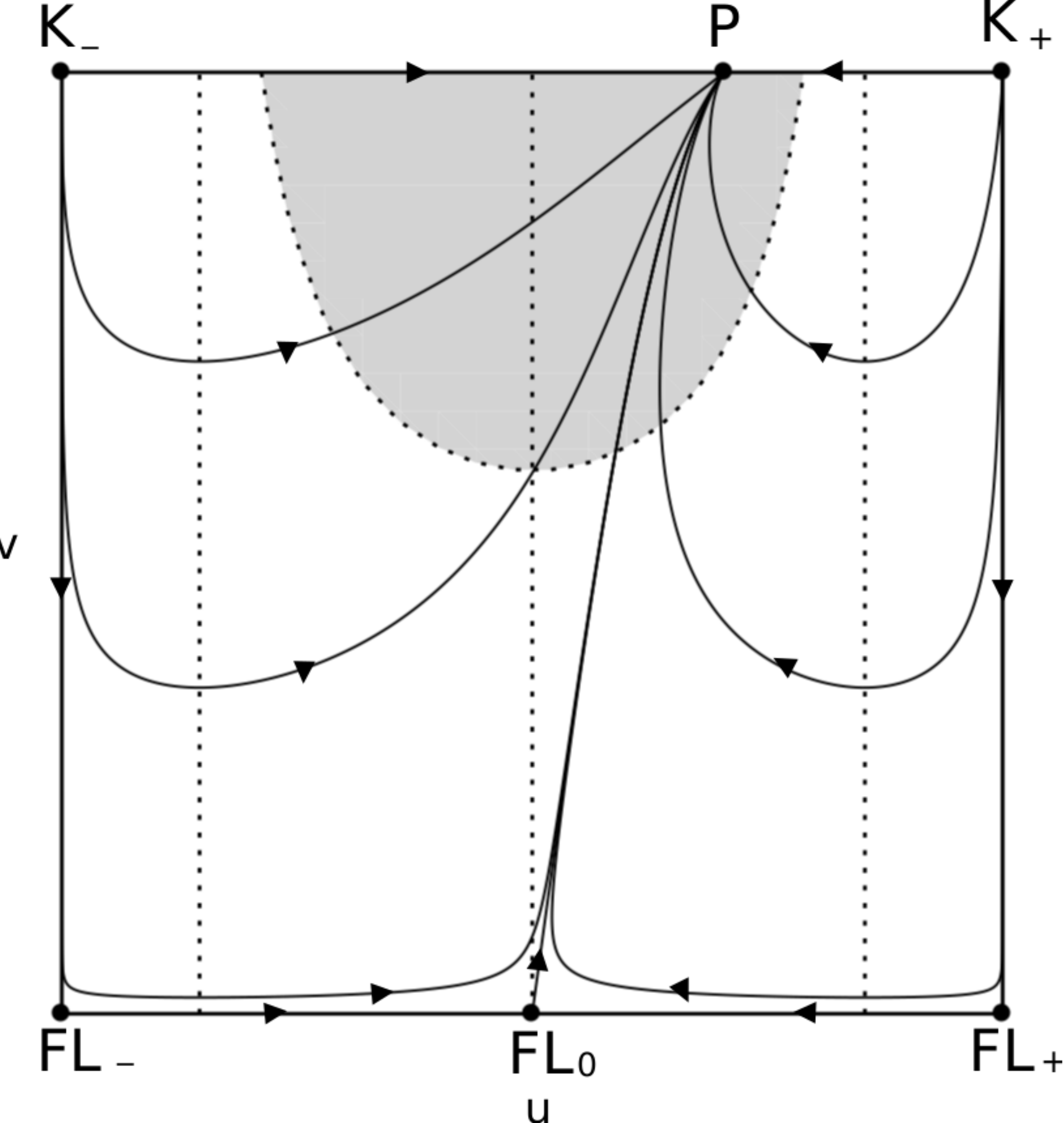}}
		\vspace{-0.5cm}		
	\end{center}
	\caption{Figure (a) illustrates the solution space on the potential independent boundaries of the
        $w_{\gamma\mathrm{m}}=1/3$ boundary (the $w_{\gamma\mathrm{m}}=0$ boundary is similar):
		$v=0$ ($\Omega_\varphi=0$, $\Omega_\gamma=1$) results in $\bar{\varphi} = \bar{\varphi}_* = \mathrm{constant}$
        (and hence that the scalar field $\varphi = \varphi_*$ is frozen),
        while $u=\pm\sqrt{2}$ ($w_\varphi = 1$; $\Omega_V=0$) leads to that $v$ is monotonically
		decreasing and that $\bar{\varphi}$ is monotonically increasing (decreasing)
		when $u=\sqrt{2}$ ($u= - \sqrt{2}$).
        Figure (b) describes a projection onto $(\bar{\varphi},w_{\gamma\mathrm{m}})$-space of the $v=0$
        boundary, where $\bar{\varphi} = \bar{\varphi}_* = \mathrm{constant}$.
        Figures (c) and (d) illustrate the solution space structure on the $\bar{\varphi} = \pm 1$ boundaries,
        where the grey regions correspond to acceleration, {\it i.e.} when $q<0$.}
	\label{SSTentBox}
\end{figure}

The energy conservation of radiation and matter,
$\rho_\gamma^\prime = -4\rho_\gamma$,
$\rho_\mathrm{m}^\prime = -3\rho_\mathrm{m}$, respectively, yields
\begin{equation}\label{rhorm}
\rho_\gamma = \rho_{\gamma,0}e^{-4N} = 3H_0^2\Omega_{\gamma,0}e^{-4N},\qquad
\rho_\mathrm{m} = \rho_{\mathrm{m},0}e^{-3N} = 3H_0^2\Omega_{\mathrm{m},0}e^{-3N},
\end{equation}
which leads to
\begin{equation}\label{wrmsol}
w_{\gamma\mathrm{m}} =
\frac13\left(\frac{1}{1 + \left(\frac{\Omega_{\mathrm{m},0}}{\Omega_{\gamma,0}}\right)e^{N}}\right).
\end{equation}
Thus, as also follows from~\eqref{wrmeq}, $w_{\gamma\mathrm{m}}$ is strictly monotonically
decreasing for all interior orbits ({\it i.e.}, solution trajectories residing in
the interior state space), which thereby originate from the invariant radiation
boundary $w_{\gamma\mathrm{m}} = 1/3$ and end at the invariant matter boundary $w_{\gamma\mathrm{m}}=0$,
where the dynamics on these two boundaries are qualitatively similar.
In particular, all fixed points are located on the boundaries of
these two boundaries, and are, for the presently considered models,
given by the same values of $\bar{\varphi}, u, v$, see Figure~\ref{SSTentBox}.

Next we note that $\bar{\varphi} = \bar{\varphi}_* = \mathrm{constant}$ on the
invariant $v=0$ ($\Omega_\varphi=0$) boundary. Furthermore, the equation for $w_{\gamma\mathrm{m}}$
decouples from those of $\bar{\varphi}$ and $u$ on this boundary, which thereby form a reduced state
space $(\bar{\varphi},u)$, which, due to the decoupling of $w_{\gamma\mathrm{m}}$, has a solution space
that is identical to that on the invariant $(w_{\gamma\mathrm{m}},v)=(\frac13,0)$ and
$(w_{\gamma\mathrm{m}},v)=(0,0)$ boundaries, see Figure~\ref{SSTentBox}.
The invariant boundaries $u=\pm \sqrt{2}$ leads to a dynamical system that is independent of the
potential, where the scalar field yields the same contribution as a stiff perfect fluid
since $w_\varphi = u^2 - 1 = 1$; as a consequence it is not difficult to explicitly solve the equations
on these boundaries, as is done in section~\ref{sec:freezing}. The features
of the solution space of these boundaries are illustrated in Figure~\ref{SSTentBox}.
%

Further insights into the dynamics are obtained by noticing
that~\eqref{rhorm} and~\eqref{wrmsol} yields
\begin{equation}
\rho_\gamma + \rho_\mathrm{m} =
3H_0^2\left(\frac{\Omega_{\mathrm{m},0}^{4}}{\Omega_{\gamma,0}^{3}}\right)
\frac{(3w_{\gamma\mathrm{m}})^3}{(1 - 3w_{\gamma\mathrm{m}})^4},\label{rhormOmrm}
\end{equation}
which together with the definitions in~\eqref{vardef1} leads to
\begin{equation}\label{intbasis}
3H^2 = \frac{V(\bar{\varphi})}{\Omega_V} =
\frac{\rho_\gamma + \rho_\mathrm{m}}{\Omega_\gamma + \Omega_m} =
3H_0^2\left(\frac{\Omega_{\mathrm{m},0}^{4}}{\Omega_{\gamma,0}^{3}}\right)
\frac{(3w_{\gamma\mathrm{m}})^3}{(1 - 3w_{\gamma\mathrm{m}})^4(1 - 3v^2)}.
\end{equation}
Due to~\eqref{qdef1}, which implies $H^\prime = -(1+q)H$, and~\eqref{qdef},
it follows that
\begin{equation}\label{3H2gen}
(3H^2)^\prime = -2(1+q)(3H^2) = -3\left[3u^2v^2 + (1+w_{\gamma\mathrm{m}})(1-3v^2)\right](3H^2).
\end{equation}
Hence, due to that $1 + q > 0$ (everywhere except at the scalar field boundary
where $1 + q \geq 0$, which requires special treatment, see section~\ref{sec:inflation}),
$3H^2$, which can be expressed in the state space variables
by means of~\eqref{intbasis}, is strictly monotonically decreasing in
the interior state space, as well as on the interior radiation and matter boundaries
$w_{\gamma\mathrm{m}}=1/3$, $w_{\gamma\mathrm{m}}=0$, respectively.
We also note from~\eqref{ueq} that $u$ is monotonically increasing when $u\leq 0$.
By taking into account these global features, and the structures on the boundaries,
including using the global results concerning the constant and exponential
potential boundaries $\bar{\varphi}=\pm 1$ derived in~\cite{alhetal23},
it is not difficult to prove (although for brevity we will refrain from explicitly
doing so) that all interior orbits originate from the following fixed points:

\begin{itemize}
\item An open set of interior orbits originates from the two `kinaton' 
(see section~\ref{SlowRoll}) sources
$\mathrm{K}_-^+$ and $\mathrm{K}_+^-$ on the radiation boundary with
$(\bar{\varphi},u,v,w_{\gamma\mathrm{m}}) = (1,-\sqrt{2},1/\sqrt{3},1/3)$ and
$(\bar{\varphi},u,v,w_{\gamma\mathrm{m}}) = (-1,\sqrt{2},1/\sqrt{3},1/3)$, respectively.
The orbits originating from $\mathrm{K}_-^+$ correspond to
solutions for which $\varphi\rightarrow\infty$ toward the past (case (i) in
section~\ref{subsec:fieldheur}) while the orbits coming from
$\mathrm{K}_+^-$ corresponds to solutions for which $\varphi\rightarrow-\infty$
toward the past (case (ii) in section~\ref{subsec:fieldheur}).
\item
A set of co-dimension one of orbits originates from the line of fixed points
$\mathrm{FL_0^{\varphi_*}}$ on the radiation boundary with
$(\bar{\varphi},u,v,w_{\gamma\mathrm{m}}) = (\bar{\varphi}_*,0,1/\sqrt{3},1/3)$
(case (iii) in section~\ref{subsec:fieldheur}).
\item
All interior orbits end at the fixed point $\mathrm{dS}^+$ ($\mathrm{P}^+$)
on the matter boundary with $(\bar{\varphi},u,v,w_{\gamma\mathrm{m}}) = (1,0,1/\sqrt{3},0)$
when $\lambda_+=0$
($(\bar{\varphi},u,v,w_{\gamma\mathrm{m}}) = (1,\lambda_+/\sqrt{3},1/\sqrt{3},0)$
when $0<\lambda_+<\sqrt{2}$).
\end{itemize}

In the next section we will show that there is a single orbit 
on the scalar field boundary,
$v = 1/\sqrt{3}$ ($\Omega_\varphi = 1$, where there thereby is no
radiation and matter content), that
originates from a de Sitter fixed point on this boundary, and 
that this orbit corresponds to a solution that belongs to case (ii)
in section~\ref{subsec:fieldheur}. Finally we note that all interior
orbits, including those on the scalar field boundary, are heteroclinic
orbits ({\it i.e} solution trajectories that originate and end at two
different fixed points) that end at $\mathrm{dS}^+$ ($\mathrm{P}$)
when $\lambda_+ =0$ ($0<\lambda_+<\sqrt{2}$).

We conclude this section by noticing
that the first and last equalities
in~\eqref{intbasis} yield the following integral
\begin{equation}\label{intbasis2}
\frac{V_-}{3H_0^2} =
\frac32\left(\frac{\Omega_{\mathrm{m},0}^{4}}{\Omega_{\gamma,0}^{3}}\right)
\frac{(3w_{\gamma\mathrm{m}})^3(2-u^2)v^2}{(1 - 3w_{\gamma\mathrm{m}})^4(1 - 3v^2)\bar{V}(\bar{\varphi})},
\end{equation}
which thereby foliates the 4-dimensional interior state space into 3-dimensional invariant subsets,
parametrized by $\frac{V_-\Omega_{\gamma,0}^{3}}{3H_0^2\Omega_{\mathrm{m},0}^{4}}$.
\section{The scalar field boundary and inflation\label{sec:inflation}}

\subsection{The scalar field boundary}

As will be shown, inflation is associated with that an open set of interior orbits in
the full state space intermediately closely shadow a special orbit,
the inflationary `attractor solution,' on the invariant scalar field boundary
$v = 1/\sqrt{3}$ during the radiation epoch. The scalar field boundary corresponds to setting
$\rho_\gamma = \rho_\mathrm{m}=0$ ($\Omega_\gamma = \Omega_\mathrm{m}=0$; $\Omega_\varphi=1$),
which leads to that the differential equation for $w_{\gamma\mathrm{m}}$ decouples from those of
$\bar{\varphi}$ and $u$, where $w_{\gamma\mathrm{m}}$ can be regarded as a test field
on a pure scalar field background. However, since inflation takes place
deep in the radiation epoch, it is the vicinity of the intersection of the radiation
and scalar field boundaries, $w_{\gamma\mathrm{m}}=1/3$ and $v = 1/\sqrt{3}$,
that is relevant when situating inflation
in the full state space. As a consequence the essentials of inflation are connected with the
state space $(\bar{\varphi},u)$ and dynamics determined by the
following 2-dimensional dynamical system, obtained by setting $v=1/\sqrt{3}$ in~\eqref{Dynsys}:
\begin{subequations}\label{Dynsyspurescalarfield}
\begin{align}
\bar{\varphi}^\prime &= \frac{1}{\sqrt{2\alpha}}(1 - \bar{\varphi}^2)u,\\
u^\prime &= \frac32\left(2 - u^2\right)\!\left(\frac{\lambda(\bar{\varphi})}{\sqrt{3}} - u\right). \label{Sigpure}
\end{align}
\end{subequations}
%
%
%
%
The state vector $(\bar{\varphi},u)$ has boundaries
given by $\bar{\varphi}=\pm 1$ and $u = \pm \sqrt{2}$. Since
\begin{equation}\label{inflq}
q= -1 + \frac32 u^2, \qquad \epsilon = 1 + q = \frac32 u^2,
\end{equation}
for this case due to~\eqref{qdef}, where $\epsilon$ is the Hubble slow-roll
parameter in an inflationary context, the boundaries $u = \pm \sqrt{2}$, which correspond
to setting $w_\varphi=1$ and $\Omega_V=0$ (and thereby $V=0$),
result in $q = 2$, $\epsilon = 1$. As a consequence
$H^\prime = -(1+q)H= -3H \Rightarrow
3H^2 = \rho =\rho_\varphi = 3H_0^2\exp(-6N) = 3H_0^2(a_0/a)^6$,
which characterizes kinaton evolution (a nomenclature
introduced in~\cite{joypro98}), {\it i.e.},
$u=\pm \sqrt{2}$ on the $v=1/\sqrt{3}$ ($\Omega_\gamma = \Omega_\mathrm{m} = 0$)
boundary yields the \emph{kinaton boundary}.

Using~\eqref{intbasis} and inserting $v=1/\sqrt{3}$ into~\eqref{wphiu} yields
\begin{equation}\label{H2Int}
3H^2 = \frac{V(\bar{\varphi})}{\Omega_V} = \frac{V(\bar{\varphi})}{1 - \frac12u^2},
\end{equation}
while $H^\prime = -(1+q)H$ in combination with~\eqref{inflq} results in
\begin{equation}\label{inflection}
(3H^2)^\prime = -3u^2(3H^2), \quad
(3H^2)''|_{u = 0} = 0,\quad (3H^2)'''|_{u = 0} = -18\lambda^2(\bar{\varphi})(3H^2).
\end{equation}
Thus, for the interior state space of the present boundary
$3H^2 = \frac{V(\bar{\varphi})}{1 - \frac12u^2}$ is monotonically decreasing
when $u \neq 0$, and since $\lambda(\bar{\varphi})>0$ when
$\bar{\varphi}\in (-1,1)$ it follows from~\eqref{inflection}
that $3H^2 = \frac{V(\bar{\varphi})}{1 - \frac12u^2}$
only goes through an inflection point when $u=0$, which can only happen once since
$u^\prime|_{u=0} > 0$. Due to this, the asymptotics of all
interior orbits reside on the boundaries $\bar{\varphi}= \pm 1$, $u=\pm \sqrt{2}$, but since
$\bar{\varphi}^\prime|_{u=\pm \sqrt{2}} = \pm \frac{1}{\sqrt{\alpha}}(1 - \bar{\varphi}^2)$,
and since the dynamics on $\bar{\varphi}=\pm 1$ are easily obtained,
it follows that all interior orbits in the state space $(\bar{\varphi},u)$ 
are heteroclinic orbits that asymptotically
originate and end at fixed points on the $\bar{\varphi}=\mp 1$ boundaries.
There are 6 fixed points on these boundaries in the state space $(\bar{\varphi},u)$:
\begin{itemize}
\item Four of these are at the corners of the
state space, the \emph{`kinaton' fixed points} with $q=2$ ($w_\varphi = 1$),
$\mathrm{K}_-^-$, $\mathrm{K}_-^+$, $\mathrm{K}_+^-$, $\mathrm{K}_+^+$,
where superscripts (subscripts) denote
the signs of the values of $\bar{\varphi}$ ($u$), {\it e.g.}, $\mathrm{K}_-^+$
corresponds to $(\bar{\varphi},u) = (+1,-\sqrt{2})$. The fixed points $\mathrm{K}_-^+$
and $\mathrm{K}_+^-$ are hyperbolic sources, while $\mathrm{K}_-^-$ and
$\mathrm{K}_+^+$ are hyperbolic saddles.
\item
There is a \emph{de Sitter fixed point} $\mathrm{dS}^-$ at $(\bar{\varphi},u) = (-1,0)$
and an additional one $\mathrm{dS}^+$ at $(\bar{\varphi},u) = (+1,0)$ when
$\lambda_+=0$ (for de Sitter fixed points, $\varphi^\prime = 0$ and $\Omega_V=1$),
forming a global sink;
if $\lambda_+>0$ $\mathrm{dS}^+$ is replaced with the global (hyperbolic) sink $\mathrm{P}$ at
$(\bar{\varphi},u) = (1,\lambda_+/\sqrt{3})$, where this fixed point corresponds to a self-similar
\emph{power law} solution. The condition $\lambda_+ = 2/\sqrt{6\alpha}<\sqrt{2}$,
and hence
\begin{equation}
\alpha > \frac13,
\end{equation}
leads to that $\mathrm{P}$ results in future eternal
acceleration, since this condition yields
$q = - 1 + \lambda_+^2/2= -1 + 1/(3\alpha) < 0$.
\end{itemize}

\begin{figure}[ht!]
	\begin{center}
		\subfigure[$\text{EC}$: $\alpha=\frac{7}{3}$, $\nu=10$.]{\label{}
			\includegraphics[width=0.30\textwidth]{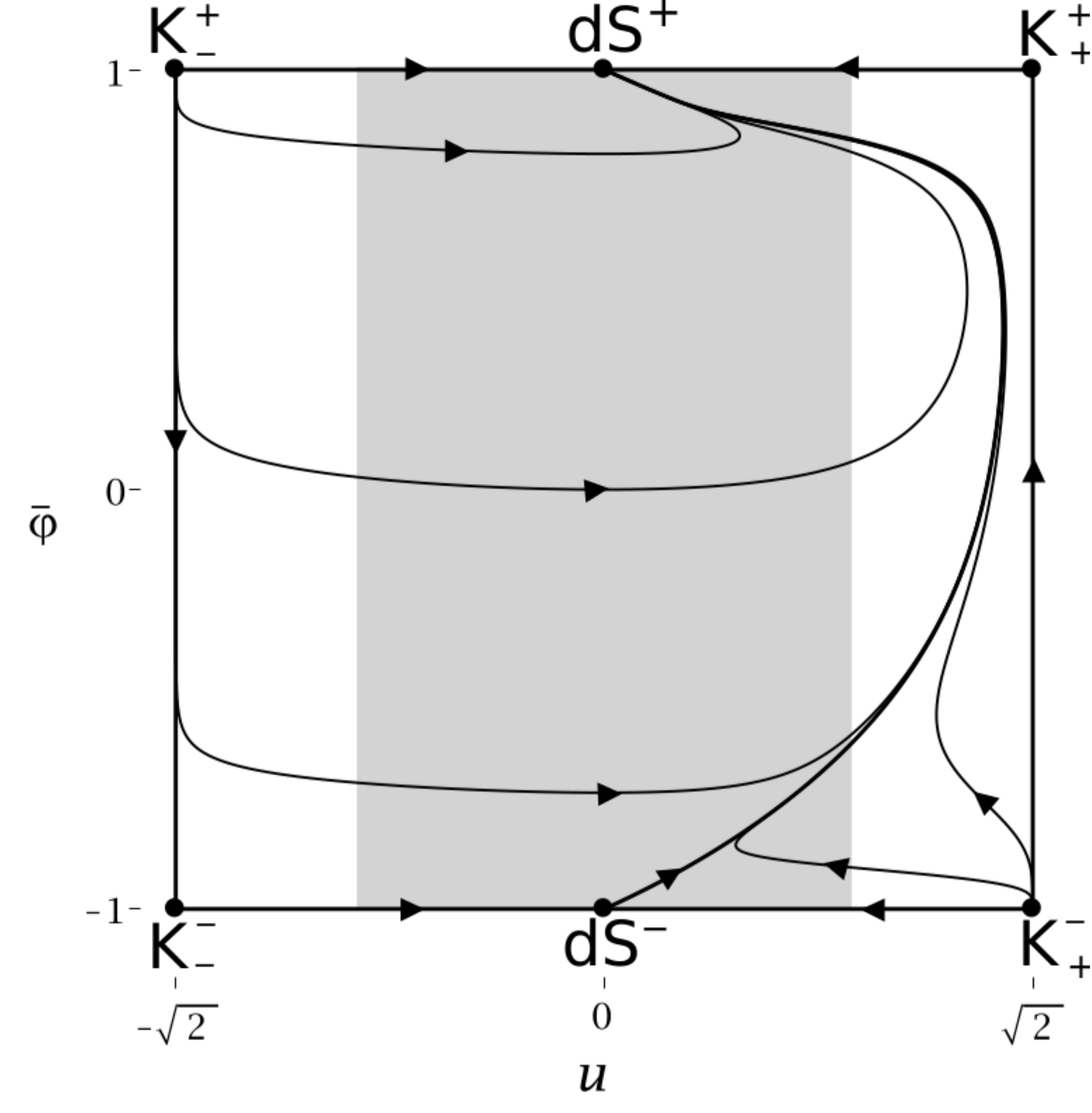}}
		\hspace{1.0cm}
		\subfigure[$\text{EC}$: $\alpha=\frac{7}{3}$, $\nu=128$.]{\label{}
			\includegraphics[width=0.30\textwidth]{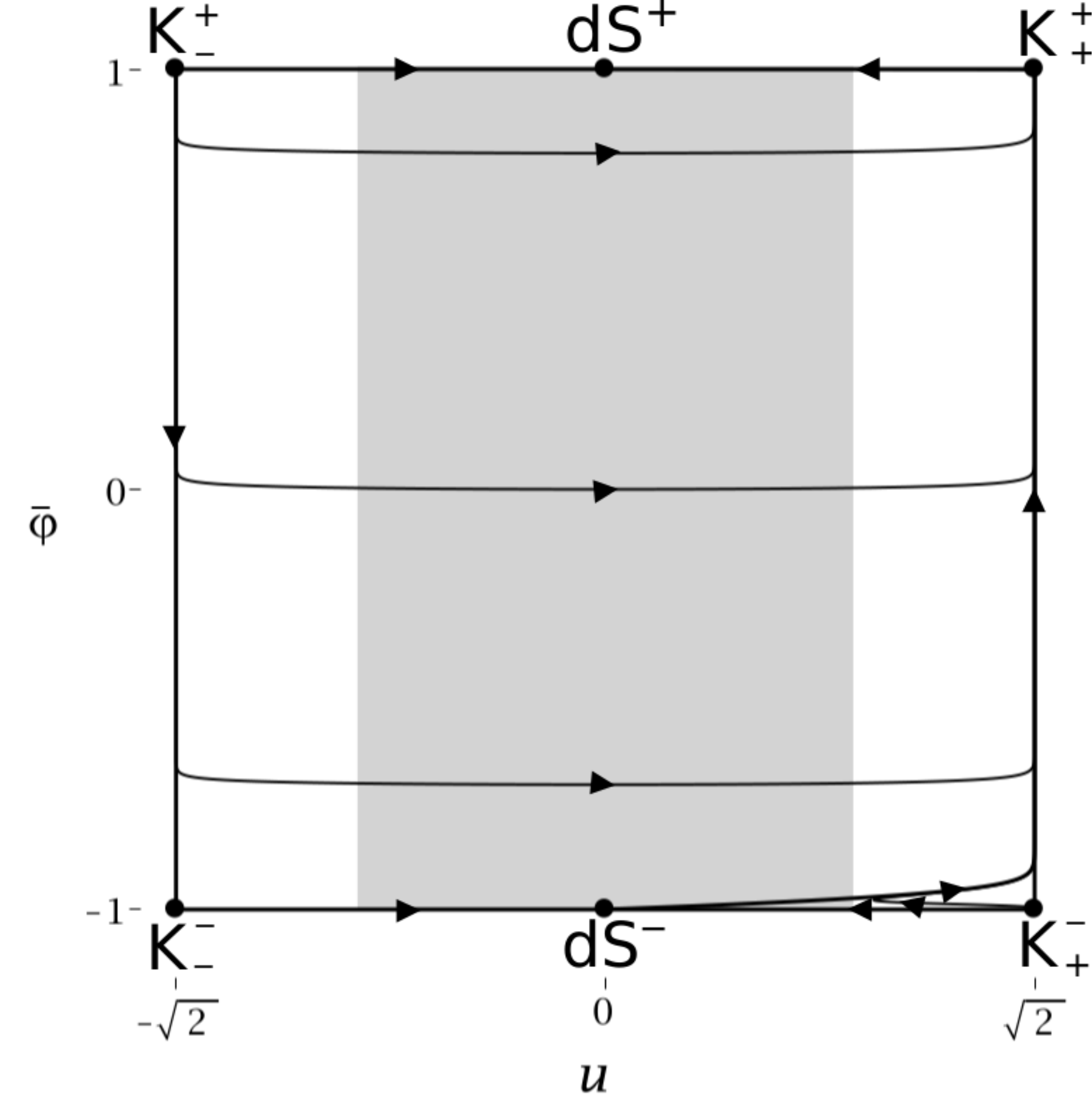}}
		\vspace{-0.5cm}
		\subfigure[$\text{ET}$: $\alpha=\frac{7}{3}$, $\nu=10$.]{\label{}
			\includegraphics[width=0.30\textwidth]{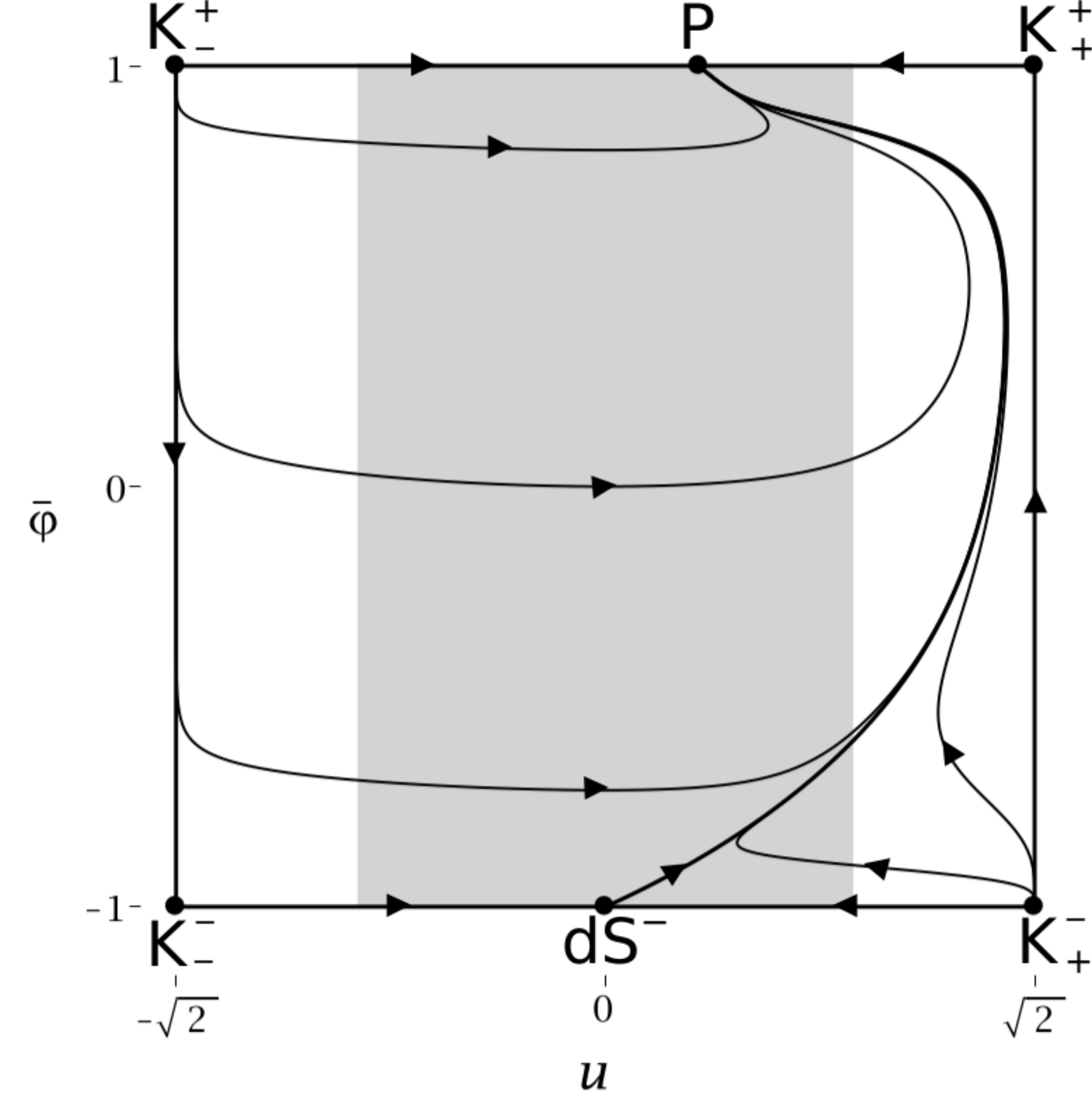}}
		\hspace{1.0cm}
		\subfigure[$\text{ET}$: $\alpha=\frac{7}{3}$,  $\nu=128$.]{\label{}
			\includegraphics[width=0.30\textwidth]{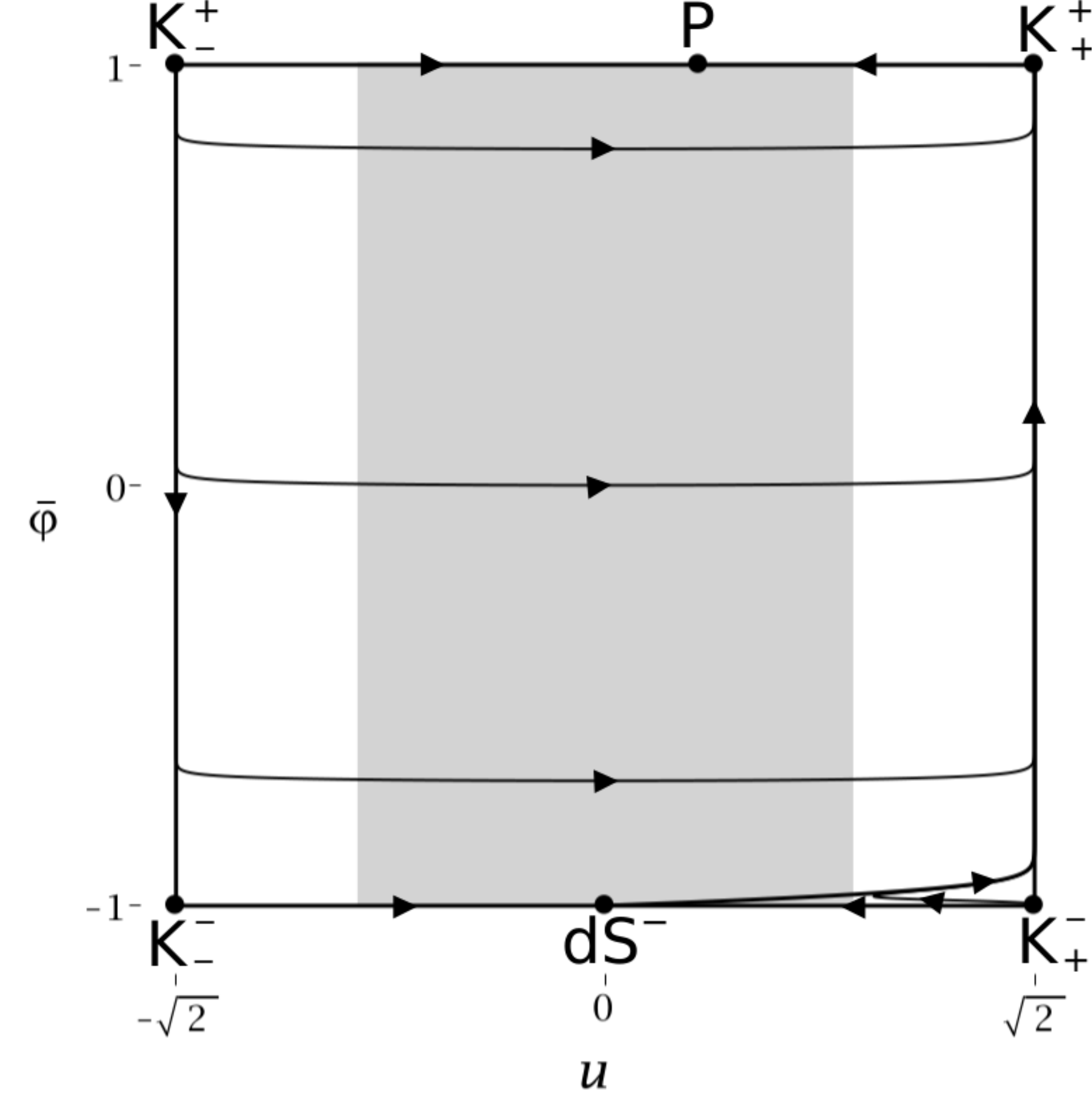}}
	\end{center}
	\caption{Examples of orbits on the $\Omega_\gamma = \Omega_\mathrm{m} = 0$; $\Omega_\varphi=1$,
    $v=1/\sqrt{3}$ scalar field boundary for the $\text{EC}$ and $\text{ET}$ models,
    with $\alpha=\frac{7}{3}$ and $\nu = 10$, $\nu = 128$, including the center manifold orbit (the
    inflationary `attractor' solution) originating from $\mathrm{dS}^-$. The grey regions
    correspond to acceleration, {\it i.e.}, $q<0$. As illustrated, the center manifold orbit for the EC and ET models
    exhibits a kinaton phase with $w_\varphi \approx 1$, {\it i.e.}, a period when it shadows
    the boundary $u = \sqrt{2}$, when $\frac{\nu}{\sqrt{6\alpha}}\gtrsim 3.5$,
    which holds for the presently considered range of values for $\alpha$ and $\nu$.
    }
	\label{Fig_SFB_ECET}
\end{figure}

The state space $(\bar{\varphi},u)$ is naturally divided into the regions $u<0$,
$u =0$ and $u>0$, where $\bar{\varphi}$
is monotonically decreasing (increasing) when $u<0$
($u>0$). Since $u^\prime >0$ when $u \leq 0$ in the interior
state space, where $\lambda(\bar{\varphi})>0$, it follows that the one-parameter
set of interior orbits that originates from the source $\mathrm{K}_-^+$ eventually
enters the invariant region $u>0$
where they stay until they, as all interior orbits,
end at $\mathrm{dS}^+$ or $\mathrm{P}$, depending on if $\lambda_+=0$ or
$\lambda_+=2/\sqrt{6\alpha}<\sqrt{2}$, respectively.
These features, taken together with
\begin{equation}\label{varphiprimscalarboundary}
\varphi^\prime = \sqrt{3}u,
\end{equation}
which is due to~\eqref{udef} and $\Omega_\varphi=1$, show that orbits originating from
$\mathrm{K}_-^+$ belong to case (i) in section~\ref{subsec:fieldheur}.
In contrast, the one-parameter set of orbits originating from the source
$\mathrm{K}_+^-$ and the single heteroclinic separatrix orbit (the
inflationary attractor solution, to be discussed below)
$\mathrm{dS}^-\rightarrow \mathrm{dS}^+/\mathrm{P}$ reside
completely in the invariant interior subset $u > 0$, and hence belong
to case (ii) in section~\ref{subsec:fieldheur}.
The solution space is illustrated in Figure~\ref{Fig_SFB_ECET}.

Equation~\eqref{barvexamplesECET} that gives $\lambda(\bar{\varphi})$ for the
EC and ET models, illustrated in Figures~\ref{EClambda} and~\ref{ETlambda},
shows that $\lambda_\mathrm{max} = \nu/\sqrt{6\alpha}$ at $\bar{\varphi} = 0$ for the
EC models, which is also close to $\lambda_\mathrm{max}$ for the ET models when $\nu$
is large; moreover, the EC and ET models have quite similar $\lambda(\bar{\varphi})$
when $\nu$ is large, until $\bar{\varphi}$ is close to one, due to the limits $\lambda_+=0$
and $\lambda_+=2/\sqrt{6\alpha}<\sqrt{2}$ in the EC and ET cases, respectively.
A large $\nu/\sqrt{6\alpha}$ leads to that $u^\prime$ becomes larger
than $\bar{\varphi}^\prime$ in most of the interior state space,
which is the reason why the orbits for such EC and ET models in
Figure~\ref{Fig_SFB_ECET} are characterized by `horizontal' trajectories
where $u$ changes much faster than $\bar{\varphi}$.

When $\frac{\nu}{\sqrt{6\alpha}} \gtrsim 3.5$, as for the presently considered models, 
the heteroclinic separatrix orbit
$\mathrm{dS}^-\rightarrow \mathrm{dS}^+/\mathrm{P}$, and nearby shadowing
orbits, exhibit a transient \emph{kinaton epoch} where they shadow a part
of the heteroclinic \emph{kinaton orbit} $\mathrm{K}_+^- \rightarrow \mathrm{K}_+^+$,
{\it i.e.}, the kinaton boundary component with $u = \sqrt{2}$ where
$\bar{\varphi}$, and $\varphi$, is increasing. This epoch starts earlier at a smaller $\bar{\varphi}$
and ends later at a larger $\bar{\varphi}$, and is
more pronounced, when $\nu/\sqrt{6\alpha}$ increases, as illustrated in
Figure~\ref{Fig_SFB_ECET}.

%
%
%
%

Inflationary descriptions are intimately connected with the slow-roll
approximation and slow-roll expansions. We will therefore briefly review this
topic and show how it is connected with the present dynamical
systems approach and the center manifold analysis of the de Sitter
fixed point $\mathrm{dS}^-$.

\subsection{Slow-roll expansions}\label{SlowRoll}

The slow-roll approximation is based on the
assumptions (see {\it e.g.} \cite{stetur84,lidlyt93,lidetal94})
\begin{subequations}
\begin{align}
\dot{\varphi}^2 &< V(\varphi),\\
\dot{\varphi} &\approx -\frac{V_{,\varphi}}{3H},\label{slowroll2main}\\
\epsilon_V &:= \frac12\lambda^2 \ll 1.
\end{align}
\end{subequations}
The first condition corresponds to that the kinetic part of the scalar field
energy density is dominated by the potential part, which can be expressed as
$u^2 < 1$ while the second condition corresponds to
$u \approx \lambda/\sqrt{3}$. We then note that at the de Sitter fixed points
$\mathrm{dS}^\pm$ we have $\lambda=0$, $u=0$ and hence $\Omega_V=1$,
{\it i.e.} these fixed points represent the limit of the slow-roll
approximation.

According to~\cite{lidetal94}, the potential slow-roll parameters
\begin{equation}\label{PSR}
\epsilon_V = \frac{1}{2}\left(\frac{V_{,\varphi}}{V}\right)^2 = \frac12\lambda^2, \qquad
\eta_V := \frac{V_{,\varphi\varphi}}{V} = \lambda^2 - \lambda_{,\varphi},
\end{equation}
can be used to approximate the Hubble slow-roll parameter $\epsilon = 1 + q$
to second order as follows:
\begin{equation}\label{HSRtoPSR}
\epsilon \approx \epsilon_V\left[1 - \frac{2}{3}\!\left(2\epsilon_V - \eta_V\right)\right],
\end{equation}
which when expressed in $\lambda$ yields
\begin{equation}\label{HSRtoPSRlambda}
\epsilon \approx \frac12\lambda^2\left(1-\frac{2}{3}\lambda_{,\varphi}\right).
\end{equation}
Since $\epsilon = 1 + q = \frac12(\varphi^\prime)^2$ (we will express things in $\varphi^\prime$ as well
as in $u$ since $\varphi^\prime$ provides a more natural contact with slow-roll approximations while
$u$ is needed for connecting with the present state space approach),
it follows that
\begin{equation}\label{Sigslowroll}
\varphi^\prime = \sqrt{3}u \approx \lambda\left(1-\frac{1}{3}\lambda_{,\varphi}\right).
\end{equation}
Since $2\epsilon_V - \eta_V = \lambda_{,\varphi}$ needs to be small in order for the slow-roll approximations
to be accurate, we can improve the second order slow-roll approximation in~\eqref{Sigslowroll}
by the following $[1,1]_{\varphi^\prime}$ Pad{\'e} slow-roll approximant (see {\it e.g.}~\cite{alhugg15}
and~\cite{lidetal94} for a discussion and references on Pad{\'e} approximants):
\begin{equation}\label{slowrollPade}
\varphi^\prime = \sqrt{3}u \approx \frac{\lambda}{1+\frac{1}{3}\lambda_{,\varphi}}.
\end{equation}

Let us next Taylor expand~\eqref{Sigslowroll} at $\bar{\varphi}=-1$
up to second order. To do so we first define
\begin{equation}
x := 1 + \bar{\varphi},
\end{equation}
and note that
\begin{equation}
\lambda_{,\varphi} = \left(\frac{d\bar{\varphi}}{d\varphi}\right)\lambda_{,\bar{\varphi}} =
\frac{1}{\sqrt{6\alpha}}\left(1 - \bar{\varphi}^2\right)\lambda_{,\bar{\varphi}} =
\frac{x}{\sqrt{6\alpha}}\left(2-x\right)\lambda_{,\bar{\varphi}}.
\end{equation}
We then introduce
\begin{equation}
\ell_n = \lambda^{(n)}_- = \left.\frac{d^n\lambda}{d\bar{\varphi}^n}\right|_{\bar{\varphi}=-1} =
\left.\frac{d^n\lambda}{d\bar{\varphi}^n}\right|_{x=0},
\end{equation}
where we use $\ell_n$, $n=1,2,3,\dots$ to obtain succinct notation.
A Taylor expansion of $\lambda(\bar{\varphi})$ at $\bar{\varphi}=-1$
($x=0$) thereby yields
\begin{equation}\label{taylorlu}
\lambda(x) = \sum^{+\infty}_{k=1} \frac{\ell_k}{k!}x^k
\end{equation}
(recall that $\lambda_- = \lambda(x=0) = 0$).
Expanding the expression in~\eqref{Sigslowroll} to second order
in $x = 1 + \bar{\varphi}$ yields the following expressions:
\begin{equation}\label{secasympSig}
\begin{split}
\varphi^\prime = \sqrt{3}u &\approx \ell_1x\left\{1 - \left[\left(\frac{2}{3\sqrt{6\alpha}}\right)\ell_1 - \frac12\left(\frac{\ell_2}{\ell_1}\right)\right]x\right\}\\
&= \ell_1(1 + \bar{\varphi})\left\{1 - \left[\left(\frac{2}{3\sqrt{6\alpha}}\right)\ell_1 - \frac12\left(\frac{\ell_2}{\ell_1}\right)\right](1 + \bar{\varphi})\right\}.
\end{split}
\end{equation}
To improve the accuracy we can use the above expression to obtain
the following $[1,1]_{\varphi^\prime}$ Pad{\'e} approximant:
\begin{equation}\label{2ndPade}
\begin{split}
\varphi^\prime = \sqrt{3}u  &\approx
\frac{\ell_1 x}{1 + \left(\frac{2}{3\sqrt{6\alpha}} \ell_1 - \frac{\ell_2}{2\ell_1} \right)x}\\
&=  \frac{\ell_1(1 + \bar{\varphi})}{1 + \left[\left(\frac{2}{3\sqrt{6\alpha}}\right)\ell_1 - \frac12\left(\frac{\ell_2}{\ell_1}\right)\right](1 + \bar{\varphi})}.
\end{split}
\end{equation}

Next, we will show that the above slow-roll approximations yield approximations
for the center manifold (CM) orbit of $\mathrm{dS}^-$ in the present
dynamical systems formulation.

\subsection{Center manifold analysis of $\mathrm{dS}^-$\label{center}}

Linearizing the dynamical system~\eqref{Dynsyspurescalarfield}
around $\mathrm{dS}^-$ at $(\bar{\varphi},u) = (-1,0)$
yields the tangent spaces
\begin{subequations}\label{linearanalysis}
\begin{align}
E^s &= \{(\bar{\varphi},u)|\, 1 + \bar{\varphi} = 0\},\\
E^c &= \left\{(\bar{\varphi},u)|\, \sqrt{3}u - (1+\bar{\varphi})\ell_1 = 0\right\}.
\end{align}
\end{subequations}
%
%
%
Adapting the variables to the center manifold,
\begin{equation}
x := 1 + \bar{\varphi}, \qquad y := \sqrt{3}u - \ell_1x,
\end{equation}
results in that~\eqref{Dynsyspurescalarfield} yields
the dynamical system
\begin{subequations}\label{backgroundeqcuv}
\begin{align}
x^\prime &= \frac{1}{\sqrt{6\alpha}}(2-x)x(y + \ell_1 x), \label{xEq}\\
y^\prime &= \frac12\left[6 - (y + \ell_1 x)^2\right](\lambda(x) - y - \ell_1 x) - \ell_1 x^\prime .
\end{align}
\end{subequations}
The center manifold $W^c$ is obtained as the graph $y = h(x)$ near
$(y,x) = (0,0)$ ({\it i.e.}, use $x$ as an independent variable), where $h(0) =0$
(fixed point condition) and $\left.\frac{dh}{dx}\right|_{x=0} = 0$ (tangency condition).
Inserting $y=h(x)$ into~\eqref{backgroundeqcuv} and using $x$ as the
independent variable leads to
\begin{equation}\label{hx}
\begin{split}
& \frac{1}{\sqrt{6\alpha}}(2-x)x\left(h(x) + \ell_1 x\right)\left(\frac{dh}{dx} + \ell_1\right) \\
& \quad -
\frac12\left[6  - (h(x) + \ell_1 x)^2 \right](\lambda(x) - h(x) - \ell_1 x) = 0.
\end{split}
\end{equation}
This equation can be solved approximately by representing $h(x)$ as the formal power series
\begin{equation}
h(x) = \sum_{i=2}^n a_ix^i + {\cal O}(x^{n+1}) \qquad \text{as}\qquad x\rightarrow 0,
\end{equation}
and by using the Taylor series~\eqref{taylorlu} of $\lambda(x)$.
Inserting these series expansions into~\eqref{hx} and solving algebraically for the
coefficients $a_i$ results in
\begin{equation}\label{pCM}
\begin{split}
\varphi^\prime &= \sqrt{3}u \approx
\ell_1x\left\{1 - \left[\left(\frac{2}{3\sqrt{6\alpha}}\right)\ell_1 - \frac12\left(\frac{\ell_2}{\ell_1}\right)\right]x\right\},
\end{split}
\end{equation}
which is the same expression as in~\eqref{secasympSig}, obtained by an asymptotic expansion
of the slow-roll expansion up to second order, {\it i.e.} taking the
slow-roll expansion and expanding it in the regime where it is expected
to be asymptotically exact, {\it i.e.} at the de Sitter fixed point $\mathrm{dS}^-$,
yields the same result as the center manifold analysis.
Note, however, that the slow-roll approximations, which involve $\lambda(x)$, are not the same
approximations as the ones that are Taylor expanded in $x$.

Finally, a similar analysis holds for the sink $\mathrm{dS}^{+}$ and leads to closely related
results, but where $\mathrm{dS}^+$ is a center sink. In this case the expansion
also yields an approximation for the analytical center manifold orbit $W^c$, but note that
all orbits are asymptotically tangential to $E^c$ in this case.

The center subspace $W^c$ of $\mathrm{dS}^-$ thereby corresponds to the separatrix
inflationary center manifold (CM) orbit originating from that fixed point, with the tangent
$u = (1+\bar{\varphi})\lambda^{(1)}_-/\sqrt{3}$ when $\bar{\varphi} \rightarrow -1$
where $\lambda^{(1)}_->0$ for the LC, LT, EC, ET models (see
equation~\eqref{lambdaMinusPot} in Appendix~\ref{app:asymp}). The
separatrix CM orbit, originating from $\mathrm{dS}^-$, thereby enters the
interior state space in the invariant $u>0$ region.
In the case of $\lambda_+=0$,
the center subspace $W^c$ of $\mathrm{dS}^+$ corresponds to the separatrix orbit
originating from $\mathrm{dS}^-$ approaching the sink $\mathrm{dS}^+$ with the tangent
$u = -(1-\bar{\varphi})\lambda^{(1)}_+/\sqrt{3}$ when $\bar{\varphi} \rightarrow + 1$,
where $\lambda^{(1)}_+<0$ for the LC and EC models (see equation~\eqref{lambdaPlusPot}
in Appendix~\ref{app:asymp}), where all interior orbits therefore approach
$\mathrm{dS}^+$ from $u>0$, since they are tangential to $E^c$ toward $\mathrm{dS}^+$.
When $0<\lambda_+<\sqrt{2}$ the CM orbit of $\mathrm{dS}^-$ is still a
separatrix orbit, although in this case it ends at the sink $\mathrm{P}$, for
which $u = \lambda_+/\sqrt{3} >0$.

\begin{figure}[ht!]
	\begin{center}
		\subfigure[EC: $\alpha=\frac{7}{3}$, $\nu=128$.]{\label{}
			\includegraphics[width=0.35\textwidth]{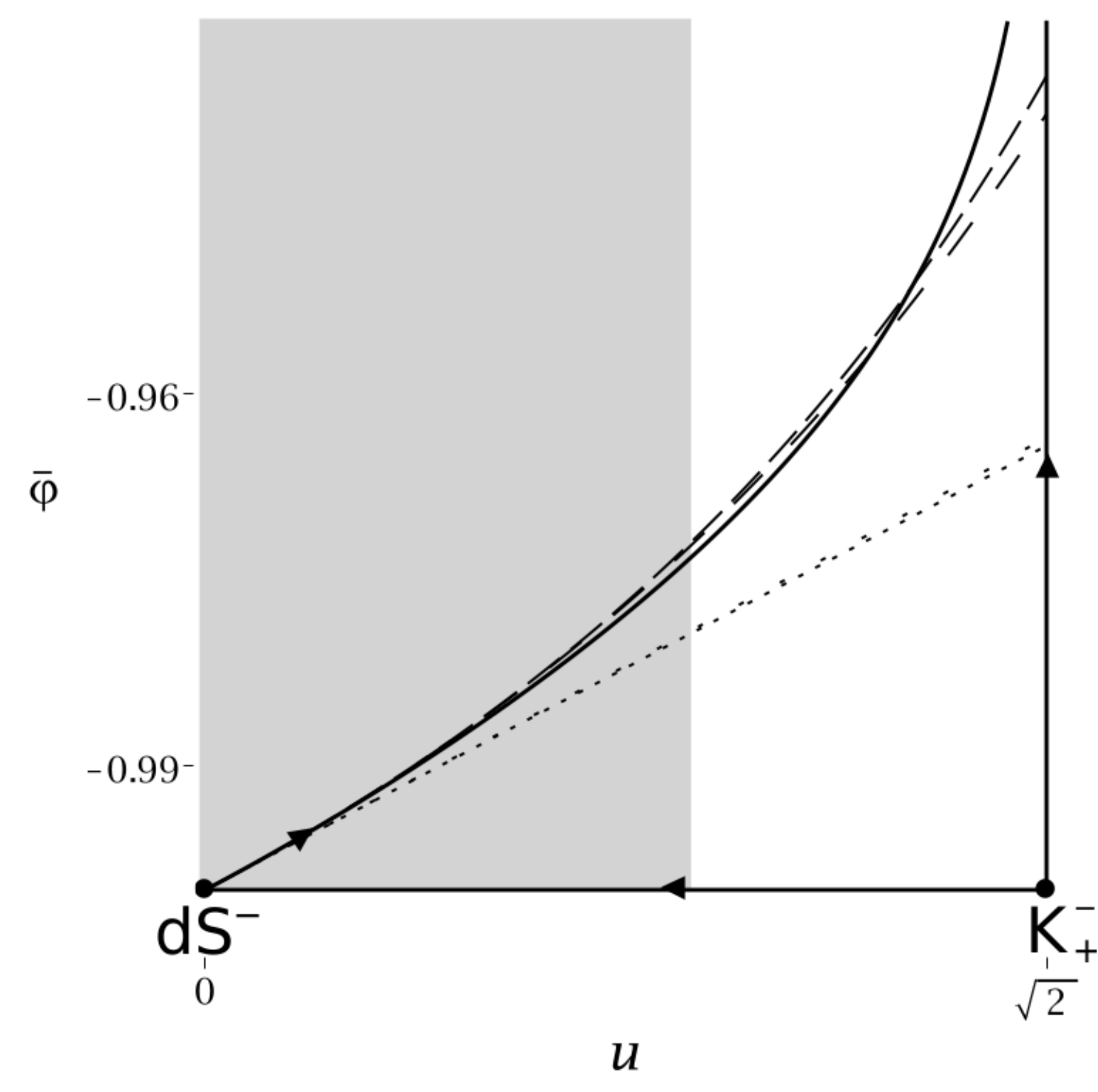}}
\hspace{1.0cm} 
		\hspace{1.0cm}
		\subfigure[ET: $\alpha=\frac{7}{3}$, $\nu=128$.]{\label{f}
			\includegraphics[width=0.35\textwidth]{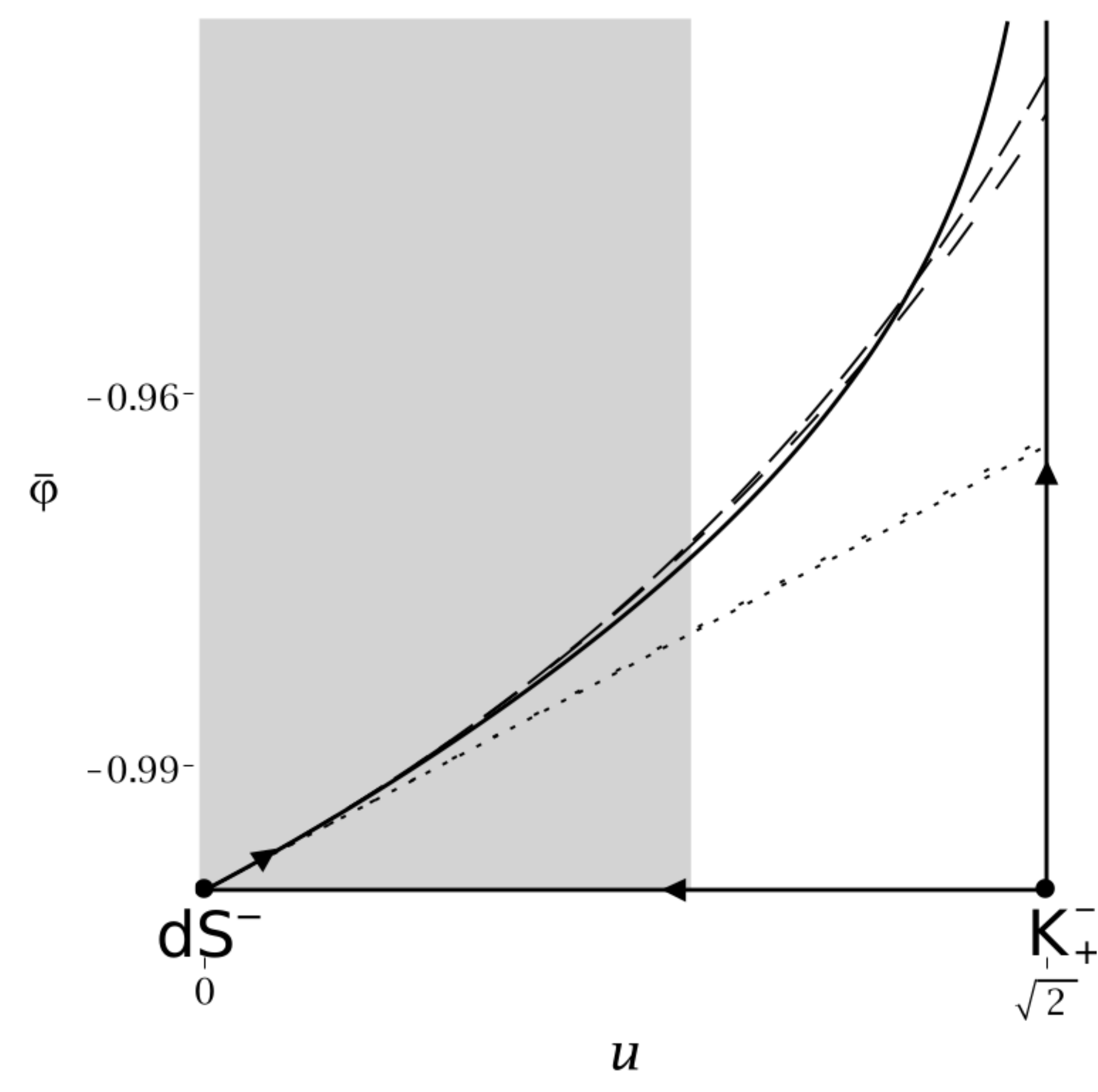}}
		\vspace{-0.5cm}
	\end{center}
	\caption{The numerically computed CM orbit is given by the continuous curve;
    the slow-roll and the 1st order center manifold approximations are given by the
    almost overlapping space-dotted and dotted curves, respectively, while
    the slow-roll and CM Pad\'e $[1,1]_{\varphi^\prime}$ approximants are represented
    by the space-dashed and dashed curves, respectively. The grey regions correspond
    to when $q<0$.}
	\label{CompPot}
\end{figure}

The accuracy of the above approximations for the CM orbit are exemplified
by a comparison with a numerical calculation of the CM orbit for the EC and ET
models with $\alpha = 7/3$ and $\nu = 128$, given in Figure~\ref{CompPot}.
%


The inflationary center manifold orbit attracts nearby orbits on the
$w_{\gamma\mathrm{m}} = 1/3$ boundary since
the fixed point $\mathrm{dS}^-$ has a center saddle structure $E^s$, $E^c$ at $\mathrm{dS}^-$,
where the stable manifold $W^s$ has co-dimension one \emph{on} the radiation boundary while the
unstable manifold $W^c$ is the CM orbit. It is the co-dimension one stable manifold
on the radiation boundary in combination with the zero eigenvalue associated with
the CM orbit that makes nearby orbits be strongly attracted to it, explaining,
from a dynamical systems perspective, why the CM orbit on the
$\Omega_\gamma = \Omega_\mathrm{m}=0$
($\rho_\gamma = \rho_\mathrm{m}=0$), $\Omega_\varphi=1$ boundary
is known as the inflationary attractor solution.

Before considering some aspects concerning horizon crossing and
the end of inflation, let us review some restrictions on model parameters
motivated by inflationary considerations.

\subsection{Parameter restrictions from the inflationary epoch}

As discussed earlier, arguments in~\cite{dimowe17,dimetal18,akretal18,akretal20}
suggest that $\alpha$ and $\nu$ take values in the
range given in equation~\eqref{alphanuapprox}, {\it i.e},
\begin{equation}\label{alphanuapprox2}
\alpha \sim {\cal O}(1),\qquad \nu \sim {\cal O}(100).
\end{equation}

By using the slow-roll approximation during the inflationary epoch,
Akrami {\it et al.}~\cite{akretal20} determine the inflationary plateaux value $V_-$
of the potential, which according to equation (2.17) in Akrami {\it et al.}~\cite{akretal20}
is given by
\begin{equation}\label{inflV}
V_- \approx \frac{(12\pi)^2\alpha\Delta N {\cal A}_s}{(2\Delta N - 3\alpha)^3},
\end{equation}
where ${\cal A}_s$ is the amplitude of the power spectrum of primordial scalar
perturbations. As follows from equation (2.2) in~\cite{akretal20},
\begin{equation}\label{inflDeltaN}
\Delta N \approx \frac{2}{1 - n_s}
\end{equation}
is the number of $e$-foldings during inflation through the COBE/Planck
normalization, where $n_s$ is the (primordial scalar tilt) spectral index.
Using values based on $\Lambda$CDM, we obtain the following from
Table 1 in Akrami {\it et al.}~\cite{akretal20}:
\begin{equation}
{\cal A}_s = 2.12605\cdot 10^{-9},\qquad n_s = 0.966.
\end{equation}
Inserting these values into~\eqref{inflDeltaN} and~\eqref{inflV}
yields\footnote{Based on $\Delta N\gg\alpha$, equation~\eqref{inflV} can be
approximated by $V_- \approx 18\pi^2\alpha{\cal A}_s/(\Delta N)^2 = \frac92\pi^2\alpha{\cal A}_s(1-n_s)^2$,
which for the above values of ${\cal A}_s$ and $n_s$ yields $V_- \approx 10^{-10}\alpha$.}
\begin{equation}
V_- \approx \frac{6.985919\cdot 10^{-10}\alpha}{(4 - 0.102\alpha)^3},
\end{equation}
which for $\alpha = \frac73$ results in
\begin{equation}\label{Vmvalue}
V_- \approx 3.061568\cdot 10^{-10}.
\end{equation}
%



\subsection{Horizon crossing and the end of inflation\label{app:horinfl}}

Inflationary considerations, {\it e.g.} in~\cite{dimowe17}, are based on the first order
expansion of the slow-roll approximation, {\it i.e.} the first order term in
equation~\eqref{pCM}, given by
\begin{equation}\label{varphipslowroll}
\varphi^\prime = \lambda_-^{(1)}(1 + \bar{\varphi}) = \frac{2\nu}{\sqrt{6\alpha}}(1 + \bar{\varphi})
\end{equation}
for the EC model, which is also used as an approximation for the ET model for which
$\lambda_-^{(1)} = \frac{2\nu}{\sqrt{6\alpha}(1 - e^{-2\nu})} \approx \frac{2\nu}{\sqrt{6\alpha}}$,
due to that $\nu \gg 1$. This, in combination with
\begin{equation}\label{barvarphivarphiapprox}
\bar{\varphi}\approx -1 + 2e^{\frac{2\varphi}{\sqrt{6\alpha}}}
\end{equation}
for both the EC and ET models, as follows from~\eqref{phivarphi} and~\eqref{barvarphidef},
yields
\begin{equation}\label{varphiend}
\varphi^\prime = \frac{4\nu}{\sqrt{6\alpha}}e^{\frac{2\varphi}{\sqrt{6\alpha}}}.
\end{equation}

At the end of inflation, {\it i.e.} at $q=0$ on the $v=1/\sqrt{3}$ boundary
where $u=\sqrt{2/3}$, it follows, since $\varphi^\prime = \sqrt{3}u$, that
\begin{equation}
\varphi^\prime_{\mathrm{end}} = \sqrt{2}.
\end{equation}
Inserting $\varphi^\prime_{\mathrm{end}} = \sqrt{2}$
into~\eqref{varphipslowroll}, and also using~\eqref{barvarphivarphiapprox}, gives,
\begin{equation}\label{phiend}
\bar{\varphi}_\mathrm{end} \approx -1 + \frac{\sqrt{3\alpha}}{\nu},\qquad
\varphi_\mathrm{end} \approx -\frac{\sqrt{6\alpha}}{2}\ln{\left(\frac{2\nu}{\sqrt{3\alpha}}\right)}.
\end{equation}

Using that $\varphi_{\mathrm{cross}} \ll \varphi_{\mathrm{end}}$ and~\eqref{varphiend}, \eqref{barvarphivarphiapprox}
leads to
\begin{equation}
\begin{split}
\Delta N &= N_{\mathrm{end}} - N_{\mathrm{cross}} =
\int^{\varphi_\mathrm{end}}_{\varphi_\mathrm{cross}}\left(\varphi^{\prime}\right)^{-1}d\varphi \\
&\approx \frac{\sqrt{6\alpha}}{4\nu}\int^{\varphi_\mathrm{end}}_{\varphi_\mathrm{cross}}e^{-\frac{2\varphi}{\sqrt{6\alpha}}}d\varphi
\approx \frac{3\alpha}{4\nu}e^{-\frac{2\varphi_\mathrm{cross}}{\sqrt{6\alpha}}} \approx \frac{3\alpha}{2\nu}(1 + \bar{\varphi}_\mathrm{cross})^{-1},
\end{split}
\end{equation}
which results in the following leading order approximations:
\begin{equation}\label{AppCross}
\bar{\varphi}_\mathrm{cross} \approx -1 + \frac{3\alpha}{2\nu\Delta N},\qquad \varphi_\mathrm{cross}
\approx -\frac{\sqrt{6\alpha}}{2}\ln{\left(\frac{4\nu\Delta N}{3\alpha}\right)},
\end{equation}
which together with~\eqref{varphipslowroll} or, alternatively, \eqref{varphiend} yields
\begin{equation}
\varphi^{\prime}_\mathrm{cross} \approx \frac{\sqrt{6\alpha}}{2\Delta N}.
\end{equation}
Taking $\Delta N = 60$ for $\alpha=7/3$ and $\nu=128$ results in that the above equations
give $\bar{\varphi}_\mathrm{end}\approx -0.979330$, $\varphi_\mathrm{end}\approx -8.553845$,
and $(\bar{\varphi}_\mathrm{cross},\varphi^\prime_{\mathrm{cross}})\approx(-0.999544,0.0311805)$,
$\varphi_\mathrm{cross}\approx-15.951660$. Let us now connect this with our state
space description.

To have around $\Delta N =  N_\mathrm{end} - N_\mathrm{cross} \approx 60$ $e$-folds
between horizon crossing and the end of inflation at $q=0$ for the CM orbit
and for orbits that intermediately shadow the CM orbit during their evolution
implies that horizon crossing initial data have to be \emph{extremely}
close to the de Sitter fixed point $\mathrm{dS}^-$.
For example, a numerical computation for the CM orbit for the EC and ET models with
$\alpha = 7/3$ and $\nu=128$ results in that $\Delta N = 60$ yields
$(\bar{\varphi}_\mathrm{cross},u_\mathrm{cross}) \approx (-0.999542,0.018106)$, where
$\bar{\varphi}_\mathrm{cross}$ and $u_\mathrm{cross}$
correspond to $\varphi_\mathrm{cross}=-15.678944$ and $\varphi^\prime_\mathrm{cross}\approx 0.0313613$,
respectively, while $q=0$ leads to $u_\mathrm{end} = \sqrt{2/3}$ and, numerically,
$\bar{\varphi}_\mathrm{end}\approx-0.973204$, which corresponds to
$\varphi_\mathrm{end}\approx -8.04298889$, while $\varphi_\mathrm{end}^\prime = \sqrt{2}$
(note that the previous results based on the slow-roll approximation are in
good agreement with these numerical results).

Finally, one might ask: What is natural with a region of initial data for
inflation that is \emph{extremely} close to the fixed point
$\mathrm{dS}^-$ on the radiation boundary? To obtain the present
variables we have performed a non-canonical transformation, and one might argue that
this region should be translated into a phase space region where a symplectic geometry
measure can be applied, but should it be for $\phi$, $\varphi$ or $\bar{\varphi}$
and does this matter? For discussions about the naturalness of inflationary data,
see, {\it e.g.}, \cite{remcar13,remcar14,gruslo16}.

\section{Connecting the end of inflation with scalar field freezing\label{sec:freezing}}

Let us now turn to some aspects concerning reheating and connecting
the end of inflation with scalar field freezing, which subsequently
initiates quintessence evolution, with a focus on the dynamical systems
perspective. Setting $q=0$ in~\eqref{inflq} to obtain the end of inflation
results in $u^2 = 2/3$ and inserting this into~\eqref{H2Int} leads to
\begin{equation}
H^2_\mathrm{end} = \frac12 V(\bar{\varphi}_\mathrm{end}) =
\frac12V_-\bar{V}(\bar{\varphi}_\mathrm{end}).
\end{equation}
Using~\eqref{lambdaECmain} for the EC models, which also provides an approximation for the present ET models,
and~\eqref{phiend} then gives
\begin{equation}
H^2_\mathrm{end} = \frac12V_-e^{-\nu(1 + \bar{\varphi}_\mathrm{end})} \approx \frac12 V_-e^{-\sqrt{3\alpha}}.
\end{equation}
Inserting~\eqref{inflV} into this expression and using~\eqref{inflDeltaN} leads to
\begin{equation}\label{H2endns}
H^2_\mathrm{end} = \frac{(12\pi)^2{\cal A}_s\alpha e^{-\sqrt{3\alpha}}\Delta N}{2(2\Delta N - 3\alpha)^3} \approx
\frac{(12\pi)^2 {\cal A}_s\alpha e^{-\sqrt{3\alpha}}}{16\Delta N^2}
= \frac{1}{64}(12\pi)^2 {\cal A}_s\alpha e^{-\sqrt{3\alpha}}(1 - n_s)^2
\end{equation}
when $\Delta N \gg \alpha$.

Following the reasoning of Dimopoulos and Owen in~\cite{dimowe17},
the least efficient reheating mechanism is \emph{gravitational reheating}:
\begin{equation}
(\Omega_{\gamma,\mathrm{end}})_\mathrm{gr} \lesssim \Omega_{\gamma,\mathrm{end}} \lesssim 1,
\end{equation}
where
\begin{equation}
(\Omega_{\gamma,\mathrm{end}})_\mathrm{gr} = \frac{q_* g_{*,\mathrm{end}}}{(12\pi)^2} H^2_\mathrm{end},
\end{equation}
where $q_* \sim 1$ is an efficiency factor and $g_{*,\mathrm{end}} = {\cal O}(100)$ is the number of
effective relativistic degrees of freedom at the energy scale of inflation. Together with~\eqref{H2endns}
this leads to
\begin{equation}
(\Omega_{\gamma,\mathrm{end}})_\mathrm{gr} = \frac{1}{64}q_* g_{*,\mathrm{end}} {\cal A}_s\alpha e^{-\sqrt{3\alpha}}(1 - n_s)^2.
\end{equation}
Setting $q_* = 1$, 
${\cal A}_s = 2.12605\cdot 10^{-9}$, $n_s = 0.966$ gives
%
%
%
\begin{equation}
(\Omega_{\gamma,\mathrm{end}})_\mathrm{gr} = 3.84g_{*,\mathrm{end}}\,\alpha e^{-\sqrt{3\alpha}}\cdot 10^{-11}.
\end{equation}

After some additional arguments concerning reheating, Dimopoulos and Owen~\cite{dimowe17}
turn to when $V(\varphi) \approx 0$ and obtain the following approximation
$\varphi_* \approx \varphi_\mathrm{end} + \sqrt{\frac23}[1 - \frac32\ln(\Omega_{\gamma,\mathrm{end}})]$.
Their analysis corresponds to an approximation based on the
$w_{\gamma\mathrm{m}} = 1/3$, $u = \sqrt{2}$ ($\Omega_V=0$ and hence $V=0$, $w_\varphi = 1$) boundary,
which we now turn to from a dynamical systems perspective.

The equations on this boundary are easily solvable, especially if one
uses $\varphi$ instead of $\bar{\varphi}$. Let us begin with the case
$u=\sqrt{2}$ (the case $u = - \sqrt{2}$ is easily obtained by means of a
discrete symmetry, but this is not relevant in the present context)
and $w_{\gamma\mathrm{m}} = 1/3$, which leads to
\begin{subequations}\label{Dynsysstiffrad}
\begin{align}
\varphi^\prime &= 3\sqrt{2}v,\\
v^\prime &= -(1 - 3v^2)v, \label{vstiffrad}
\end{align}
\end{subequations}
from which we obtain
\begin{equation}
\frac{dv}{d\varphi} = - \frac{1}{3\sqrt{2}}\left(1 - 3v^2\right).
\end{equation}
The solution to this equation is given by
\begin{equation}\label{varphistiffradsol}
\varphi = \varphi_\mathrm{i} +
\sqrt{\frac32}\ln\left[\frac{(1 - \sqrt{3}v)(1 + \sqrt{3}v_\mathrm{i})}{(1 - \sqrt{3}v_\mathrm{i})(1 + \sqrt{3}v)}\right],
\end{equation}
where $(\varphi_\mathrm{i},v_\mathrm{i})$ denotes an initial point, which
determines a particular orbit. The limit $v\rightarrow 0$ yields
the following freezing value $\varphi_*$ for $\varphi$:
\begin{equation}
\varphi_* = \lim_{v\rightarrow 0}\varphi = \varphi_\mathrm{i} - \sqrt{\frac32}\ln\left[\frac{1 - \sqrt{3}v_\mathrm{i}}{1 + \sqrt{3}v_\mathrm{i}}\right]
= \varphi_\mathrm{i} - \sqrt{\frac32}\ln\left[\frac{1 - \sqrt{1 - \Omega_{\gamma,\mathrm{i}}}}{1 + \sqrt{1 - \Omega_{\gamma,\mathrm{i}}}}\right],
\end{equation}
where we have used that $\Omega_\gamma = 1 - \Omega_\varphi = 1 - 3v^2$
on this boundary (for the orbit structure, see Figure~\ref{fig:Boxstatespace}).
When $\Omega_{\gamma,\mathrm{i}}$ is close to one, the
above exact expression yields the following approximation:
\begin{equation}
\varphi_* \approx \varphi_\mathrm{i} + \sqrt{\frac32}\left[2\ln 2 - \ln(\Omega_{\gamma,\mathrm{i}})\right].
\end{equation}
Setting the initial values to the of inflation values for
$\varphi_\mathrm{i} = \varphi_\mathrm{end}$ and $\Omega_{\gamma,\mathrm{i}} = \Omega_{\gamma,\mathrm{end}}$
yield the following approximate freezing value $\varphi_*$:
\begin{equation}
\varphi_* \approx \varphi_\mathrm{end} + \sqrt{\frac32}\left[2\ln 2 - \ln(\Omega_{\gamma,\mathrm{end}})\right].
\end{equation}
We note that this approximation differs slightly from the less rigorously derived approximation in~\cite{dimowe17},
$\varphi_* \approx \varphi_\mathrm{end} + \sqrt{\frac23}[1 - \frac32\ln(\Omega_{\gamma,\mathrm{end}})]$,
where the difference between the two expressions is given by $\sqrt{\frac32}(2\ln 2) - \sqrt{\frac23} \approx 0.88$.

It is even possible to give the solution on the $u=\pm \sqrt{2}$ boundaries that include both radiation and matter
in terms of a quadrature. We first note that due to~\eqref{udef}
\begin{equation}
\varphi^\prime = \pm\sqrt{6}\sqrt{\frac{\rho_\varphi}{3H^2}},
\end{equation}
which in combination with
\begin{equation}
\rho_\varphi = \rho_{\varphi,\mathrm{i}}e^{-6N},\qquad
3H^2 = \rho_{\varphi,\mathrm{i}}e^{-6N} + \rho_{\gamma,\mathrm{i}}e^{-4N} + \rho_{\mathrm{m},\mathrm{i}}e^{-3N},
\end{equation}
where $N$ is the number of $e$-folds from some convenient initial time $t_\mathrm{i}$, gives
\begin{equation}
\varphi = \varphi_\mathrm{i} \pm \sqrt{6}\int_0^N(1 + r_\gamma e^{2\tilde{N}} + r_\mathrm{m}e^{3\tilde{N}})^{-1/2}d\tilde{N},
\end{equation}
where $r_\gamma := \rho_{\gamma,\mathrm{i}}/\rho_{\varphi,\mathrm{i}}$ and
$r_\mathrm{m} := \rho_{\mathrm{m},\mathrm{i}}/\rho_{\varphi,\mathrm{i}}$. Specializing to
$\rho_{\mathrm{m},\mathrm{i}}=0$ and hence $r_\mathrm{m}=0$ leads to an explicitly
solvable integral which can be used to obtain the previous
result~\eqref{varphistiffradsol}.

\section{Inflationary $\alpha$-attractor quintessence evolution\label{sec:quintevol}}

Although arguably unfair, observations and initial data are
usually interpreted in the context of the $\Lambda$CDM paradigm, and we
will not deviate from this tradition in this paper when exploring
cosmologically observationally viable solutions. We will therefore follow~\cite{akretal18,akretal20}
and define observationally viable solutions as solutions associated with
$\Lambda$CDM compatible initial data and time developments
that are fairly close to $\Lambda$CDM during the observable quintessence epoch.
Appendix~\ref{app:LCDMrad} contains exact relations that are useful
for comparisons between models with a scalar field and $\Lambda$CDM.

To find compatibility with observations in the present context,
one has to address the problem of simultaneously choosing
(i) the inflationary plateaux-normalized potential $\bar{V}(\varphi)$, which yields
$\lambda = -V_{,\varphi}/V =  -\bar{V}_{,\varphi}/\bar{V}$, where, in
the case of the EC and ET models, specification of $\bar{V}(\bar{\varphi})$
entails choosing the parameter $\nu$, as seen from equation~\eqref{barvexamplesECET},
(ii) the parameters, $\alpha$, $V_-/3H_0^2$, which, together with $\bar{V}(\bar{\varphi})$,
form the present $\alpha$-attractor model space, and (iii) initial data
$(\bar{\varphi}_\mathrm{i}, u_\mathrm{i}, v_\mathrm{i},w_{\mathrm{rm},\mathrm{i}})$
at some initial time $N_\mathrm{i}$.

As regards initial data and solution structures we note the following:
\begin{itemize}
\item To obtain $\sim 60$ $e$-folds of inflation requires solutions to be \emph{extremely} close to the
$\mathrm{dS}^-$ fixed point on the radiation boundary $w_{\gamma\mathrm{m}}= 1/3$, as illustrated by the
CM orbit (the inflationary attractor solution) which requires, as exemplified by the EC and ET
$\alpha = 7/3$, $\nu=128$ models, $(\bar{\varphi}_\mathrm{i},u_\mathrm{i}) \approx (-0.999542,0.018106)$
({\it i.e.} $\varphi_\mathrm{i}=-15.678944$, $\varphi^\prime_\mathrm{i}\approx 0.0313613$),
where $t_\mathrm{i}$ corresponds to $t_\mathrm{cross}$.

Solutions that exhibit $\sim 60$ $e$-folds of inflation
and that hence pass extremely close to the $\mathrm{dS}^-$ fixed point on the radiation boundary
$w_{\gamma\mathrm{m}}= 1/3$ then, thanks to the strong attracting properties of the CM orbit,
shadow the CM orbit extremely closely, which in turn for models with
$\alpha \sim {\cal O}(1)$, $\nu \sim {\cal O}(100)$ leads to intermediate close shadowing of
the kinaton orbit $\mathrm{K}_+^-\rightarrow\mathrm{K}_+^+$ on the $u = \sqrt{2}$,
$\Omega_\varphi = 1$, $w_{\gamma\mathrm{m}}=1/3$ boundary.
Moreover, the CM orbit and the
shadowing orbits end at the same fixed point $\mathrm{dS}^+$ when $\lambda_+=0$ ($\mathrm{P}$
when $0<\lambda_+<\sqrt{2}$) on the matter boundary, $w_{\gamma\mathrm{m}} = 0$.

%
\item As pointed out in~\cite{alhetal23}, a long radiation and matter dominated
epoch requires that orbits come very close to the $v=0$ ($\Omega_\varphi=0$)
boundary. This epoch, is initiated before primordial nucleosynthesis
and lasts until $\rho_\varphi$ makes a significant contribution, thereby commencing the quintessence epoch,
which is characterized by $\Omega_\varphi > 0.03$, as suggested in~\cite{alhetal23} and
references therein. Moreover, the radiation and matter dominated pre-quintessence time interval
is required to satisfy $\Delta N \approx 25$. In order to have a pre-quintessence
epoch with $\Omega_\varphi < 0.03$ ($v>0.1$,) and a time interval $\Delta N \approx 25$,
numerical calculations show that the EC and ET models with $\alpha = \frac73$ $\nu = 128$
requires $v_\mathrm{min} \approx 10^{-8}$ (at $u=1$ with
$\Omega_{\varphi,\mathrm{min}} \approx 10^{-15}$), which due to the closeness to
the $v=0$ boundary yields an approximately frozen value of $\varphi = \varphi_*$.

The radiation and matter dominated pre-quintessence epoch corresponds to orbits that
subsequently shadow quintessence models associated with the unstable manifold of the
line of fixed points $\mathrm{FL}_0^{\varphi_*}$ on the radiation boundary
$w_{\gamma\mathrm{m}} = 1/3$ at $\varphi = \varphi_*$, described below,
when $\Omega_\varphi \geq 0.03$ and $N \gtrsim -1.5$, cf.~\cite{alhetal23}.
\end{itemize}

Due to the uncertainties associated with reheating, we follow~\cite{akretal18} and
divide the evolution in an inflationary period from $N \sim -85$ to the end of inflation at
$N \sim -25$, described by the CM orbit, or rather, an open set of orbits that closely
shadows it, and then make a temporal jump to $N=-15$ in the radiation
dominated epoch, at a freezing value $\varphi = \varphi_*$ at the line of fixed points
$\mathrm{FL}_\mathrm{0}^{{\varphi}_*}$ at the $w_{\gamma\mathrm{m}}=1/3$ boundary, {\it i.e.}
$(\bar{\varphi},u,v,w_{\gamma\mathrm{m}}) = (\bar{\varphi}_*,0,1/\sqrt{3},1/3)$. It is the unstable
manifold of this line that yields the quintessence evolution we,
as in~\cite{akretal20}, will now focus on.



As pointed out in~\cite{akretal20}, one can view the present models as quintessence models,
with inflationary motivated parameter ranges, and only consider
the quintessence epoch, which is the problem we will now address and
situate in the present state space setting.
Since the observable quintessence epoch is finite in time, $\varphi$ only increases with a finite
amount from $\varphi_*$. This makes it more convenient to use $\varphi$ rather than $\bar{\varphi}$
when describing this epoch, although we will use $\bar{\varphi}$ to describe the entire
quintessence epoch, which also includes the future $N\in (0,\infty)$.

Next we present a new method for obtaining initial values for quintessence evolution.
Quintessence evolution is described by the unstable manifold of $\mathrm{FL}_0^{{\varphi}_*}$
on the $w_{\gamma\mathrm{m}}=1/3$ radiation boundary.
This is due to that quintessence begins when $\Omega_\gamma + \Omega_\mathrm{m}\approx 1$
in the radiation dominated regime where $\Omega_\gamma \gg \Omega_\mathrm{m}$,
and since each fixed point on $\mathrm{FL}_0^{{\varphi}_*}$ \emph{on} the $w_{\gamma\mathrm{m}}=1/3$
boundary state space ($\bar{\varphi},u,v$) has a single positive eigenvalue, corresponding
to the unstable manifold of each fixed point of $\mathrm{FL}_0^{{\varphi}_*}$,
a single zero eigenvalue corresponding to
the line of fixed points, and one negative eigenvalue. As a consequence
orbits near $\mathrm{FL}_0^{{\varphi}_*}$ are pushed toward the
orbits of the unstable manifold which they thereafter shadow. Due to this, the
quintessence evolution of the shadowing orbits are approximately described by the unstable
$\mathrm{FL}_0^{{\varphi}_*}$ orbits, which form the `unstable \emph{quintessence attractor submanifold}'.
Note, however, that the attracting property toward the unstable manifold orbits is not as
strong as that of the CM orbit, which is due to that the eigenvalue associated with the
unstable quintessence attractor manifold is positive while
that for the CM orbit is zero.

The future development of an orbit on the unstable manifold of
the line of fixed points $\mathrm{FL}_0^{{\varphi}_*}$ on the $w_{\gamma\mathrm{m}}=1/3$
boundary is obtained by picking a
point on the unstable manifold, or very near the unstable manifold, for some initial
time $N_\mathrm{i}$, which we in agreement with~\cite{akretal18}
set to $N_\mathrm{i}=-15$.
However, in contrast to~\cite{akretal18,akretal20}, where $\varphi^\prime=0$ ($u=0$) is used
as an initial value, which yields an orbit that shadows the unstable manifold, and by varying
constants in the EC and ET potentials to obtain $H=H_0$ at the present time, we use a
local analysis of the line of fixed points $\mathrm{FL}^{\varphi_*}_0$,
$(\bar{\varphi},u,v,w_{\gamma\mathrm{m}}) = (\bar{\varphi}_*,0,1/\sqrt{3},1/3)$
in the state space $(\bar{\varphi},u,v,w_{\gamma\mathrm{m}})$, which yields
the following initial values
\begin{subequations}
\begin{align}
\varphi_\mathrm{i} &\approx \varphi_*,\\
u_\mathrm{i} &\approx \frac35 \lambda_*v_\mathrm{i},
\end{align}
\end{subequations}
where $\lambda_* = \lambda(\varphi_*)$.
Second, we use the exact relation for $w_{\gamma\mathrm{m}}$ in eq.~\eqref{wrmsol}
to obtain an initial value for $w_{\gamma\mathrm{m}}$ at $N_\mathrm{i}$
connected to the present values $\Omega_{\gamma,0}$, $\Omega_{\mathrm{m},0}$
of $\Omega_\gamma$ and $\Omega_\mathrm{m}$ (alternatively one can use the
linearized version of this equation):
\begin{equation}
(w_{\gamma\mathrm{m}})_\mathrm{i} =
\frac13\left(\frac{1}{1 + \left(\frac{\Omega_{\mathrm{m},0}}{\Omega_{\gamma,0}}\right)e^{N_\mathrm{i}}}\right).
\end{equation}

To connect the unstable $\mathrm{FL}_0^{\varphi_*}$ orbits
from the radiation boundary $w_{\gamma\mathrm{m}} = 1/3$ to quintessential
$\alpha$-attractor inflation we need to ensure that the values
of $V_-/3H_0^2$ are those associated with the inflationary epoch. To do so
we can use the integral~\eqref{intbasis2}, but is more convenient to
do this in the near vicinity of the $\mathrm{FL}_0^{\varphi_*}$ line of
fixed points and use a leading order approximation
of this integral, which results in
\begin{equation}\label{integralconstrainlocal}
\frac{V_-}{3H_0^2} \approx \frac{3\Omega_{\gamma,0}v_\mathrm{i}^2e^{-4N_\mathrm{i}}}{\bar{V}(\varphi_*)},
\end{equation}
where a given value of $V_-/3H_0^2$ allows one to solve for $v_\mathrm{i}$
in terms of a given $\varphi_*$. Then $\varphi_*$ has to be varied
so that $H$, $\Omega_\gamma$, $\Omega_\mathrm{m}$ obtains their present day values
(thanks to the integral and the exact expression for $w_{\gamma\mathrm{m}}$ only one of these
variables is independent of the other two).\footnote{We follow
Akrami {\it et al.}~\cite{akretal20} and set
$\Omega_{\mathrm{m},0} = 0.32$, $\Omega_{\gamma,0} = 10^{-5}$,
and $h=0.72$ in
$$
H_0 = 100\, h\,\mathrm{km}\,\mathrm{s}^{-1}\mathrm{Mpc}^{-1} =
3.248\, h\cdot 10^{-18}\text{s}^{-1}=1.747\, h\cdot 10^{-61} \,t^{-1}_{\mathrm{Pl}},
$$
which leads to
$H_0 = 1.258\cdot 10^{-61}\text{t}^{-1}_{\mathrm{Pl}}$ and hence, using eq.~\eqref{Vmvalue}
for $V_-$ and thereby restricting to $\alpha = 7/3$,
$$
\frac{V_-}{3H^2_0} = 3.061568\cdot 10^{-10}\frac{m_\mathrm{Pl}^2}{8\pi\cdot3H^2_0}\approx
\frac{1.2182\cdot10^{-11}m^2_\mathrm{Pl}}{3\cdot (1.258\cdot 10^{-61}\text{t}^{-1}_{Pl})^2}\approx 2.5658\cdot 10^{110}.
$$
\label{H0}}


In Akrami {\it et al.}~\cite{akretal20} for the $\mathrm{EC}$ and $\mathrm{ET}$ models
with $\alpha=7/3$, and $\varphi_\mathrm{ini}=10$, they obtained the best fit
values of $\nu\approx 127.871$ and $\nu \approx 127.645$ respectively,
while we obtain the best fit values $\nu\approx 127.876$ and $\nu\approx 127.650$.
This small discrepancy is due to the slightly different initial data.
Akrami {\it et al.}~\cite{akretal20} used $\Omega_{\gamma,0}$, $\Omega_{\mathrm{m},0}$, $\varphi_*$,
and $\varphi^\prime=0$ and hence $u=0$ as their initial data. Strictly speaking, using
$u=0$ as initial data is incompatible with the inflationary paradigm since the inflationary
attractor solution resides in the $u>0$ region and since the kinaton epoch corresponds to
$u\approx + \sqrt{2}$, while models with $u=0$ as initial data originate from $\mathrm{K}^+_-$
at $(\bar{\varphi},u,v,w_{\gamma\mathrm{m}}) = (1,-\sqrt{2},\frac{1}{\sqrt{3}},\frac13)$.
The reason our results and those of Akrami {\it et al.}~\cite{akretal20} are in such good
agreement is due to the attracting nature of the unstable manifold of
$\mathrm{FL}_0^{\varphi_*}$ on the radiation boundary $w_{\gamma\mathrm{m}}=1/3$.

\begin{figure}[ht!]     
	\begin{center}
		\subfigure[$\mathrm{EC}$ model.]{\label{fig:VarphiEC}
			\includegraphics[width=0.32\textwidth]{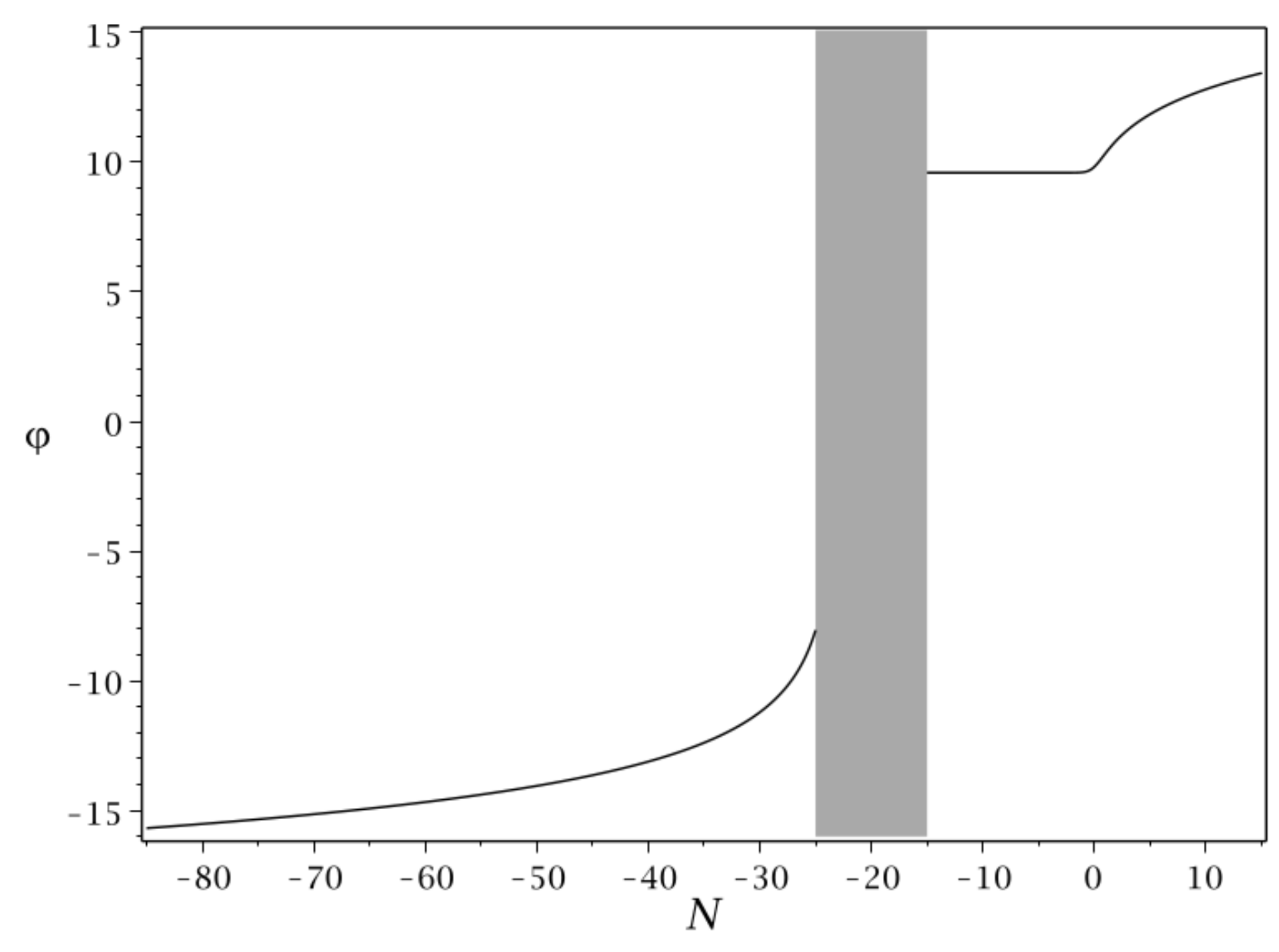}}
		\hspace{1.0cm}
		\subfigure[$\mathrm{EC}$ model.]{\label{fig:3DEC}
			\includegraphics[width=0.32\textwidth]{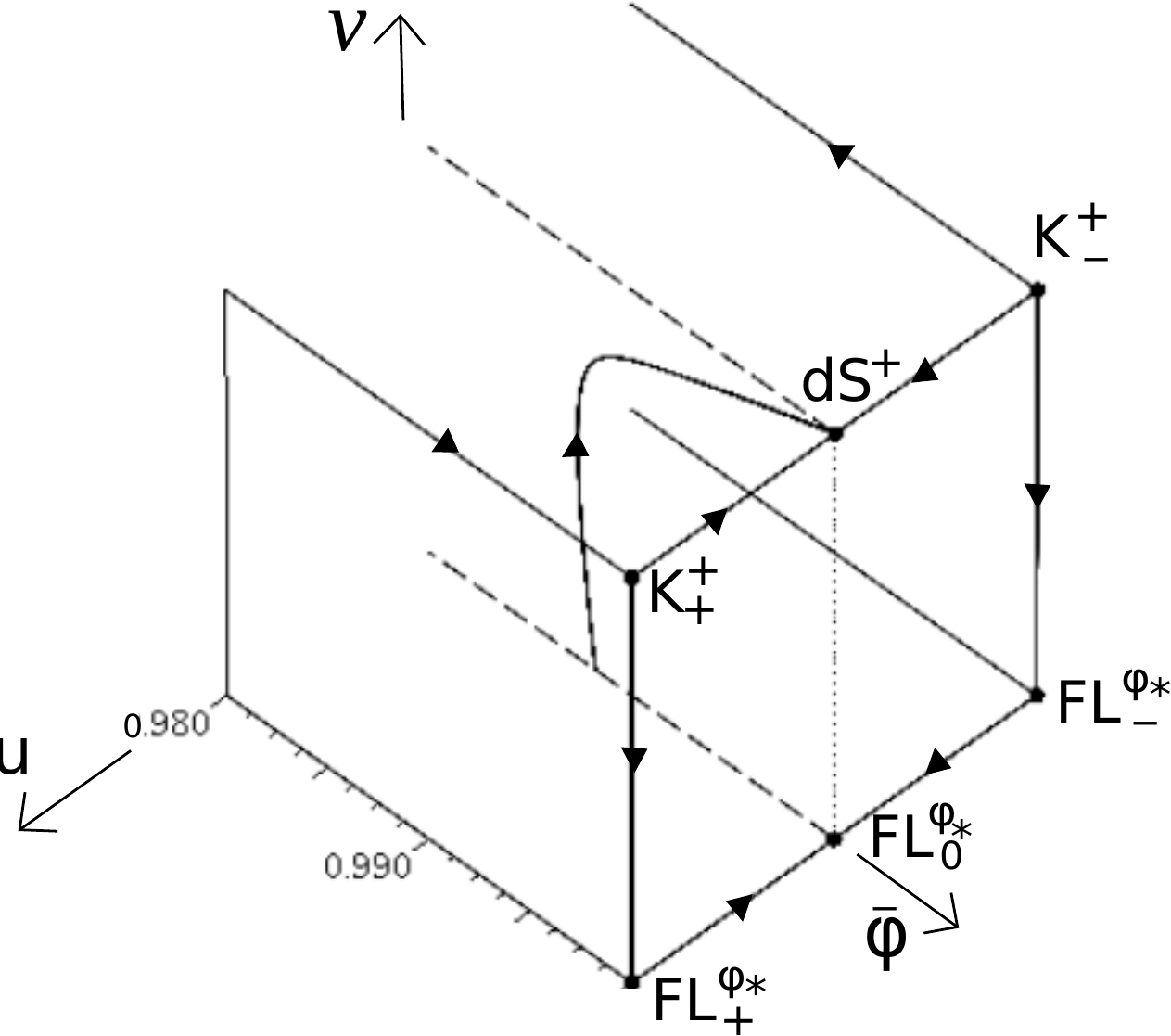}}
         \\
				\subfigure[$w_\varphi(N)$ and $w_\mathrm{eff}(N)$.]{\label{Figb_wDEEC}
			\includegraphics[width=0.32\textwidth]{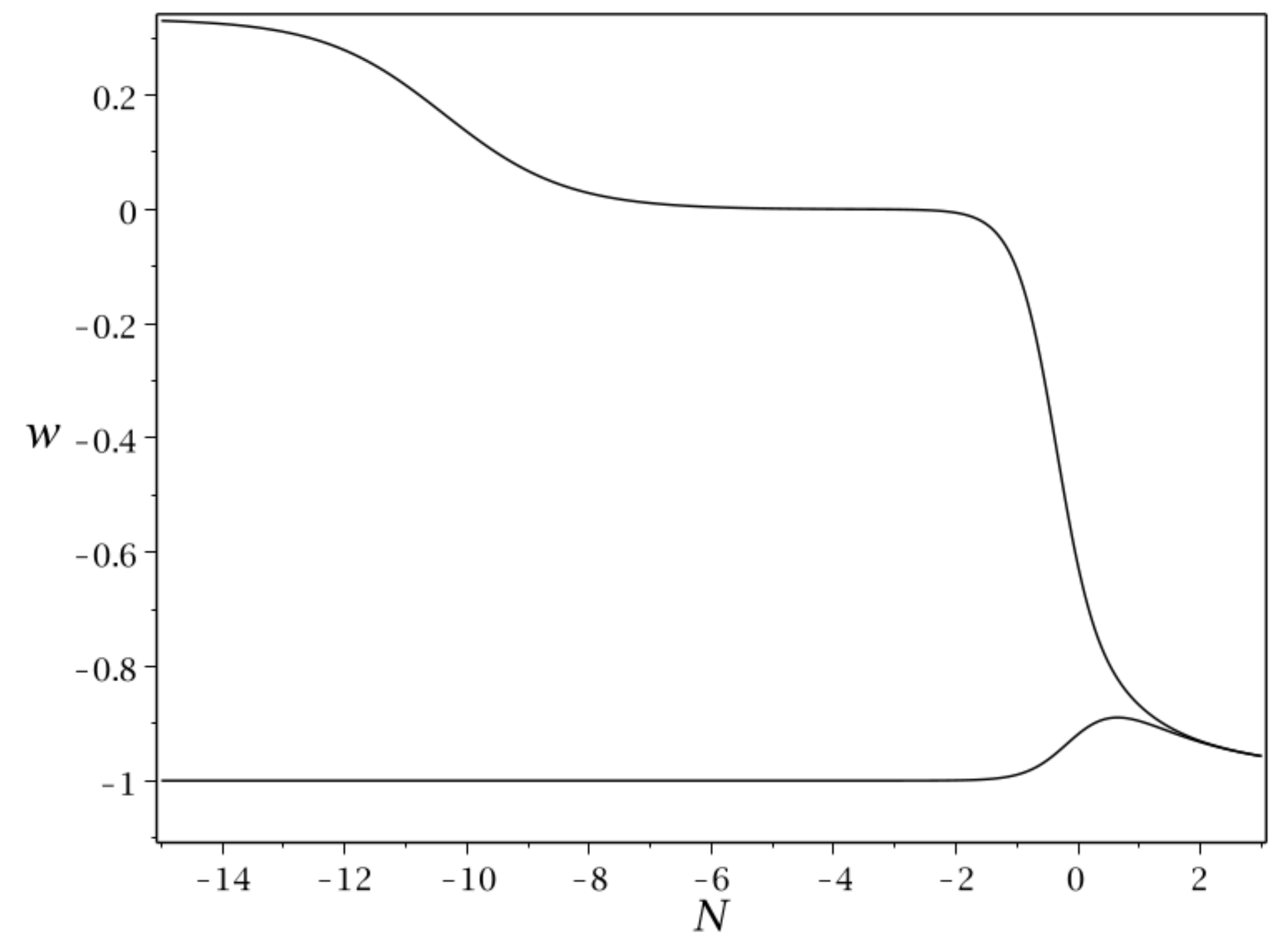}}
		\hspace{1.0cm}
		\subfigure[$\frac{H_{\varphi}}{H_\Lambda}(N)$.]{\label{Figb_EC}
			\includegraphics[width=0.32\textwidth]{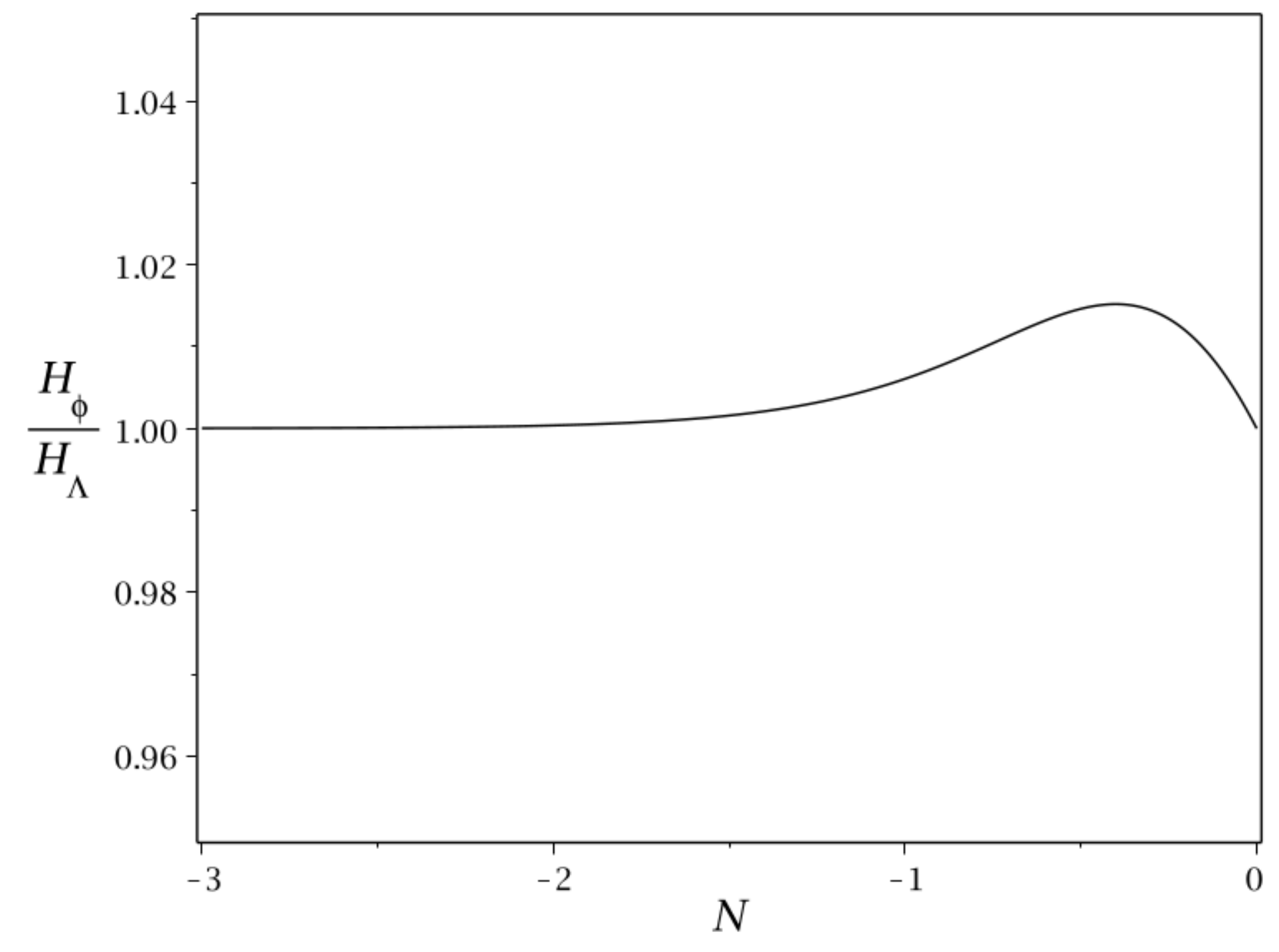}}
		\vspace{-0.5cm}
	\end{center}
	\caption{The EC model with $\alpha=\frac73$ and $\nu=128$. Figure (a) depicts the evolution in $\varphi$ during the
     inflationary and quintessence epochs, for an observationally viable freezing value of $\varphi$.
      The state space representation of the quintessence epoch of this solution, projected onto
      $(\bar{\varphi},u,v)$, is depicted in figure (b), while the associated graphs $w_\varphi(N)$
      (lower curve) and $w_\mathrm{eff}(N)$ (upper curve)
      are shown in figure (c); finally, $\frac{H_\varphi}{H_{\Lambda}}(N)$ is shown in figure (d).
      Note that $w_{\gamma\mathrm{m}}=0$ when $\bar{\varphi} = 1$.}\label{fig:5}
\end{figure}

\begin{figure}[ht!]     
	\begin{center}
		\subfigure[$\mathrm{ET}$ model.]{\label{fig:VarphiET}
			\includegraphics[width=0.32\textwidth]{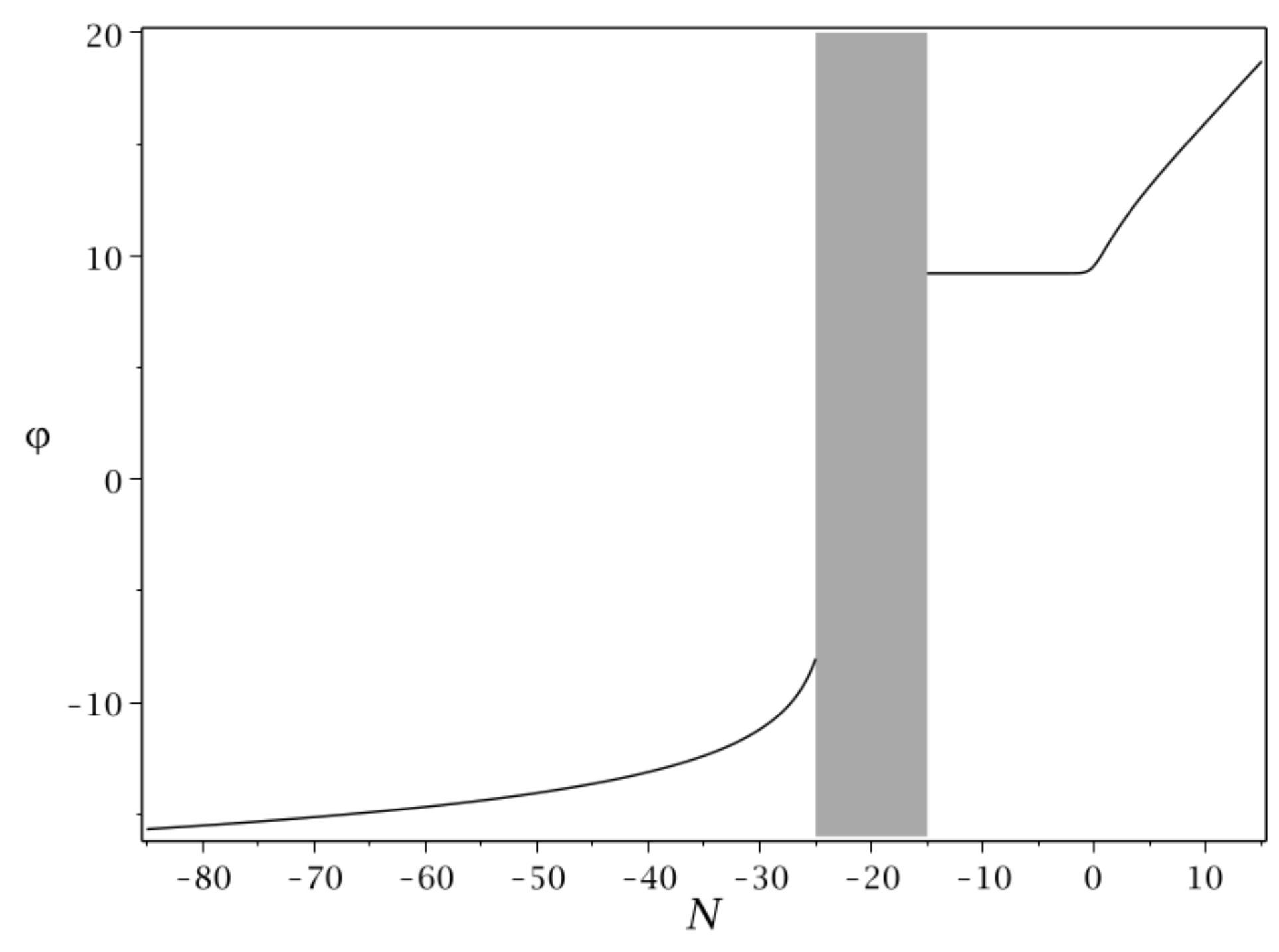}}
		\hspace{1.0cm}
		\subfigure[$\mathrm{ET}$ model.]{\label{fig:3DET}
			\includegraphics[width=0.32\textwidth]{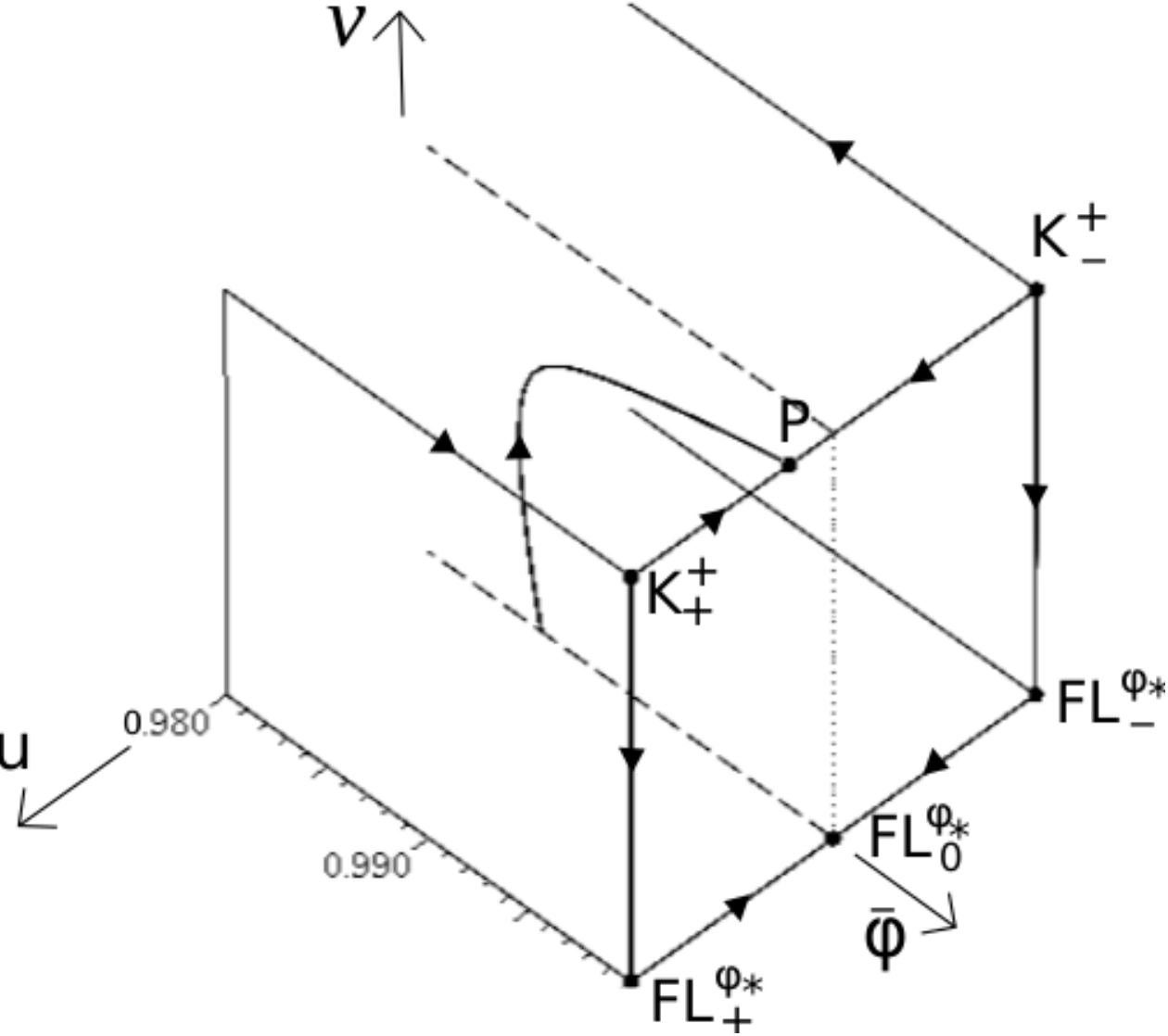}}
		\\
		\subfigure[$w_\varphi(N)$ and $w_\mathrm{eff}(N)$.]{\label{Figb_wDE}
			\includegraphics[width=0.32\textwidth]{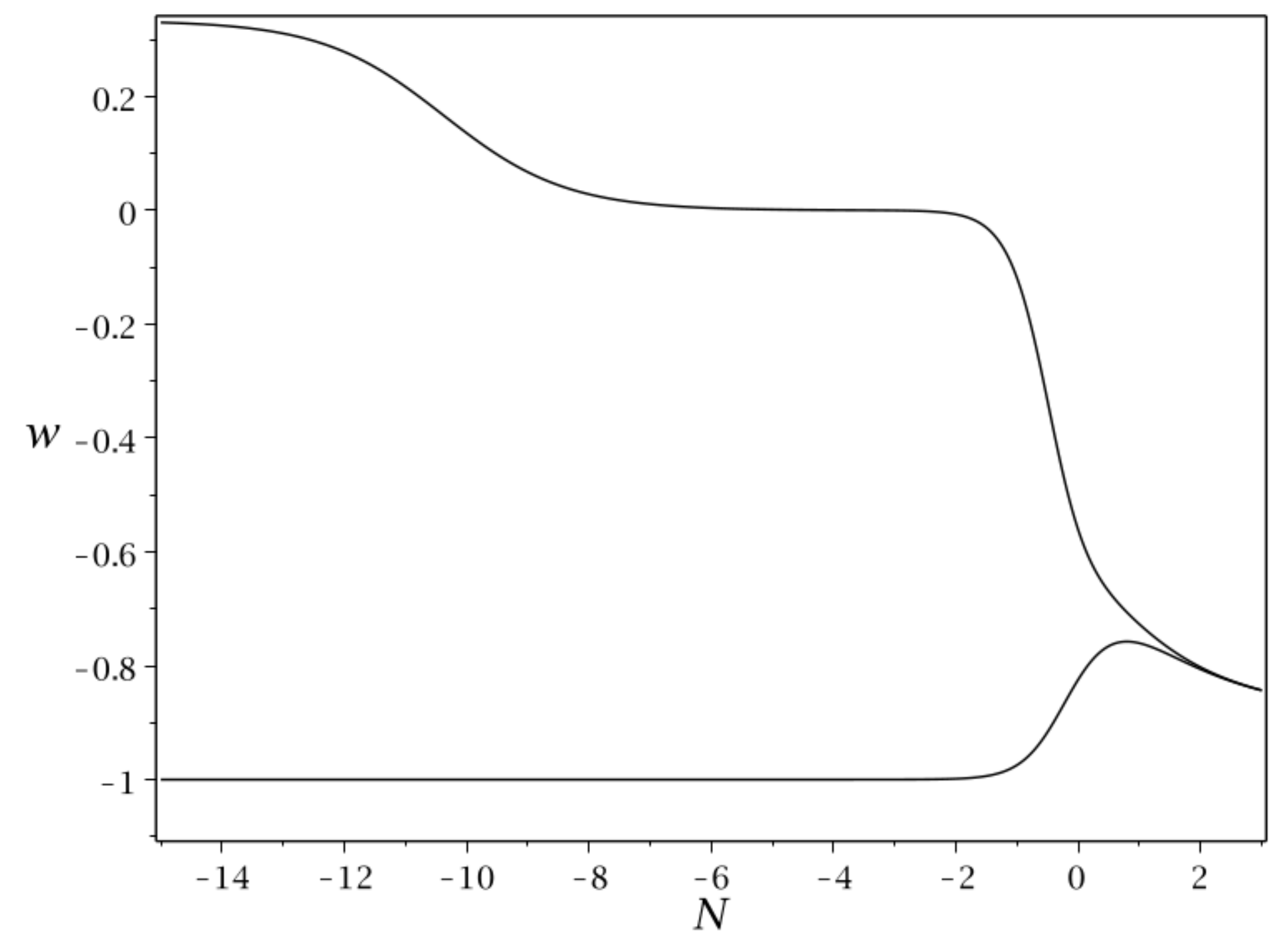}}
		\hspace{1.0cm}
		\subfigure[$\frac{H_{\varphi}}{H_\Lambda}(N)$.]{\label{Figb_E}
			\includegraphics[width=0.32\textwidth]{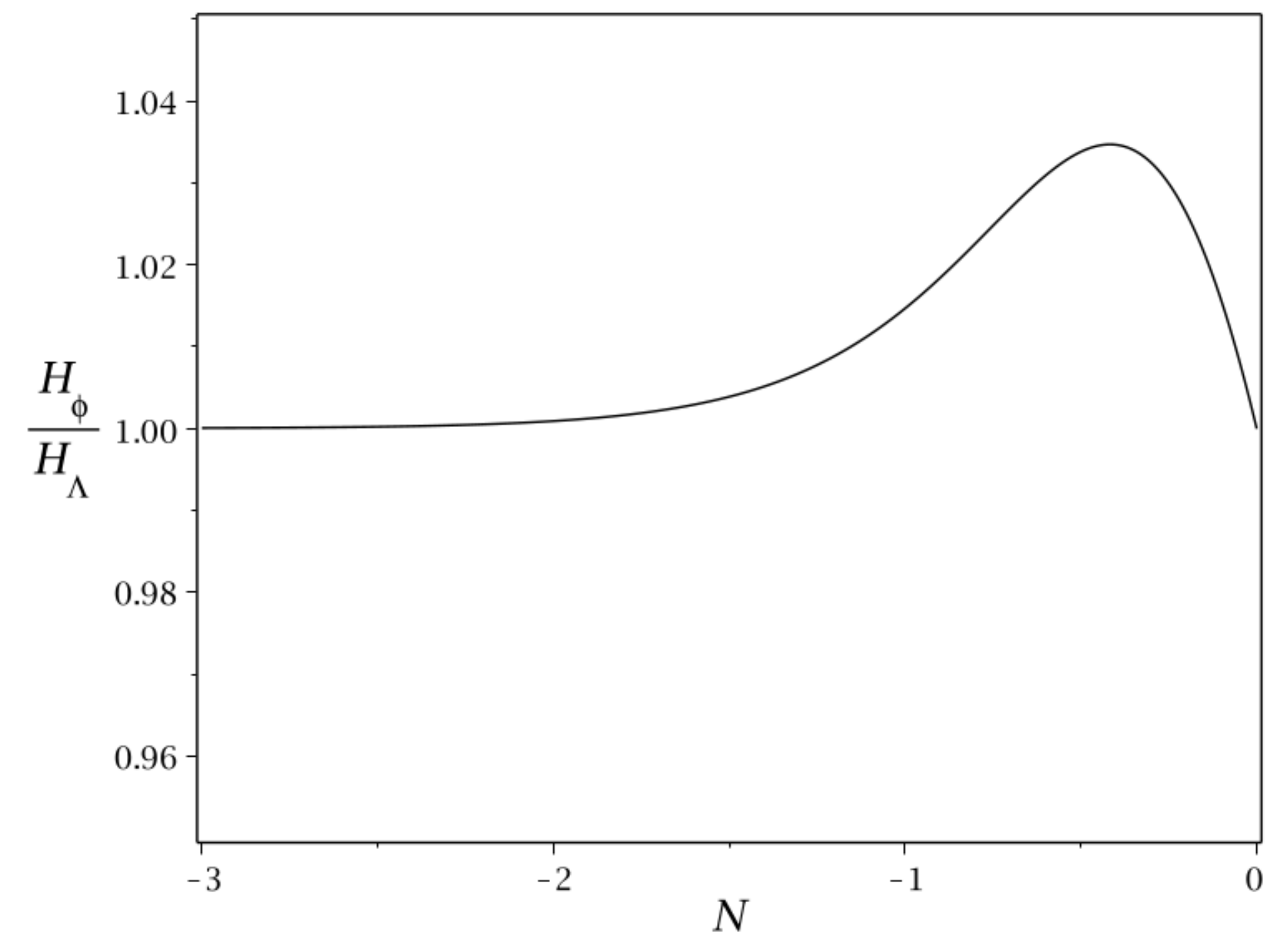}}
		\vspace{-0.5cm}
	\end{center}
	\caption{The ET model with $\alpha=\frac73$ and $\nu=128$. Figure (a) depicts the evolution in $\varphi$
     during the inflationary and quintessence epochs, for an observationally viable freezing value of $\varphi$.
     The state space representation of the quintessence epoch of this solution, projected onto
     $(\bar{\varphi},u,v)$, is depicted in figure (b), while the associated graphs $w_\varphi(N)$
     (lower curve) and $w_\mathrm{eff}(N)$ (upper curve)
     are shown in figure (c); finally, $\frac{H_\varphi}{H_{\Lambda}}(N)$ is shown in figure (d).
}\label{fig:6}
\end{figure}

As another example, we now fix $\nu=128$, and keep $\alpha=7/3$. For the $\mathrm{EC}$ and $\mathrm{ET}$
models we obtain the best fit values $\varphi_*\approx 9.594$ ($\bar{\varphi}_*\approx0.988$)
and $\varphi_*\approx 9.214$ ($\bar{\varphi}_*\approx0.986$), respectively.
Figures~\ref{fig:5} and~\ref{fig:6} show the respective orbits in the state space
(together with the inflationary attractor CM orbit evolution in $\varphi$
in Figures~\ref{fig:VarphiEC} and~\ref{fig:VarphiET}), and
the corresponding graphs for $w_\varphi$, and $H_\varphi/H_{\Lambda}$, where we have used
the relevant comparison formulas between $\Lambda$CDM and radiation and the present models
given in Appendix~\ref{app:LCDMrad}.

%

\section{Discussion\label{sec:disc}}

In this paper we have introduced a regular dynamical system
for quintessential $\alpha$-attractor inflation on a compact state space.
This has the advantage that powerful dynamical systems methods can be
applied and this enabled us to obtain (i) a qualitative
description of the associated entire solution space and its
properties, including asymptotic features, (ii) approximations of solutions,
(iii) an analytically based systematic quantitative numerical investigation
of the quintessential $\alpha$-attractor EC and ET global
solution space structure. For the present models the latter involves
two tasks: (a) Connecting orbits to the fixed points from which
the physically viable solutions originate and end by varying
constants in expressions obtained from local analysis of such fixed points.
(b) Identifying the cosmological viable solutions in the global state space.

In particular, this has enabled us to construct a new method for
obtaining, in a systematic manner, initial data for viable quintessence
evolution. We have also clarified how slow-roll approximations are connected with
approximations based on center manifold theory, and in a future paper
we will give new accurate and simple approximations for quintessence
evolution, using similar dynamical systems based approximation methods
as for inflationary evolution, which will further clarify and simplify
the identification of observational viable quintessence models and
evolution.

\subsection*{Acknowledgments}
We thank Yashar Akrami for helpful comments concerning the numerical calculations in~\cite{akretal20}.
A. A. is supported by FCT/Portugal through CAMGSD, IST-ID, projects UIDB/04459/2020 and UIDP/04459/2020.

\begin{appendix}

\section{Asymptotic features for the $\alpha$-attractor inflation potentials\label{app:asymp}}

The potentials in~\eqref{pot2exp} for the LC, LT, EC, ET models are characterized by
\begin{subequations}\label{barvexamples}
\begin{alignat}{3}
\text{LC}\!:\quad \bar{V} &= \frac{1 + \nu(1 - \bar{\varphi})}{1 + 2\nu}, &\qquad
V_+ &= \frac{V_-}{1 + 2\nu}, &\qquad \lambda &= \frac{\nu}{\sqrt{6\alpha}}
\left(\frac{1-\bar{\varphi}^2}{1 + \nu(1 - \bar{\varphi})}\right),\\
\text{LT}\!:\quad \bar{V} &= \frac12 (1 - \bar{\varphi}), &\qquad
V_+ &= 0, &\qquad \lambda &= \frac{1}{\sqrt{6\alpha}}(1 + \bar{\varphi}),\\
\text{EC}\!:\quad \bar{V} &= e^{-\nu(1 + \bar{\varphi})}, &\qquad
V_+ &= e^{-2\nu}V_-, &\qquad
\lambda &= \frac{\nu}{\sqrt{6\alpha}}\left(1 - \bar{\varphi}^2\right), \label{lambdaEC}\\
\text{ET}\!:\quad \bar{V} &= \frac{e^{-\nu(1 + \bar{\varphi})} - e^{-2\nu}}{1 - e^{-2\nu}}, &\qquad
V_+ &= 0, &\qquad \lambda &= \frac{\nu}{\sqrt{6\alpha}}
\left(\frac{1 - \bar{\varphi}^2}{1 - e^{-\nu(1 - \bar{\varphi})}}\right),\label{lambdaET}
\end{alignat}
\end{subequations}
when given in $\bar{\varphi}$, which leads to the following asymptotic potential features
\begin{subequations}\label{lambdaMinusPot}
\begin{alignat}{2}
\text{LC}\!:\quad \lambda^{(1)}_{-} &= \frac{2\nu}{\sqrt{6\alpha}(1+2\nu)}, &\qquad
\lambda^{(2)}_{-} &= -\frac{\lambda^{(1)}_{-}}{(1+2\nu)}, \\
\text{LT}\!:\quad	\lambda^{(1)}_{-} &= \frac{1}{\sqrt{6\alpha}} , &\qquad \lambda^{(2)}_{-} &= 0, \\
\text{EC}\!:\quad	\lambda^{(1)}_{-} &= \frac{2\nu}{\sqrt{6\alpha}},  &\qquad
\lambda^{(2)}_{-} &= -\lambda^{(1)}_{-},\\
\text{ET}\!:\quad\lambda^{(1)}_{-} &= \frac{2\nu}{\sqrt{6\alpha}(1 - e^{-2\nu})}, &\qquad
\lambda^{(2)}_{-} &= -\left(1 - \frac{2\nu e^{-2\nu}}{1 - e^{-2\nu}}\right)\lambda^{(1)}_{-},
\end{alignat}
\end{subequations}
and
\begin{subequations}\label{lambdaPlusPot}
\begin{alignat}{3}
\text{LC}\!:\quad \lambda_+ & =0 , &\qquad \lambda^{(1)}_{+} &= -\frac{2\nu}{\sqrt{6\alpha}}, &\qquad
\lambda^{(2)}_{+} &= (1+2\nu)\lambda^{(1)}_{+}, \\
\text{LT}\!:\quad \lambda_+ &= \frac{2}{\sqrt{6\alpha}}, &\qquad \lambda^{(1)}_{+} &= \frac{1}{\sqrt{6\alpha}}, &\qquad
\lambda^{(2)}_{+} &= 0, \\
\text{EC}\!:\quad \lambda_+ &=0 , &\qquad \lambda^{(1)}_{+} &= -\frac{2\nu}{\sqrt{6\alpha}} , &\qquad
\lambda^{(2)}_{+} &= \lambda^{(1)}_{+}, \\
\text{ET}\!:\quad \lambda_+ &= \frac{2}{\sqrt{6\alpha}} , &\qquad \lambda^{(1)}_{+} &= \frac{1-\nu}{\sqrt{6\alpha}} , &\qquad \lambda^{(2)}_{+} &= -\frac{(1-\frac{\nu}{3})\nu}{\sqrt{6\alpha}}.
\end{alignat}
\end{subequations}
%

\section{Exact relations and $\Lambda$CDM comparisons\label{app:LCDMrad}}

In order to compare a scalar field model with the  $\Lambda$CDM model we, for simplicity,
identify the models at the present time as regards the rate of expansion and the matter content.
Specifically we require that
\begin{equation}
H_{\varphi,0} = H_{\Lambda,0} = H_0,\qquad
\Omega_{\Lambda,0} = \Omega_{\varphi,0} = 1 - \Omega_{\gamma,0} - \Omega_{m,0},
\end{equation}
where $\Omega_{\Lambda,0} := \Lambda/3H_0^2$, while
$H_0$, $\Omega_{\gamma,0}$ and $\Omega_{m,0}$ are the observed Hubble parameter and
the dimensionless Hubble-normalized radiation and matter densities at the present time.
%
%
Common for scalar field models and $\Lambda$CDM with radiation is also
$\rho_\gamma = 3H_0^2\Omega_{\gamma,0} e^{-4N}$,
$\rho_\mathrm{m} = 3H_0^2\Omega_{\mathrm{m},0} e^{-3N}$ and hence
\begin{equation}
w_{\gamma\mathrm{m}} = \frac13\left(\frac{1}{1 + \left(\frac{\Omega_\mathrm{m,0}}{\Omega_{\gamma,0}}\right)e^{N}}\right).
\end{equation}

We then define
\begin{equation}
E_\varphi(N) := \frac{H_\varphi(N)}{H_0},\qquad E_\Lambda(N) := \frac{H_\Lambda(N)}{H_0},
\end{equation}
and note that
\begin{equation}\label{EE}
E_\varphi^2 = \frac{\Omega_{\gamma,0} e^{-4N} + \Omega_{\mathrm{m},0}e^{-3N}}{1 - \Omega_\varphi},\qquad
E_\Lambda^2 = \Omega_{\gamma,0}e^{-4N} + \Omega_{\mathrm{m},0}e^{-3N} + \Omega_{\Lambda,0},
\end{equation}
where we recall that $\Omega_\varphi = 3v^2$.
Due to eq.~\eqref{EE} it follows that
\begin{equation}
\left(\frac{H_\varphi}{H_\Lambda}\right)^2 = \left(\frac{E_\varphi}{E_\Lambda}\right)^2 =
\frac{\Omega_{\gamma,0} e^{-4N} + \Omega_{\mathrm{m},0}e^{-3N}}
{(1 - \Omega_\varphi)(\Omega_{\gamma,0}e^{-4N} + \Omega_{\mathrm{m},0}e^{-3N} + \Omega_{\Lambda,0})}.
\end{equation}

To obtain a feeling for the initial value of $N = N_\mathrm{i}$, note
that
\begin{equation}
N = \ln\left[\left(\frac{\rho_\mathrm{m}}{\rho_\gamma}\right)
\left(\frac{\Omega_{\gamma,0}}{\Omega_{\mathrm{m},0}}\right)\right].
\end{equation}
Thus, setting, {\it e.g.}, $\Omega_{\gamma,0} = 10^{-5}$ and $\Omega_{\mathrm{m},0}=0.32$,
results in
\begin{subequations}
\begin{align}
\frac{\rho_\mathrm{m}}{\rho_\gamma} &\in [10^{-2},10^2] \quad \Rightarrow \quad
N \in [-14.98,-5.77],\\
\frac{\rho_\mathrm{m}}{\rho_\gamma} &= 1 \quad \Rightarrow \quad N = N_\mathrm{eq} = - 10.37.
\end{align}
\end{subequations}

The cosmological redshift, $z$, is kinematically determined by
\begin{equation}
z = \frac{a_0}{a} - 1 = e^{-N} - 1 \qquad \Rightarrow \qquad N = - \ln(1+z).
\end{equation}
Thus, {\it e.g.}, $z=0.4$, $z=1$, $z=1.5$, $z=10$, $z=147.4$, $z=1100$, $z=10^{10}$ correspond to
$N=-0.34$, $N=-0.69$, $N=-0.92$, $N=-2.40$, $N=-5.00$, $N=-7.00$, $N=-23.03$, respectively.

\end{appendix}

\bibliographystyle{unsrt}
\bibliography{../Bibtex/cos_pert_papers}

\begin{thebibliography}{10}

\bibitem{rieetal98}
A.~G.~Riess et~al.
\newblock Observational evidence from supernovae for an accelerating universe
  and a cosmological constant.
\newblock {\em Astron. J.}, {\bf 116}:1009, 1998.

\bibitem{peretal99}
S.~Perlmutter et~al.
\newblock Measurements of omega and lambda from 42 high redshift supernovae.
\newblock {\em Astron. J.}, {\bf 517}:565, 1999.

\bibitem{planck18}
Planck Collaboration.
\newblock Planck 2018 results.{ VI}. cosmological parameters.
\newblock {\em arXiv:1807.06209 [astro-ph.CO]}, 2018.

\bibitem{plaX18}
Planck Collaboration: Y.~Akrami et~al.
\newblock Planck 2018 results. x. constraints on inflation.
\newblock {\em arXiv:1807.06211 [astro-ph.CO]}, 2018.

\bibitem{alametal17}
S.~Alam et~al.
\newblock The clustering of galaxies in the completed sdss-iii baryon
  oscillation spectroscopic survey: cosmological analysis of the dr12 galaxy
  sample.
\newblock {\em MNRAS}, {\bf 470}:2617, 2017.

\bibitem{AbbDES20}
T.~M. C.~Abbott et~al. DES~Collaboration.
\newblock Dark energy survey year 1 results: Cosmological constraints from
  cluster abundances and weak lensing.
\newblock {\em Phys. Rev. D}, 102:023509, 2020.

\bibitem{rieetal19}
A.~G. Riess, S.~Casertano, W.~Yuan, L.~M. Macri, and D.~Scolnic.
\newblock Large magellanic cloud cepheid standards provide a 1{\%} foundation
  for the determination of the hubble constant and stronger evidence for
  physics beyond $\lambda${CDM}.
\newblock {\em The Astrophysical Journal}, {\bf 876}(1):85, 2019.

\bibitem{peevil99}
P.~J.~E. Peebles and A.~Vilenkin.
\newblock Quintessential inflation.
\newblock {\em Phys. Rev. D}, {\bf 59}:063505, 1999.

\bibitem{maretal14}
J.~Martin, C.~Ringeval, and V.~Vennin.
\newblock Encyclop{\ae}dia inflationaris.
\newblock {\em Physics of the Dark Universe}, {\bf 5}:75, 2014.

\bibitem{galetal15}
M.~Galante, R.~Kallosh, A.~Linde, and D.~Roest.
\newblock Unity of cosmological inflation attractors.
\newblock {\em Phys. Rev. Lett.}, {\bf 114}:141302, 2015.

\bibitem{kallin15b}
R.~Kallosh and A.~Linde.
\newblock Escher in the sky.
\newblock {\em Comptes Rendus Physique}, {\bf 16}:914--927, 2015.

\bibitem{ferkal16}
S.~Ferrara and R.~Kallosh.
\newblock Seven-disc manifold $\alpha$-attractors and b modes.
\newblock {\em Phys. Rev. D}, {\bf 94}:0126015, 2016.

\bibitem{kaletal17}
R.~Kallosh, A.~Linde, D.~Roest, and Y.~Yamada.
\newblock D3 induced geometric inflation.
\newblock {\em JHEP}, {\bf 07}:057, 2017.

\bibitem{kaletal17b}
R.~Kallosh, A.~Linde, T.~Wrase, and Y.~Yamada.
\newblock Maximal supersymmetry and b-mode targets.
\newblock {\em JHEP}, {\bf 04}:144, 2017.

\bibitem{akretal20}
Y.~Akrami et~al.
\newblock Quintessential $\alpha$-attractor inflation: forecasts for stage iv
  galaxy surveys.
\newblock {\em JCAP}, {\bf 04}:006, 2021.

\bibitem{dimowe17}
K.~Dimopoulos and C.~Owen.
\newblock Quintessential inflation with $\alpha$-attractors.
\newblock {\em J. of Cosmology and Astroparticle Physics}, {\bf 06}:027, 2017.

\bibitem{dimetal18}
K.~Dimopoulos, L.~D. Wood, and C.~Owen.
\newblock Instant preheating in quintessential inflation with
  $\ensuremath{\alpha}$-attractors.
\newblock {\em Phys. Rev. D}, {\bf 97}:063525, Mar 2018.

\bibitem{akretal18}
Y.~Akrami et~al.
\newblock Dark energy, $\alpha$-attractors, and large-scale structure surveys.
\newblock {\em JCAP}, {\bf 06}:041, 2018.

\bibitem{kaletal14}
R.~Kallosh, A.~Linde, and D.~Roest.
\newblock Large field inflation and double $\alpha$-attractors.
\newblock {\em J. High Energ. Phys.}, {\bf 08}:052, 2014.

\bibitem{kallin3b}
R.~Kallosh and A.~Linde.
\newblock Universality class in conformal inflation.
\newblock {\em J. of Cosmology and Astroparticle Physics}, {\bf 2013}:1475,
  2013.

\bibitem{caretal15a}
J.~J.~M. Carrasco, R.~Kallosh, and A.~Linde.
\newblock Cosmological attractors and initial conditions for inflation.
\newblock {\em Phys. Rev. D}, {\bf 92}:063519, 2015.

\bibitem{caretal15b}
J.~J.~M. Carrasco, R.~Kallosh, and A.~Linde.
\newblock $\alpha$-attractors: Planck, lhc and dark energy.
\newblock {\em J. High Energ. Phys.}, {\bf 147}, 2015.

\bibitem{lin15}
A.~Linde.
\newblock Single-field $\alpha$-attractors.
\newblock {\em J. of Cosmology and Astroparticle Physics}, {\bf 2015}:003,
  2015.

\bibitem{lin17}
A.~Linde.
\newblock Gravitational waves and large field inflation.
\newblock {\em J. of Cosmology and Astroparticle Physics}, {\bf 2017}:006,
  2017.

\bibitem{lin17b}
A.~Linde.
\newblock Random potentials and cosmological attractors.
\newblock {\em J. of Cosmology and Astroparticle Physics}, {\bf 2017}:028,
  2017.

\bibitem{Sta80}
A.A. Starobinsky.
\newblock A new type of isotropic cosmological models without singularity.
\newblock {\em Physics Letters B}, 91(1):99--102, 1980.

\bibitem{BC88}
John~D. Barrow and S.~Cotsakis.
\newblock Inflation and the conformal structure of higher-order gravity
  theories.
\newblock {\em Physics Letters B}, 214(4):515--518, 1988.

\bibitem{BS08}
Fedor Bezrukov and Mikhail Shaposhnikov.
\newblock The standard model higgs boson as the inflaton.
\newblock {\em Physics Letters B}, 659(3):703--706, 2008.

\bibitem{alhetal19b}
A.~Alho, C.~Uggla, and J.~Wainwright.
\newblock Dynamical systems in perturbative scalar field cosmology.
\newblock {\em Class. Quantum Grav.}, {\bf 37}(22):225011, 2020.

\bibitem{alhetal15}
A.~Alho, J.~Hell, and C.~Uggla.
\newblock Global dynamics and asymptotics for monomial scalar field potentials
  and perfect fluids.
\newblock {\em Class. Quant. Grav.}, {\bf 32}(14):145005, 2015.

\bibitem{alhugg15b}
A.~Alho and C.~Uggla.
\newblock Scalar field deformations of lambda-cdm cosmology.
\newblock {\em Phys. Rev. D}, {\bf 92}(10):103502, 2015.

\bibitem{alhugg17}
A.~Alho and C.~Uggla.
\newblock Inflationary $\alpha$-attractor cosmology: A global dynamical systems
  perspective.
\newblock {\em Phys. Rev. D}, {\bf 95}(8):083517, 2017.

\bibitem{alhetal19a}
A.~Alho, C.~Uggla, and J.~Wainwright.
\newblock Perturbations of the lambda-cdm model in a dynamical systems
  perspective.
\newblock {\em JCAP}, {\bf 09}:045, 2019.

\bibitem{alhetal23}
A.~Alho, C.~Uggla, and J.~Wainwright.
\newblock Quintessence in a state space perspective.
\newblock {\em Physics of the Dark Universe}, {\bf 39}:101146, 2023.

\bibitem{collins71}
C.~B. Collins.
\newblock More qualitative cosmology.
\newblock {\em Comm. Math. Phys.}, {\bf 23}(2):137--158, 1971.

\bibitem{BN73}
O.~I. Bogoyavlenskii and S.~P. Novikov.
\newblock Singularities of the cosmological model of the bianchi ix type
  according to the qualitative theory of differential equations.
\newblock {\em JETP}, {\bf 37}(5):747--755, 1973.

\bibitem{waiell97}
J.~Wainwright and G.~F.~R. Ellis.
\newblock {\em Dynamical systems in cosmology}.
\newblock Cambridge University Press, 1997.

\bibitem{col03}
A.~A. Coley.
\newblock {\em Dynamical systems and cosmology}.
\newblock Kluwer Academic Publishers, Dordrecht, 2003.

\bibitem{bahetal18}
S.~Bahamonde, C.~G. B{\"o}hmer, S.~Carloni, E.~J. Copeland, Wei Fang, and
  N.~Tamanini.
\newblock Dynamical systems applied to cosmology: Dark energy and modified
  gravity.
\newblock {\em Physics Reports}, {\bf 775-777}:1--122, 2018.

\bibitem{galetal23}
Giacomo Galloni, Nicola Bartolo, Sabino Matarrese, Marina Migliaccio, Angelo
  Ricciardone, and Nicola Vittorio.
\newblock Updated constraints on amplitude and tilt of the tensor primordial
  spectrum.
\newblock {\em Journal of Cosmology and Astroparticle Physics}, 2023(04):062,
  apr 2023.

\bibitem{joypro98}
Michael Joyce and Tomislav Prokopec.
\newblock Turning around the sphaleron bound: Electroweak baryogenesis in an
  alternative post-inflationary cosmology.
\newblock {\em Phys. Rev. D}, {\bf 57}:6022--6049, 1998.

\bibitem{stetur84}
P.~J. Steinhardt and M.~S. Turner.
\newblock Prescription for successful new inflation.
\newblock {\em Phys. Rev. D}, 29:2162--2171, 1984.

\bibitem{lidlyt93}
A.~R. Liddle and D.~H. Lyth.
\newblock The cold dark matter density perturbations.
\newblock {\em Physics Reports}, {\bf 231}:1--105, 1993.

\bibitem{lidetal94}
A.~R. Liddle, P.~Parsons, and J.~D. Barrow.
\newblock Formalizing the slow-roll approximation in inflation.
\newblock {\em Phys. Rev. D}, {\bf 50}:7222, 1994.

\bibitem{alhugg15}
A.~Alho and C.~Uggla.
\newblock Global dynamics and inflationary center manifold and slow-roll
  approximants.
\newblock {\em Journal of Mathematical Physics}, {\bf 56}(012502), 2015.

\bibitem{remcar13}
G.N. Remmen and S.M. Carroll.
\newblock Attractor solutions in scalar-field cosmology.
\newblock {\em Phys. Rev. D}, {\bf 88}:083518, 2013.

\bibitem{remcar14}
G.N. Remmen and S.M. Carroll.
\newblock How many e-folds should we expect from high-scale inflation?
\newblock {\em Phys. Rev. D}, {\bf 90}:063517, 2014.

\bibitem{gruslo16}
R.~Grumitt and D.~Sloan.
\newblock Measures in mutlifield inflation, 2016.

\end{thebibliography}

\end{document}